\documentclass[
 reprint,=
 amsmath,amssymb,
 aps,unsortedaddress,prb]{revtex4-1}
\usepackage{xcolor}
\usepackage{graphicx}% Include figure files
\usepackage{dcolumn}% Align table columns on decimal point
\usepackage{bm}% bold math
\usepackage{amsmath}
\usepackage{subfigure}

\newcommand{\be}{\begin{equation} }
\newcommand{\ee}{\end{equation} }
\newcommand{\ba}{\begin{eqnarray} }
\newcommand{\ea}{\end{eqnarray} }

\newcommand{\mac}{\mathcal}

\newcommand{\bit}{\begin{itemize}}
\newcommand{\eit}{\end{itemize}}

\begin{document}

\title{Multipolar Topological Field Theories: Bridging Higher Order Topological Insulators and Fractons}

\author{Yizhi You}
\affiliation{Princeton Center for Theoretical Science, Princeton University, 
NJ, 08544, USA}

\author{F.~J. Burnell}
\affiliation{Department of Physics, University of Minnesota Twin Cities, 
MN, 55455, USA}

\author{Taylor L. Hughes}
\affiliation{Department of Physics and Institute for Condensed Matter Theory, University of Illinois at Urbana-Champaign, Illinois 61801, USA}

\date{\today}
\begin{abstract}
Two new recently proposed classes of topological phases, namely fractons and higher order topological insulators (HOTIs), share at least superficial similarities.
The wide variety of proposals for these phases calls for a universal field theory description that captures their key characteristic physical phenomena. In this work, we construct topological multipolar response theories that capture the essential features of some classes of fractons and higher order topological insulators. Remarkably, we find that despite their distinct symmetry structure, some classes of fractons and HOTIs can be connected through their essentially identical topological response theories.
 More precisely, we propose a topological quadrupole response theory that describes both a 2D symmetry enriched fracton phase and a related bosonic quadrupolar HOTI with strong interactions. Such a topological quadrupole term encapsulates the protected corner charge modes and, for the HOTI, predicts an anomalous edge with fractional dipole moment. In 3D we propose a dipolar Chern-Simons theory with a quantized coefficient as a description of the response of both second order HOTIs harboring chiral hinge currents, and of a related fracton phase. This theory correctly predicts chiral currents on the hinges and anomalous dipole currents on the surfaces. We generalize these results to higher dimensions to reveal a family of multipolar Chern-Simons terms and related $\theta$-term actions that can be reached via dimensional reduction or extension from the Chern-Simons theories.

\end{abstract}

\maketitle

\tableofcontents

\section{Introduction}
A decade of intense effort has resulted in a thorough classification and characterization of symmetry protected topological materials. For a refined classification of topological insulators and superconductors, along with their bosonic analogs, the concept of symmetry protection has been extended to include spatial symmetries~\cite{fu2011topological,hughes2011,hsieh2012topological,cheng2016translational,ando2015topological,slager2013space,hong2017topological,qi2015anomalous,huang2017building,teo2013existence,song2017topological,watanabe2017structure,po2017symmetry,isobe2015theory}. In addition to protected gapless boundary modes, some topological crystalline phases admit gapped edges or surfaces separated by gapless corners or hinges. Exemplifying a much richer bulk-boundary correspondence, insulators of this type are now termed higher-order topological 
insulators (HOTIs)~\cite{benalcazar2017quantized,benalcazar2017electric,schindler2017higher,langbehn2017reflection,song2017d,song2017d}. Realizations of quadrupole HOTIs have appeared in a variety meta-material contexts\cite{serra2018,peterson2018quantized,imhof2018}, and there is tantalizing evidence for the existence of a 3D HOTI with hinge states in bismuth as well\cite{schindler2018higher}. 

While the classification and characterization of strongly interacting higher order topological phases (or topological crystalline phases, broadly defined) has been widely-explored\cite{isobe2015theory,song2017interaction,song2017topological,you2018higher,rasmussen2018intrinsically,rasmussen2018classification,thorngren2018gauging,benalcazar2018quantization,araki2019mathbb,tiwari2019unhinging,you2018higher,wieder2018axion}, the connection between HOTIs and topological response phenomena is still nebulous. In the prominent examples of topological insulators, the 2D Chern-Simons response predicts a quantized Hall conductivity in $\mathcal{T}$-breaking insulators, while 3D $\mathcal{T}$-invariant TIs can exhibit quantized axion electrodynamics\cite{axion1,axion2}, e.g., a topological magnetoelectric effect. These topological response coefficients are quantized signatures of the symmetry protected topology, and are thus robust to any perturbations as long as symmetry is preserved and there are no phase transitions. Thus, quantized topological response coefficients can be treated as smoking-gun experimental characteristics of TIs. 

Remarkably, the topological field theory descriptions of TIs in various dimensions, and in various symmetry classes, are connected through a beautiful dimensional hierarchy, e.g., the boundary of a 3D TI with axion electrodynamics contains a 2D Chern-Simons theory at level $1/2$, suggesting the (previously-known) existence a single Dirac cone at the boundary\cite{axion2}. Meanwhile, the 2D Chern-Simons response on a thin torus can be viewed as the charge polarization response of a 1D TI\cite{axion2}, and the Laughlin gauge argument\cite{laughlin1981} for the 2D Hall current maps onto Thouless charge pumping\cite{thouless1983} in the 1D system. It remains unclear to what extent such a dimensional hierarchy can be generalized to HOTIs and their interacting descendants.

In this work, we address these open issues by introducing a set of topological multipole field theories that can be viewed as the higher order generalization of the topological electromagnetic response in TIs. Parallel to the relationship between the 1D charge polarization and 2D Chern-Simons theory, the HOTIs we consider exhibit a topological 2D quadrupolar response (polarization of dipoles) term\cite{wheeler2018many,kang2018many,trifunovic2019geometric,ono2019difficulties}, and a 3D dipolar Chern-Simons term that can be interpreted as the topological response of  a chiral hinge insulator. These topological dipole responses exactly match the microscopic phenomenology of the protected gapless modes at the corner/hinge, which exemplify the underlying bulk-boundary correspondence in 2D and 3D HOTIs respectively.
Additionally, despite the fact that the edge/surface of a HOTI is gappable, our field theory implies that such gapped boundary regions are still anomalous in certain symmetry classes.
Finally, we complete the analogy by briefly considering an analogy to a 4D axion dipole electrodynamics.

To be more explicit, our 2D quadrupole response theory for HOTIs describes fractionalized dipoles on the edges that are anomalous under certain symmetry conditions (comparable to fractional charge at the ends of a 1D system). Such an anomalous edge carrying a fractional dipole moment is linked to the existence of protected corner modes (or fractional corner charges depending on the symmetry), both of which are characteristic signatures  of the 2D HOTI, and can persist in the presence of arbitrarily strong interactions. 

Likewise, the dipolar Chern-Simons term in 3D predicts a current anomaly on the hinges of a sample, which exactly matches the phenomenology of the chiral hinge currents on a 3D 2nd order HOTI with $C_4\mathcal{T}$ symmetry. The dipolar Chern-Simons response theory produces a transverse charge current ($J_z$) in response to some configurations of electric field gradients ($\partial_x E_x+\partial_y E_y)$, as well as dipole currents $(J_{x}^y, J_{y}^{x})$ in the presence of an electric field $(E_z)$.  We prove that the level of the dipole Chern-Simons theory is quantized, thus we expect the coefficient to be robust even in the presence of strong interactions. Indeed, just as the 2D Hall conductance is dimensionless, so is the 3D dipolar Chern-Simons coefficient, having units of $e^2/h$.
Aside from the transverse current response, our dipolar Chern-Simons term also predicts that a 3D HOTI will have a bulk magnetic quadrupole moment in an electrostatic potential\cite{raab2005,shitade2018,gao2018}. The magnetic quadrupole moment manifests in the bound, circulating hinge currents, and can be probed by placing the system in a magnetic field gradient. 

Interestingly, we can also make a connection between our 2D quadrupole response and our 3D dipolar Chern-Simons response.  In a thin-annulus limit, the 3D dipolar Chern-Simons response dimensionally reduces to the 2D charge quadrupole response, and the anomalous hinge currents when a flux is threaded through the periodic direction correspond to a shift of the 2D quadrupole moment, i.e., a they are the result of an effective dipole current. Such a dimensional reduction structure stemming from the dipolar Chern-Simons theory sheds light on the connection between the 2D quadrupole insulator and the 3D HOTI with chiral hinge currents, that has already been understood in free fermion systems\cite{benalcazar2017electric}.

In order to propose our topological response theories we make a connection between HOTIs and parallel research in fracton phases. Recently, distinct long-range entangled states, transcending the conventional TQFT paradigm, and termed fracton phases, have been discovered and intensively studied via exactly solvable models, including quantum stabilizer codes and higher rank gauge theories \cite{Haah2011-ny,Halasz2017-ov,Vijay2016-dr,Vijay2015-jj,Chamon2005-fc,shirley2018fractional,Slagle2017-ne,Ma2017-qq,Hsieh2017-sc,Vijay2017-ey,Slagle2017-gk,Williamson2016-lv,Ma2017-cb,you2019emergent}. Earlier literature shows that topological fracton order shares many features of topological order, including long-range entangled ground states, and non-trivial braiding statistics. At the same time, fracton phases have a subextensive ground-state degeneracy depending on the system size and lattice topology, which is beyond the paradigm of topological quantum field theory. Fracton phases have quasiparticles with restricted mobility such that they move only within lower-dimensional manifolds \cite{Ma2017-qq,Slagle2017-gk,Ma2017-cb,pretko2017fracton,ma2018fracton,bulmash2018higgs,Slagle2017-la,shirley2017fracton,prem2018pinch,slagle2018symmetric,gromov2017fractional,pai2018fractonic,pretko2017subdimensional,ma2018higher,pretko2017generalized,pretko2017finite,you2018symmetric}. The subdimensional nature of fracton excitations is a consequence of subsystem conservation laws whose associated charges are conserved on a sub-manifold such as planes, lines, or fractals. Such strongly constrained motion gives rise to unconventional features including glassiness and subdiffusive dynamics \cite{Chamon2005-fc,prem2017-ql}.

The most salient, unifying property in fracton phases is the immobility of individual charged particles.  This immobility results from requiring that the charge be conserved in some set of subsystems, such as lines, fractals, or planes, rather than in the system as a whole.  Consequently, charge motion is frozen in fracton models (at least in some directions), and there is no charge conductivity.  Instead, for phases with $U(1)$ subsystem symmetries the leading order transport responses are necessarily dipolar. %For a symmetry enriched fracton phase, e.g., with $U(1)$ subsystem symmetries and time reversal,  the topological response manifests at the dipole level.  
Similarly, in a HOTI, the topological charge response in such insulators is absent -- even though the charge is not immobile. Instead, there appears to be a topological dipole response which describes the phenomenology associated to gapped boundaries, and gapless corner or hinge modes. Since the dipole response is most natural in a fracton phase, we propose a response theory for 2D systems with $U(1)$ subsystem symmetry, and then connect the resulting theory to HOTIs when the subsystem symmetry is broken down to a global $U(1)$ symmetry. In this sense, the HOTI can be regarded as a trivial charge insulator but a topological dipole insulator. We further show that a similar framework leads to a topological dipolar response theory of the 3D HOTI with chiral hinge modes.

The remainder of the paper is structured as follows. In Section II, we discuss subsystem protected SPTs with a quantized quadrupolar response and relate them to bosonic 2D quadrupolar HOTIs. The quantized bulk quadrupole moment engenders a protected corner mode and a fractional edge dipole which is anomalous under $C_4 \times \mathcal{T}$. In particular, we propose that the quadrupolar HOTI-type phenomenon can exist in symmetry enriched Fracton system where the corner mode is protected by subsystem U(1) symmetry. The resultant field theory contains a topological quadrupole moment $\theta$-term constructed from a higher rank gauge field. In Section III, we propose a dipolar Chern-Simons theory for 3D HOTIs with gapless chiral hinge currents and compare it with a related Fracton theory described by a rank-2 tensor gauge theory. Several experimental signatures including a transverse topological dipole and current response and bulk magnetic quadrupole moment can be explicitly obtained from this dipolar Chern-Simons term. In section IV, we study the generalization and extensions of the aforementioned topological multipole responses including topological octupolarization and multipolar Chern-Simons and axion electrodynamics in higher dimensions. We also have several appendices containing some of the detailed calculations.

\section{2D Quadrupolar Response}

In this section, we introduce two classes of 2D gapped bosonic models that each host protected corner modes. The first class can be interpreted as a `symmetry enriched fracton phase' (or alternatively a subsystem symmetry protected HOTI)\cite{you2018subsystem,you2018symmetric,devakul2018strong,devakul2018fractal}. This phase has gapless corner modes that are protected by the combination of $U(1)$ subsystem symmetry, and time-reversal ($\mathcal{T}$). The second class we study was introduced in Refs.~\onlinecite{you2018higher,2018arXiv180709781D} and describes a crystalline bosonic HOTI protected by $ U(1) \times \mathcal{T}  \times C_4$ symmetry, where here $U(1)$ is global boson charge conservation, and $C_4$ is a discrete rotation symmetry.  We will show that, despite the difference in symmetry protection, both classes support gapless corner modes. Furthermore, they both give rise to a 2D charge response with a quantized quadrupole moment $q_{xy}$\cite{benalcazar2017quantized,wheeler2018many,kang2018many}.

\subsection{Symmetry enriched fracton phase with corner modes}

The most well-studied higher order topological phases require spatial symmetries such as spatial rotations or reflections for protection. Without spatial symmetry one can typically hybridize and remove the spatially separated corner(hinge) modes through an edge(surface) phase transition without the bulk gap closing. In this subsection, we introduce a new type of interacting higher order topological phase that does not require spatial symmetry, but instead relies on subsystem symmetry for protection.

In $D$ spatial dimensions, a subsystem symmetry consists of independent symmetry operations acting on a set of d-dimensional subsystems with $0<d<D$. In $D=2$, the subsystems in question can be lines ($d=1$) or fractals \cite{Xu2004-oj,you2018subsystem,devakul2018fractal,Vijay2016-dr}. Here we will be interested in linear subsystems consisting of all lines parallel to the $\hat{x}$ and $\hat{y}$ axes on a square lattice. The corresponding $d=1$ subsystem symmetry is associated with a  quantum number that is conserved separately on each line, leading to interesting new possibilities for both symmetry-breaking \cite{Xu2004-oj} and symmetry-protected topological \cite{you2018subsystem,devakul2018fractal,devakul2018strong,shirley2018foliated,you2019higher} phases.  Typically subsystem symmetry protected topolgogical phases contain gapless degrees of freedom  along the edges of a 2D lattice.  However, here we describe a model for a new type of subsystem SPT, which has protected gapless modes only at the corners of the 2D square lattice.  Though both types of subsystem SPTs have connections to HOTI phases \cite{you2018subsystem}, the example we present here is illuminating because it allows for a direct field-theoretic connection between subsystem SPT's and HOTIs having $U(1)$ and $\mathcal{T}$ symmetries.

%and therefore strongly restrict the type of couplings in a Hamiltonian. 
%In particular, the subsystem symmetry protected HOTI that we describe exists only in strongly interacting systems without a non-interacting counterpart. 

The specific model we will consider is essentially a 2D generalization of the 1D AKLT chain\cite{affleck1988valence} -- or more accurately, a generalization of a dimerized spin-1/2 chain. In a 1D chain of this type, neighboring spins are entangled between unit cells such that, in the ground state, each boundary has an effective free spin-1/2. Since quadratic spin interactions are incompatible with 1D subsystem symmetry in a 2D system, we instead use quartet interactions that entangle spin-1/2's between four different corners of a square plaquette. We will see that, in the ground state, each corner of the lattice contains an odd number of free spin-1/2 degrees of freedom that are decoupled from the bulk.  This leads to a higher-order topological phase whose gapless corner modes are protected by a combination of U(1)  subsystem, and global time reversal, symmetries.

\begin{figure}[h]
  \centering
      \includegraphics[width=0.25\textwidth]{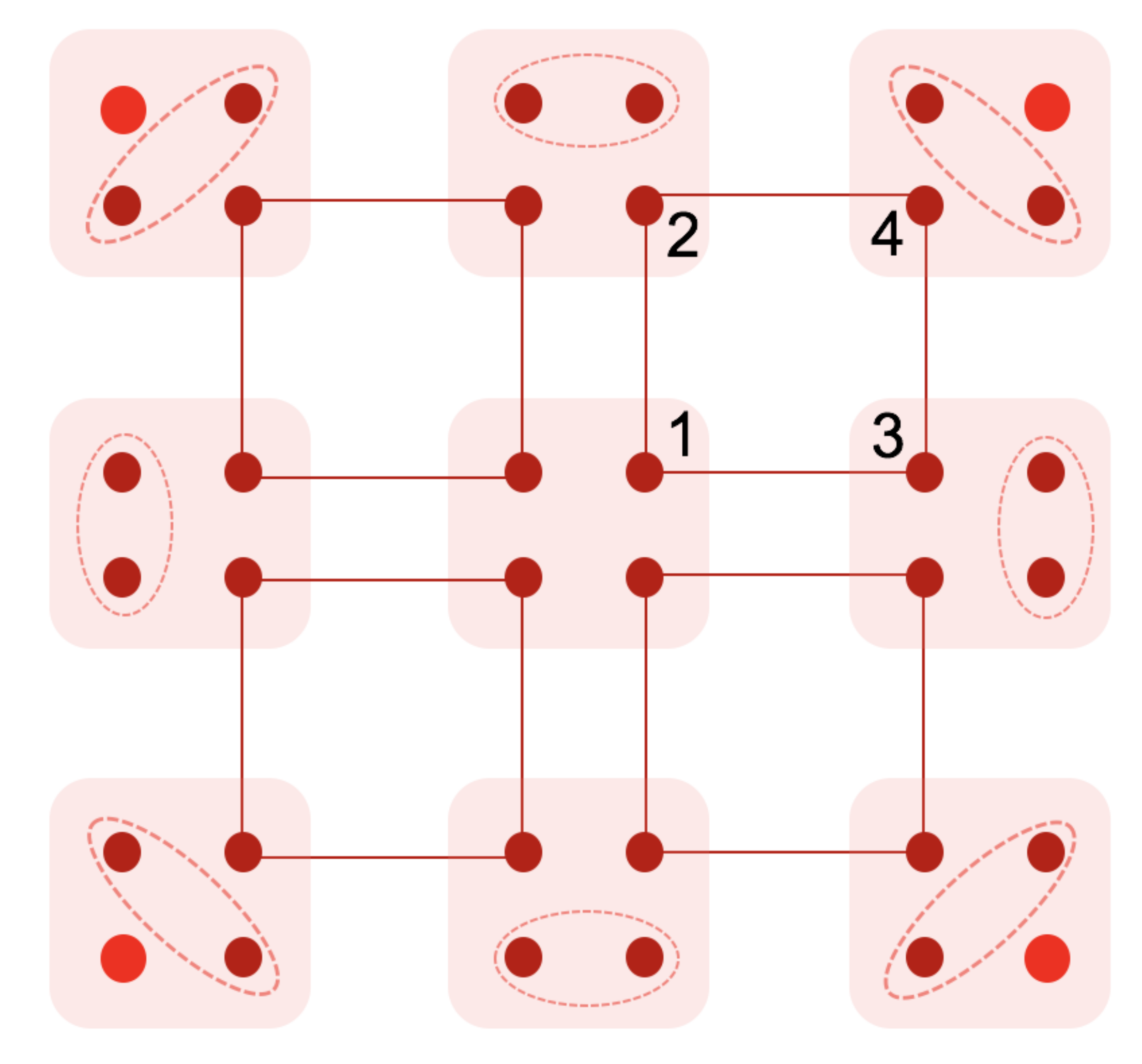}
  \caption{There are four bosons/spins (red dots) in each unit cell (shaded island), and each of them participates in one  plaquette ring-exchange term. Each ring-exchange term involves four bosons living at the four corners of the plaquette interacting at the quartic level. Bosons at the edges can also be gapped using quadratic terms, but only within the unit cell. A single boson zero mode (alternatively free spin-1/2) survives on each of the corners and is protected by $\mathcal{T}$ and U(1) subsystem symmetry.} 
\label{f2}
\end{figure}

More precisely, consider spins arranged on a square lattice as shown in Fig.~\ref{f2}. There are four spins (red dots) per unit cell (shaded island), and each spin independently interacts with one of the four plaquette clusters adjacent to the site via a ring-exchange coupling,
\begin{align} 
&H_{Q}=-\sum_{{\bf{R}}}\left(  S^{+}_{{\bf{R}},1} S^{-}_{{\bf{R}}+e_x,3} S^{+}_{{\bf{R}}+e_x+e_y,4} S^{-}_{{\bf{R}}+e_y,2}+h.c \right),
\label{hhh1}
\end{align}
where $S^{\pm}=\sigma^x \pm i \sigma^y,$ ${\bf{R}}$ labels each unit cell, $e_x, e_y$ represent unit distances between cells in the $x$ and $y$ directions, and $1,2,3,4$ are the degrees of freedom in each cell as labeled in Fig.~\ref{f2}.
This Hamiltonian has a global time-reversal symmetry $\mathcal{T}$ that flips each spin, and a subsystem U(1) symmetry for in-plane spin rotations in each row or column
\begin{align} 
&\mathcal{T}= \mathcal{K} i\sigma^y, \nonumber\\
& U^{sub}(1): \prod_{j \in row}e^{i \frac{\theta}{2} \sigma^z_j}.
\end{align}
 The subsystem U(1) symmetry acts only on the four spins inside each unit cell in each row or column, and rotates the spin around the $S_z$ axis.  Hence, subsystem U(1) symmetry preserves the total $S_z$ charge along any row or column, and forbids any spin-bilinear coupling \emph{between} unit cells. Furthermore, the global $\mathcal{T}$ symmetry forbids terms in the Hamiltonian that  polarize the spins, and would be spontaneously broken in the presence of magnetic order.
 
To gain intuition for this model, we can interpret the Pauli spin operators in terms of hardcore bosons:
\begin{align} 
\sigma^x+i \sigma^y=a^{\dagger},\sigma^x-i \sigma^y=a, \sigma^z=a^{\dagger}a-1/2.
\end{align}
Each hardcore boson has a restricted filling  $a^{\dagger}a=0,1$, and the states $|0 \rangle $ and $|1 \rangle = a^\dag |0 \rangle$ carry a $U(1)$ charge of $-1/2$ and $1/2$, respectively.  The boson creation operator $a^{\dagger}$ creates an $S_z = 1$  magnon excitation by changing $S_z\rightarrow S_z+1$. The Hamiltonian in Eq.~\ref{hhh1} can be translated into the hardcore boson language as,
\begin{align}  
H_{Q}=-\sum_{{\bf{R}}}\left(  a^{\dagger}_{{\bf{R}},1} a^{\phantom{\dagger}}_{{\bf{R}}+e_x,4} a^{\dagger}_{{\bf{R}}+e_x+e_y,3}a^{\phantom{\dagger}}_{{\bf{R}}+e_y,2}+h.c\right).
\label{hhh}
\end{align}
In this language, the subsystem U(1) symmetry becomes a phase rotation for the bosons on specific rows/columns,
\begin{align} 
& U^{sub}(1): a_j \rightarrow e^{i\theta } a_j,\;\; j \in \text{row}.
\end{align} Additionally, $\mathcal{T}$ acts as a particle-hole symmetry for the hardcore bosons,
\begin{align} 
\mathcal{T}:&|1 \rangle \rightarrow | 0 \rangle, |0 \rangle \rightarrow  - |1 \rangle \\
& a \rightarrow -a^{\dagger}, a^{\dagger} \rightarrow -a. 
\end{align}

We now investigate the properties of the ground state of (\ref{hhh}).  This Hamiltonian generates a boson ring-exchange interaction among 4 of the hardcore bosons around each plaquette, as shown in in Fig. \ref{f2}. The interaction term projects the four bosons around the plaquette to a maximally entangled state $|0_1 1_2 1_3 0_4\rangle+|1_10_2 0_3 1_4\rangle$.  Since terms on different plaquettes involve different boson flavors at each site, all plaquette operators commute and can be simultaneously minimized.  Thus the system is gapped with periodic boundary conditions, with a unique ground state given by
\begin{equation}
|\Psi_0 \rangle = \prod_P \left(|0_1 1_2 1_3 0_4\rangle_P+|1_10_2 0_3 1_4\rangle_P \right),
\end{equation}
where the subscript $P$ labels plaquettes on the square lattice which are coupled by the ring exchange terms in Eq. \ref{hhh}.

It is straightforward to see that $|\Psi_0\rangle$ preserves both subsystem U(1) and $\mathcal{T}$. First, in each configuration that makes up ground state, the net charge along any row or column of the square lattice is exactly zero.  In fact, the dipole moment of each plaquette in this ground state is also exactly zero.
% Generically the subsystem U(1) symmetry preserves the charge along each row/column, so any boson bilinear between unit cells, e.g., single-particle tunneling interactions, is always prohibited.
 Second, since $\mathcal{T}$ interchanges the states $|1\rangle$ and $|0 \rangle$, and introduces a minus sign, the ground state of each plaquette is invariant under $\mathcal{T}$ since there are an even number of empty states that are flipped to filled states.  Thus the $\mathcal{T}$ symmetry (i.e., the particle-hole symmetry for the bosons) is manifest.

 The quasiparticle excitations above this gapped ground state are fractonic in the sense that they exhibit restricted mobility. Since the U(1) charge is preserved on each row and column, an isolated charge is immobile. Additionally, while dipoles can move, they are restricted to move only in the direction transverse to their dipole moment. In contrast, quadrupolar excitations can be fully mobile in 2D.

Now we can explore the properties of the edges and corners of the ground state $|\Psi_0\rangle$. If we have an edge termination as shown in Fig.~\ref{f2}, we find that each edge contains two dangling spins  (or hardcore bosons) per unit cell. The two spins in the \emph{same} unit cell can be paired into a $|01\rangle-|10\rangle$ state while preserving subsystem U(1) and $\mathcal{T}$. Interestingly, pairing spins \emph{between} different unit cells with bilinear spin couplings along the edge is forbidden by subsystem symmetry.

At the corner unit cell, there are three dangling spins. Two of the spins can be gapped into a symmetric bilinear state $|01\rangle-|10\rangle,$ which leaves one spin-1/2 as a zero mode with two-fold degeneracy. 
This corner spin-1/2 is protected by subsystem U(1) and global $\mathcal{T}$ symmetry, and is robust under any symmetry allowed perturbation. One typical way to gap out the corner mode is to hybridize two corner spin-1/2's on neighboring corners via an edge phase transition (while preserving the bulk gap)\cite{you2018higher}. However, since total spin  $S_z$ on each row/column is conserved, such corner spin hybridization violates the subsystem U(1) symmetry.
Thus, unlike HOTIs, whose corner modes require some spatial symmetry to be protected, the subsystem-SPT model in Eq.~\ref{hhh} only requires subsystem U(1) and global $\mathcal{T}$ symmetry. However, as subsystem symmetry implements a spatially dependent U(1) transformation with charge conservation on submanifolds, it is not exactly an internal symmetry. Instead the U(1) charge conservation is intertwined with the spatial degrees of freedom. 
We note that because of the fracton-like nature of its excitations, our subsystem proteced HOTI can be viewed as a symmetry enriched fracton order with protected corner modes.

Now let us consider how the topology is destablized if we remove symmetry. If we allow for $\mathcal{T}$-breaking then we can just polarize the corner spin to remove the degeneracy. On the other hand, if we break the subsystem U(1) down to a global U(1) symmetry, one can merge the two corner modes along the edge and hybridize them via an edge transition. We could prevent this if we required a spatial symmetry, e.g., a $C_4$ rotation symmetry. Having $C_4$ symmetry, the system would have four identical quadrants, each containing an odd number of spin-1/2s on the boundary (which includes the corner). As each equivalent quadrant contains an odd number of spin-1/2's, it is impossible to have a unique $\mathcal{T}$ invariant ground state due to Kramers' degeneracy. Here the rotational symmetry ensures that the unpaired spin-1/2's are separated by a distance on the order of the linear system size, and hence cannot be coupled by any local interaction. Thus, we find that if we demote the U(1) subsystem symmetry to a conventional global symmetry, we can retain protected modes if we require $C_4$ symmetry. This would give rise to a bosonic HOTI with $C_4 \times \mathcal{T} \times U(1)$ symmetry as discussed Refs. \onlinecite{you2018higher,dubinkin2018higher}. We will describe this second type of HOTI in more detail in Sec. \ref{sec:C4HOTI}, but for now we move on to the topological response properties for our model with subsystem symmetry.

\subsection{Topological Quadrupolar Response to External Higher Rank Fields}

The robustness of the corner zero modes, and the symmetry protected topological phase in general, can be made apparent using a low-energy effective response theory exhibiting quantized quadrupole moment density (which we denote quadrupolarization). Since the hardcore boson model in Eq.~\ref{hhh} has subsystem U(1) symmetry, we can introduce a higher rank (rank 2) background gauge field $A_{xy}$ that minimally couples to the ring exchange term\cite{you2018symmetric,pretko2017fracton,pretko2017subdimensional,Vijay2016-dr}:
\begin{align}  \label{Eq:H2Gauge}
&H_{Q}=\nonumber\\&- \sum_{{\bf{R}}}\left[ a^{\dagger}_{{\bf{R}},1} a^{\phantom{\dagger}}_{{\bf{R}}+e_x,4}a^{\dagger}_{{\bf{R}}+e_x+e_y,3}a^{\phantom{\dagger}}_{{\bf{R}}+e_y,2}e^{i A_{xy}({\bf{R}})}+h.c \right].\nonumber\\
\end{align}
We denote $A_{xy}$ as the gauge field living in the center of each plaquette that couples with the dipole current $J_{xy}({\bf{R}})=(i a^{\dagger}_{{\bf{R}},1} a^{\phantom{\dagger}}_{{\bf{R}}+e_x,4} a^{\dagger}_{{\bf{R}}+e_x+e_y,3}a^{\phantom{\dagger}}_{{\bf{R}}+e_y,2}+h.c)$. The dipole current is exactly the ring-exchange term on the plaquette, and can be viewed as a dipole oriented along $x$ hopping along $y,$ or vice versa. We also introduce a time component $A_{0}$ of the gauge field, that couples with the total charge density $\rho=\sigma^z+1/2$. These gauge fields transform under gauge transformations $\alpha$ as
\begin{align}
    &A_{xy}\rightarrow A_{xy}+\partial_x\partial_y \alpha\nonumber\\
    &A_{0}\rightarrow A_{0}+\partial_t\alpha.
    \end{align}
There is a single gauge-invariant combination of these fields, 
\begin{equation}
    E_{xy} =   \partial_x \partial_y A_0 -\partial_t A_{xy}.
\end{equation}

In the gapped subsymmetry protected phase we expect to find a generalization of the Goldstone-Wilczek (polarization)\cite{goldstone1981fractional,qi2011topological} response that is found in 1D SPTs with protected end modes. The 1D polarization response is characterized by
\begin{equation} \label{Eq:1dTheta}
    \mathcal{L}_{P}=\frac{\theta}{2\pi}\left(\partial_x A_0-\partial_t A_x\right)=\frac{\theta}{2\pi}E_x,
    \end{equation} 
where the polarization is $P=\tfrac{\theta}{2\pi}.$  Because $\theta$ is periodic modulo $2 \pi$, a term of the form (\ref{Eq:1dTheta}) is allowed even in the presence of spatial or internal symmetries under which the electric field is odd, such as reflection or particle-hole symmetry.  In the presence of such symmetries $\theta$ is quantized, and can take only the values $\theta=0,\pi.$  A non-trivial 1D symmetry protected topological phase can be characterized by a background $\theta=\pi$ configuration, and thus has a quantized polarization $P=1/2.$

Our rank-2 gauge theory in 2D admits a generalization of this topological response theory, described by the quadrupolarization $\theta$-term,
\begin{align} 
&\mathcal{L}_{Q}=\frac{\theta}{2\pi} (\partial_x \partial_y A_0-\partial_t A_{xy})=\frac{\theta}{2\pi}E_{xy}.
\label{qua}
\end{align}
As we will show at the end of this subsection, $\theta$ is defined modulo $2 \pi$.  Hence even though the electric field $E_{xy} =\partial_x \partial_y A_0-\partial_t A_{xy} $ is odd under 
 $\mathcal{T}$ symmetry 
 \begin{equation*}
     \mathcal{T}: A_0 \rightarrow -A_0, A_{xy} \rightarrow  -A_{xy}, (x,y,t)\rightarrow (x,y,t)
 \end{equation*} (which here acts as a PH symmetry), 
 such a term is symmetry-allowed provided that $\theta = 0, \pi$.  Hence, the action (\ref{qua}) describes the response of a subsystem-SPT, with a coefficient $\theta$ that is quantized in the presence of $\mac{T}$ symmetry.

We now consider the nature of the charge response described by (\ref{qua}).  The action describes the following charge and dipole current response:
\begin{align}
    j_0&=\frac{1}{2\pi}\partial_x \partial_y \theta\\
    j_{xy}&=\frac{1}{2\pi}\partial_t \theta.
    \end{align} 
By comparing these equations with the expected properties of a quadrupole moment $q_{xy}$ \cite{raab2005,benalcazar2017quantized}, we can identify the bulk quadrupole moment to be $q_{xy}=\frac{\theta}{2\pi}.$  Similar to the dipole moment of the 1D topological insulator, we find that in the rank-2 case, $q_{xy}$ is quantized in the presence of time-reversal symmetry.  
Interestingly, no matter what the space-time dependence of $\theta$ is, our response theory does not predict any bulk charge currents, i.e., no response terms for $j_i,$ and thus no changes in the bulk polarization.  The action (\ref{qua}) does predict fractional charge on corners at the intersection between edges with normal vectors $\hat{x}$ and $\hat{y}.$ At such a corner, which we can heuristically model as a product of step functions in the $\theta$ field, we have $j_0=\partial_x \partial_y \frac{\theta}{2\pi} =\pm \tfrac{\theta}{2\pi} \delta(x-x_0)\delta(y-y_0)$,  where $(x_0, y_0)$ is the position of the corner.  For $\theta = \pi$, this indicates the presence of a half-charge localized to the corner.  %As the hardcore boson creates a magnon excition with integer spin, 
In the spin language, the corner response with half the charge of local bulk excitations corresponds to an unpaired spin-$1/2$ zero mode. %Such fractionalized spinon excitation is absent in the bulk or on the edge and only appears as a localized  zero mode at the corner.
The $\mathcal{T}$ symmetry, having effectively $\mathcal{T}^2=-1$ for the corner spin, ensures the Kramers' spin degeneracy at the corner cannot be lifted by a Zeeman field.  These results all match the expected phenomena from the microscopic model. 

Finally, since  $j_{xy}$ is a dipole current (i.e., it is the current that minimally couples to the gauge field $A_{xy}$),  Eq.~\ref{qua} indicates that  adiabatically changing $\theta$ in a cycle from $0$ to $2\pi$ as a function of time creates a bulk dipole current. This process has the net effect of pumping a dipole between two parallel boundaries, i.e., an $x$-oriented dipole in between two $y$-boundaries or vice-versa. This is the dipole analog\cite{benalcazar2017electric} of Thouless' adiabatic charge pumping\cite{thouless1983}; indeed, the charge pumping effect is captured by the original Goldstone-Wilczek response. From conventional electromagnetic arguments, a dipole current is connected to a change in the quadrupole moment\cite{wheeler2018many}, just as we find here, and it confirms our identification of the quadrupole moment as $q_{xy}=\tfrac{\theta}{2\pi}.$

We can also argue that this response is quadrupolar by carefully keeping track of the boundary responses arising from Eq. \ref{qua}. We find that the charge density
\begin{equation}
    j_0=\frac{1}{2\pi}(\partial_x\partial_y\theta - \partial_y \theta\vert_x-\partial_x \theta\vert_{y}+\theta\vert_{x,y})
    \end{equation} where $\vert_{x_i}$ indicates evaluation on the edge perpendicular to $x_i,$ and $\vert_{x,y}$ indicates evaluation on the corner. The two edge contributions are arising from edge polarizations that we can identify $P^x_{y-edge}=\frac{\theta}{2\pi}=P^{y}_{x-edge}.$ Thus we can use the definition of the quadrupole moment\cite{benalcazar2017quantized} to find
    \begin{equation}
        q_{xy}\equiv P^x_{y-edge}+P^{y}_{x-edge}-Q_{corner}=\frac{\theta}{2\pi},
    \end{equation} as claimed.

Now let us return to the issue of the periodicity of $\theta$ modulo $2 \pi$. %This requires two elements: (i) $\theta$ must have a shift ambiguity, and (ii) we must impose a symmetry under which $\theta$ is odd.
%First, we point out that the value of $\theta$ is equivalent under a $2\pi$ shift. 
First, we can give a physical argument by observing that changing $\theta$ by $2\pi$ does not change the low energy topological properties of our response theory.  Specifically, taking $\theta \rightarrow \theta + 2\pi$ corresponds to adding an integer $S_z$  charge to the corner, which forms a representation under $\mathcal{T}$ that does not support Kramers' degeneracy. Such a representation can always be gapped out while preserving symmetry, so we expect that shifting $\theta$ by $2 \pi$ does not change the nature of our subsystem-SPT.

To see how the periodicity of $\theta$ arises at the field theoretic level, we first review the analogous situation in 1D. 
If $\theta$ is a constant in spacetime, the 1D Goldstone-Wilczek action (\ref{Eq:1dTheta}) is a total derivative, so naively one expects the action to be $0$ with periodic boundary conditions (PBCs) in spacetime.  However, in a compact $U(1)$ gauge theory the gauge fields need not be strictly single-valued under PBC's; rather, $A_x (t)$ and $A_x(t+ T)$ (where $T$ is the periodicity of the system in the time direction) may differ by $2 \pi n /L_x$ for integer $n$. This amounts to allowing a large gauge transformation (LGT) to occur at some point in time. The LGT changes $\int A_x dx$ by $2\pi$, and is thus a $2 \pi$ flux insertion that leaves all physical quantities single-valued in time.  
%over a time period $T$, the gauge connection has a global shift so the total path integral of the polarization term is changed by $\theta$:
However, in the presence of such large gauge transformations, the integral of the electric field need not be zero; rather, we have 
\begin{align} 
&\frac{\theta}{2\pi} \int_{0}^{L_x} dx \int_0^T dt  E_x= \theta n.
\end{align}
Thus shifting $\theta$ by $2\pi$ shifts the action by $2 \pi$, at most, and this  does not change the partition function of our theory.   

A similar argument holds for our 2D response theory. Since subsystem symmetry ensures that the charge is conserved on each line, then LGTs for the rank-2 gauge field can change the line integral of $A_{xy}$ along a specific row or column of plaquettes, e.g.,
\begin{align} 
\int A_{xy}(x,y) dx&=2\pi \delta(y_i),\nonumber\\ \text{or}~~\int A_{xy}(x,y) dy&=2\pi \delta(x_i).
\end{align}
Again, if such a LGT occurs at some point in time, the spacetime integral of the higher-rank electric field is not zero; rather it is quantized in integer multiples of $2 \pi$,  %path integral arising from Eq. \ref{qua} is changed by
such that
\begin{align} 
\frac{\theta}{2\pi}\int dx dy  \int_0^T  dt  (\partial_x \partial_y A_0-\partial_t A_{xy})=\theta n.
\end{align}
Thus, shifting $\theta$ by $2\pi$ does not affect the path integral, and as such we have $\theta \equiv \theta + 2 \pi$. In Appendix \ref{app:ambiguity} we provide more discussion of this ambiguity/periodicity. 

%Now, to quantize $\theta$ we need a symmetry. We find that the quadrupolarization term in Eq.~\ref{qua} is odd under $\mathcal{T}$ (particle-hole symmetry in the boson language). 
Analogous to the 1D TI, the periodicity of $\theta$ implies that in the presence of  $\mathcal{T}$ symmetry (under which $E_{xy}$ is odd), $\theta$ is quantized and must take one of the two values $0,\pi.$ Hence the associated quadrupole moment $q_{xy}=\tfrac{\theta}{2\pi}$ is quantized.  
We have argued above that the non-trivial SSPT  phase has $\theta = \pi$, while the trivial subsystem-symmetric insulator has $\theta =0$.  
Since the combination of $U(1)$ subsystem and $\mathcal{T}$ symmetry ensures that $\theta$ cannot change continuously, the existence of a term Eq.~\ref{qua} in the response thus describes a many-body property that cannot change without a phase transition in the presence of these symmetries. Our response theory thus characterizes features of the boson SSPT HOTI phase that are robust to symmetric perturbations
of the Hamiltonian in Eq.~\ref{hhh} away from the exactly solvable limit.  
%the topological quadrupole response will survive unchanged as long as we keep the $\mathcal{T}$ and subsysytem U(1) symmetries. Indeed, a change of $\theta$ from $0$ to $\pi$ always implies a bulk gap closing in the presence of these symmetries.

\subsection{Breakdown to global U(1) symmetry}\label{sec:C4HOTI}
%Besides, as is pointed out by earlier literature in Ref.~\cite{wheeler2018many,kang2018many}, Eq.~\ref{qua} is also odd under $C_4$ symmetry.

So far we have discussed an unconventional SSPT HOTI phase with a quadrupolar response. Now we will explore how this model is connected to a more conventional quadrupole-like model,  which is essentially a bosonic version of the original quadrupole model from Ref. \onlinecite{benalcazar2017quantized}. 
Our starting point will be the Hamiltonian of Eq.~\ref{hhh},  which was introduced in Refs.~\onlinecite{you2018higher,dubinkin2018higher} as a model for a bosonic HOTI protected by $C_4 \times \mathcal{T} \times U(1)_{\text{global}}$ symmetry.  Since the exactly solvable Hamiltonian (\ref{hhh}) is fine-tuned to a point with subsystem symmetry, we must understand the effect of perturbations that break the subsystem symmetry to a global one, while preserving $\mathcal{T}$.  As noted above, to ensure that the gapless corner modes are robust, we will also require $\mathcal{C}_4$ rotation symmetry.   Thus, for example, we now allow 
spin bilinear interactions to Eq. \ref{hhh} that break subsystem U(1), but still maintain $C_4 \times \mathcal{T} \times U(1)$. 
We will find that allowing such terms preserves the quantized quadrupole moment and gapless corner modes ---but that the physical content of the two models differs in some subtle, but interesting, ways. 

A quadratic Hamiltonian with the correct symmetries is given by: 
\begin{align} 
&H_{Q*}=\kappa\sum_{{\bf{R}}}\left(a^{\dagger}_{{\bf{R}},1} a^{\phantom{\dagger}}_{{\bf{R}}+e_x,3}+a^{\dagger}_{{\bf{R}},2} a^{\phantom{\dagger}}_{{\bf{R}}+e_x,4}\right.\nonumber\\ &\left.+a^{\dagger}_{{\bf{R}},1} a^{\phantom{\dagger}}_{{\bf{R}}+e_y,4}+a^{\dagger}_{{\bf{R}},2} a^{\phantom{\dagger}}_{{\bf{R}}+e_y,4}+h.c\right).
\label{bosona}
\end{align}
This Hamiltonian is a bosonic version of the quadrupole band insulator described in Ref. \onlinecite{benalcazar2017quantized}, with global (but not subsystem) U(1) symmetry, as well as $C_4$ and $\mathcal{T}.$  If we add a small $H_{Q*}$ to the Hamiltonian (\ref{hhh}), and preserve both $\mathcal{T}$ and $C_4$ rotations, the gapless corner modes remain robust, since the combination of $\mathcal{T}$ and $C_4$ symmetries prevents a boundary phase transition, and the bulk remains gapped.  The resulting phase is a HOTI, which is closely related to the subsystem SPT described in the previous section.  Indeed, we can tune the parameter $\kappa$ until the system is well-described by the quadratic Hamiltonian $H_{Q*}$ without closing the bulk gap and while preserving the symmetries\cite{dubinkin2019}; thus $H_{Q*}$ can be used as our basic model for the HOTI phase. We note that since the model preserves $C_4$ symmetry the bulk dipole moment of the system remains quantized throughout this process, and since it starts with a vanishing value the polarization remains vanishing.

We will now leverage the relationship between $H_{Q*}$ and $H_Q$  to derive a topological response theory describing the phase realized by $H_{Q*}.$ We know that this model has a quantized bulk quadrupole moment $q_{xy}=1/2$, and symmetry-protected gapless corner modes\cite{you2018higher,dubinkin2018higher}, 
but once the subsystem symmetry is explicitly or spontaneously broken, the tensor gauge field $A_{xy}$ in Eq.~\ref{qua} no longer exists.  Thus, in order to describe a response theory, we must couple the hardcore bosons  with a rank-1 vector U(1) gauge field $A_{\mu}$. To understand how to do this, it is useful first to ask how a rank-1 gauge field couples to dipole currents of the type that occur in our subsystem symmetric model.  Such dipolar currents do not couple to the gauge field itself, but rather to its gradients: for example, a current of dipoles with dipole moment $a \hat{x}$ ($a$ being the lattice constant) moving parallel to the $y$-axis couples to $\delta_x A_y \equiv A_y(x+a,y) - A_y(x,y)$, and similarly for $a\hat{y}$ dipole moving parallel to the $x$-axis which will couple to $\delta_y A_x.$  Thus, we could in principle replace the rank-2 gauge field $A_{xy}$ in the subsystem symmetric Hamiltonian (\ref{Eq:H2Gauge}) with a linear combination of $\delta_x A_y$ and $\delta_y A_x.$  In the absence of magnetic fields all such combinations are equivalent, since on each lattice plaquette
\begin{equation}
    \delta_x A_y = \delta_y A_x + B  .
\end{equation}
Moreover, the dipole currents allowed by subsystem symmetry do not couple to magnetic fields, since processes in which a single charge hops around a plaquette are prohibited. Thus, we expect a response theory describing subsystem symmetric systems to be compatible with taking $B=0$.
We use this liberty to choose a combination of gauge field gradients that preserves $C_4$ symmetry, i.e.,
\begin{equation} \label{Eq:Asub}
    A_{xy}\rightarrow \tfrac{1}{2}(\partial_x A_y+\partial_y A_x),
\end{equation} where
\begin{align} 
&C_4: (x, y) \rightarrow (-y, x); (A_x, A_y) \rightarrow (-A_y, A_x). %\nonumber\\
%& \frac{\theta}{2\pi} (\partial_x \partial_y A_0-\partial_t (\partial_x A_y+\partial_y A_x)/2)\nonumber\\
%\rightarrow &-\frac{\theta}{2\pi} (\partial_x \partial_y A_0-\partial_t (\partial_x A_y+\partial_y A_x)/2).
\end{align}

%A current $J_{xy}$ that couples to this combination of  rank-1 gauge fields
%via $   A_0 \rho -  \frac{1}{2}( \partial_x A_y+ \partial_y A_x) J_{xy} $
%obeys the current conservation law
%\begin{equation}
%    \partial_x \partial_y J_{xy} = \partial_t \rho
%\end{equation}
%which in turn ensures that with periodic boundary conditions in $x$ and $y$, the net charge along each line is conserved.  

Interestingly, making the substitution (\ref{Eq:Asub}) into  $\mathcal{L}_Q$, we recover the   standard coupling of the quadrupole moment to a gradient of the electric field as proposed by Refs.  \cite{wheeler2018many,kang2018many} to describe the quadrupole response of higher-order insulators:
\begin{align} 
\mathcal{L}_{Q*}&=\frac{\theta}{2\pi} (\partial_x \partial_y A_0-\partial_t (\partial_x A_y+\partial_y A_x)/2)\nonumber\\&=\frac{\theta}{4\pi}(\partial_x E_y+\partial_y E_x),
\label{quad}
\end{align}
\noindent where we have identified $q_{xy}=\tfrac{\theta}{2\pi}.$ Since $\mathcal{L}_{Q*}$ is odd under $\mathcal{C}_4$ rotations, requiring this symmetry will restrict $\theta$ to take on the values $0, \pi$, as before, {\it provided} that we can show that $\theta$ is periodic modulo $2 \pi$.  Such periodicity in this case is more subtle than for the higher rank theory since, unlike $A_{xy}$, we expect the electric fields $E_x$ and $E_y$ to be single-valued, such that for a closed system $\int d^2 x \mathcal{L}_{Q*} = 0$, rather than an arbitrary integer.   
However, for neutral, unpolarized systems, we argue in Appendix \ref{app:ambiguity} that 
%the action is similarly quantized and not necessarily $0$. 
%In other words, provided that we have a system whose net charge and net dipole moment are fixed to be $0$ at all times, 
$\theta$ is periodic modulo $2 \pi$ in this case as well.  Thus, we identify the non-trivial HOTI phase with the response for $\theta=\pi$, and the trivial one with $\theta=0$. 

We note that since $\mathcal{L}_{Q*}$ is also odd under  $\mathcal{T}$ symmetry (which acts as PH symmetry for hardcore boson), this symmetry is also sufficient to restrict $\theta$ to the values $0,\pi$, and therefore to quantize the quadrupole moment.  Thus one might naively conclude that the HOTI requires either $\mathcal{T}$ {\it or} $C_4$ rotations for stability.  However, it is easy to see that this cannot be the case:
On one hand, in the absence of $C_4$ symmetry, we can add 1D SPT chains only on, say, the two parallel $x$-boundaries. This adds a second free spin-$1/2$ degree of freedom to each corner, which can hybridize with the existing corner-bound spin-$1/2$, leading to fully gapped corner modes. On the other hand, in the absence of $\mathcal{T}$ symmetry, the degeneracy of the free spin-1/2 mode at the boundary can be lifted by an external Zeeman field. In the hadrcore boson langauge, this Zeeman field is akin to a background chemical potential that breaks the PH symmetry.  Thus, both $\mathcal{T}$ \emph{and} $C_4$ symmetry are essential to protecting the corner modes and HOTI phase. This is consistent with the situation for non-interacting quadrupole HOTIs where it is known that a quantized quadrupole moment by itself does not require corner zero modes\cite{benalcazar2017quantized}. 
%As we will discuss in more detail shortly, if we require only one of these symmetries, additional terms may be added to the action that destroy the $0$-energy degeneracy of our corner modes. 
 We will return to this point shortly.

Let us take a moment to consider other possible topological response terms that are compatible with the physical constraints and symmetries. First, in order for a system to exhibit a well-defined quadrupole moment, its net dipole moment must vanish -- meaning that we must consider response theories describing a system at energy scales below its charge gap.  In other words, our response must describe dipole currents, and have vanishing bulk polarization, i.e., our response Lagrangian should not have any terms proportional to the (rank 1) electric field, and instead should at most have terms proportional to derivatives of the electric fields.  Of the possible leading-order terms of this type, two are odd under $\mathcal{C}_4$ rotations (and thus may exhibit quantized coefficients):
 Eq. \ref{quad} which we are already considering, and
$\mathcal{L}_{Q**} = \frac{ \theta}{2 \pi} ( \partial_x E_x - \partial_y E_y).$ The latter term can be obtained by rotating $\mathcal{L}_{Q*}$ by 45 degrees and represents a quadrupole component $q_{x^2-y^2}$. Since we are considering response theories motivated by lattice models with discrete rotation symmetry, it is unlikely that both $\mathcal{L}_{Q*}$ and $\mathcal{L}_{Q**}$ would appear as response terms in the same lattice action since a 45 degree rotation is not an allowed lattice rotation symmetry. The remaining possible actions at this order are either fixed to be 0 once we require the total charge, dipole, and magnetic field to be zero, or do not exhibit quantized coefficients.

Let us now review the phenomena associated with the HOTI response  $\mathcal{L}_{Q*}$ at $\theta = \pi$. As in the subsystem symmetric case, at the edges of the HOTI the value of $\theta$ jumps from $\pi$ to $0,$ and for a corner located at $(x_0, y_0)$ we have $\tfrac{1}{2\pi}\partial_x \partial_y \theta=(\pm 1/2)\delta(x-x_0)\delta(y-y_0)$, indicating a half $U(1)$  charge, or a free spin-$1/2.$  In addition to the corner charge, and unlike $\mathcal{L}_{Q},$ this response can exhibit charge currents whenever $\partial_t\partial_i\theta \neq 0$ (derived from the variation of the response action with respect to $A_i$), and bound polarization density whenever $\partial_i \theta\neq 0$ (derived from the variation of the response action with respect to $E_i$).  These two properties are also consistent with the interpretation of $\theta$ as a 2D quadrupole moment\cite{raab2005}. Thus, this response has richer features than the SSPT case because the conservation laws are much less constraining.

It is interesting to contrast the response theory (\ref{quad}) with that of just a trivial 2D bulk system with 1D SPTs glued on the system's boundary.  Indeed, on a system with open boundaries, $\int d^3x \mathcal{L}_{Q*}$ is a pure boundary term itself and can be decomposed into a $C_4$ symmetric combination of 1D topological polarization $\theta$ terms localized on the four edges:
\begin{align} 
&\int d^3x\frac{\theta}{4\pi} (\partial_x E_y +\partial_y E_x)\nonumber\\
&=\int dy dt\left[\frac{\theta}{4\pi} E_y|_{x=0}-\frac{\theta}{4\pi} E_y|_{x=L}\right]\nonumber\\
&+\int dx dt\left[\frac{\theta}{4\pi} E_x|_{y=0}-\frac{\theta}{4\pi}E_x|_{y=L}\right].
\label{edge}
\end{align}
When the $\theta=\pi$, the globally $C_4$ symmetric edge polarization terms at each of the four boundaries carry a dipole moment with magnitude $1/4$ (c.f. Eq. \ref{Eq:1dTheta}).  In contrast, any $\mathcal{T}$- invariant 1D system we add to the boundary will have the response $\int dt dx_{edge}\frac{\theta_{1D}}{2\pi}E_{edge},$ where $\theta_{1D}$ takes the discrete values $0,\pi$. 
Hence,  despite the fact that our response theory describes the same type of boundary polarization response that a $1D$ SPT would generate when added to the edge, the action in Eq.~\ref{edge} is incompatible with U(1) and $\mathcal{T}$ symmetry in a purely 1D system, and the fractional edge dipole moment implied by Eq.~\ref{edge} can exist only at the boundary of a non-trivial bulk SPT, e.g., our HOTI phase.
%To state this another way, in the presence of $C_4$ and $\mathcal{T}$ symmetry, one cannot create a $\theta=\pi$ quadrupole moment by decorating 1D polarized chains on the four edges of a trivial insulating square. 
Indeed, adding non-trivial SPT chains on the four edges in a $C_4$ invariant way is equivalent to shifting the bulk value of $\theta$ by $2\pi,$ and hence does not change the phase.  Correspondingly, the additional \emph{pair} of zero modes added at each corner in this process can be gapped out locally in a symmetry preserving manner.

We can also revisit the topological symmetry protection by using our response theory to understand why both  $\mathcal{T}$ and $C_4$ symmetry are necessary to protect gapless corner modes. 
First, we observe that in the absence of $\mathcal{T}$ symmetry, the polarization of any 1D SPTs added to the edges are no longer required to be quantized. Hence, since the amount of 1D polarization that can be added to the edge can take any value, the 1/4 polarization on the edge induced according to $\mathcal{L}_{Q*}$ is no longer anomalous since it can be removed by a suitable choice of 1D system. 
It follows that if $\mathcal{T}$ is broken at the edges of our system, we may get rid of the $P_{\text{edge}} = 1/4$ polarization response by decorating the boundary of our system with $\mathcal{T}$-broken 1D systems in a $C_4$-invariant way; hence neither the boundary polarization nor the corner modes are robust in this case.
 %To rephrase, a quantized quadrupolarization term in the bulk does not imply a higher order SPT phase with protected corner modes. As the bulk quadrupole moment engenders dipoles on the boundary, the quadrupolarization is `topological' only under certain symmetries, e.g., when the resultant fractional edge polarization is anomalous, and thus cannot be realized in purely 1D systems. If the dipole moment on the edge is anomaly free, one can always construct an equivalent theory via edge decoration on trivial bulk states. 

Conversely, suppose that we do not require $C_4$ symmetry, but only reflection ($R_x,R_y$) and  $\mathcal{T}$ symmetries. We mentioned above that one can remove the corner modes in this case by adding 1D SPT chains along, say, the two $x$-boundaries, in a reflection symmetric manner, so that the additional end modes of the SPT chains can hybridize and gap out the corner modes. From the response theory perspective, the removal of the edge polarization in one direction is equivalent to modifying the bulk action to obtain:
\begin{align}
\mathcal{L} = & \frac{\theta}{2\pi} (\xi \partial_x E_y +(1-\xi) \partial_y E_x),
\label{tri}
\end{align}\noindent where $\xi\in [0,1].$ 
Regardless of the value of $\xi$, one can still apply the arguments in App. \ref{app:ambiguity} to show that $\theta$ is eqiuvalent mod $2\pi.$ In combination with the refection symmetries $R_x,R_y$, this requires $\theta=0,\pi$. Symmetry thus allows us to add terms to our starting action to obtain an action with, e.g., an edge polarization of $P_{y,edge}=1/2$,
\begin{align}
& \mathcal{L}=\frac{1}{2} E_y|_{x=0,L}
\end{align}\noindent when $\xi=1.$
Since this boundary response can be produced by a purely 1D system, the quadrupole term in Eq.~\ref{tri} does not describe an anomalous boundary polarization in the presence of reflection symmetry. Thus, our response theory shows that, just as for free fermions \cite{benalcazar2018quantization}, in interacting bosonic systems a quantized quadrupole moment alone does not necessarily predict a bulk HOTI with protected corner modes.
%, and thus cannot always be diagnosed as a bulk invariant. Only 
Rather, a HOTI occurs when the quantized quadrupole moment requires a boundary theory that is anomalous under one or more symmetries, guaranteeing protected gapless corner modes.

In summary, the quadrupole responses described in Eqs. (\ref{qua}) and (\ref{quad})  illustrate the parallels between subsystem-symmetry protected and HOTI phases. Though their different symmetries ensure that these are distinct phases of matter, many key features of the associated topological response are closely analogous. Essentially, this is because both describe a topological dipole response with quantized quadrupolarization.
In the SSPT phase, this is the only possibility, since subsystem symmetry does not allow for mobile charges.  For the HOTI, a more conventional charge response is allowed by symmetry; however, to obtain a well-defined quadrupole moment we must assume a large bulk charge gap, and symmetries that quantize the polarization such that the bulk dipole moment is exactly zero.  This, in turn, leads to a low-energy response best described in terms of dipole phenomena; we find that the same type of topological dipole response associated with the SSPT uniquely characterizes the corresponding possible HOTI phases in the presence of $U(1), \mathcal{T}$, and $C_4$ symmetries. 
%However, due to its higher order topology with gapped bulk and edges, the low energy dynamics is only attributed to the corner modes so we still expect a topological dipole response with quantized quadrupolarization. 
In addition, unlike the SSPT phase the HOTI boundary displays fractional 1D polarization $P_{edge}=1/4$ that can never be manifested in a pure $1D$ system with $\mathcal{T}$ symmetry. Such an `anomalous edge dipole,' despite arising from a gapped edge, must be accompanied by a nontrivial 2D bulk quadrupole insulator with $q_{xy}=1/2$. The relation between the bulk quadrupolarization and `anomalous edge dipole' captures the intrinsic nature of the HOTI response and exemplifies a bulk-edge correspondence. 

We will see in Sec. \ref{Sec:Generalizations} that a similar approach can be applied to other higher order topological phases in various dimensions. However, before we move to 3D we will spend the next two subsections discussing additional descriptions of the fraction SSPT and HOTI phases that provide alternative perspectives of our models and topological responses.

\subsection{Parton construction perspective}
Our previous discussion has shown a clear relationship between SSPT HOTI and conventional HOTI quadrupole phases that is established by breaking the subsystem symmetry down to a global symmetry.  
%In particular, the bosonic HOTI can be regarded as a mean-field description of the fracton ring-exchange model when the subsystem symmetry is broken. 
Importantly the bosonic HOTI  inherits key topological features including protected corner modes and a quadrupolar response.  We also showed how this response can be reduced to a boundary response theory exhibiting a fractional polarization of $1/4$ that is incompatible with $\mathcal{T}$  symmetry in a purely 1D system.  This anomalous boundary response characterizes the HOTI phase, and in this subsection we will give an interpretation of this anomalous edge polarization in terms of partons.

We begin with the SSPT HOTI and  decompose the hardcore boson into two partons as $a=h v$. The fields $h,v$ are quasi-one-dimensional, horizontal and vertical bosons that only couple along the x or y directions respectively. Both partons carry half of the U(1) charge of the boson, and are oppositely charged under an auxilliary U(1) gauge field $b$, which emerges since taking $v \rightarrow e^{ i \alpha }v, h \rightarrow e^{ - i \alpha }h$ does not change the physical boson operator.  Substituting $a = h v$ into our ring exchange model, we obtain four $h$ and four $v$ partons per unit cell, and the ring exchange term becomes an eight-parton cluster interaction around a plaquette. 

\begin{figure}[h]
  \centering
      \includegraphics[width=0.5\textwidth]{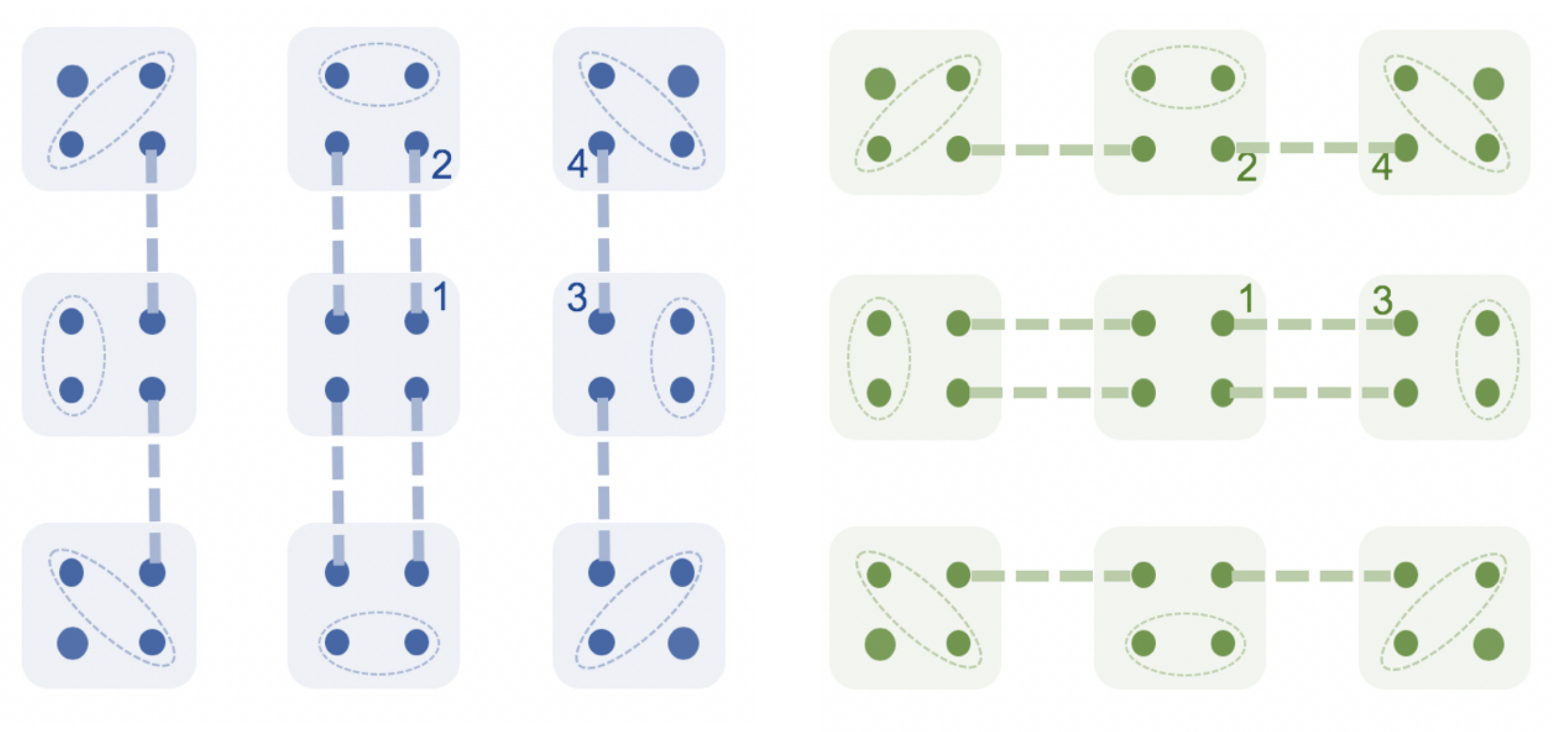}
      \caption{ Parton state for  $h$(blue/left) and  $v$(green/right) bosons. The full system has these two layers superimposed. Each parton resides in a 1D SPT chain. On the boundary rows, the dangling spins are paired within the site so there is only one non-trivial SPT chain per edge. Both parton types contribute a single protected corner mode on each corner that eventually projects to a physical boson corner mode $a=hv$.}
  \label{parton} 
\end{figure}

We choose a mean field ansatz by taking $\langle v^{\dagger}_{{\bf{R}}} v_{{\bf{R}}+e_x}\rangle =\langle h^{\dagger}_{{\bf{R}}} h_{{\bf{R}}+e_y}\rangle =t$. The ring exchange term becomes,
\begin{align} 
&H=t^3\sum_{{\bf{R}}} \left(v^{\dagger}_{{\bf{R}},1} v_{{\bf{R}}+e_x,3}+v^{\dagger}_{{\bf{R}},2} v_{{\bf{R}}+e_x,4}\right.\nonumber\\&\left.+h^{\dagger}_{{\bf{R}},1} h_{{\bf{R}}+e_y,4}+h^{\dagger}_{{\bf{R}},2} h_{{\bf{R}}+e_y,4}
+h.c\right).
\label{ptc}
\end{align}
This mean-field Hamiltonian describes a pair of decoupled 1D SPT chains running through each unit cell along the $x$ $(y)$-direction for the $v$ ($h$) boson, as illustrated in Fig.~\ref{parton}.  At the mean-field level $v$ ($h)$ bosons are mobile only along the $x$ ($y)$-direction, and thus couple only  with $b_x$($b_y$). Hence, our mean-field ansatz breaks subsystem symmetry, since the parton charge of $v$ ($h$) is only conserved on each row (column) (i.e., it couples to the a conventional rank-1 gauge field in one direction).  However, once we project onto the physical Hilbert space, where the number of $v$ and $h$ partons on each site must be equal, the physical boson charge is conserved in all rows and columns, and we expect subsystem symmetry to be restored.  

Now let us consider the system with open boundaries as in Fig. \ref{parton}. On each edge unit cell there are four $v$ partons and four $h$ partons. On vertical edges, two out of the four $v$ partons are paired to partons on adjacent boundary sites by the original Hamiltonian (\ref{ptc}), forming a single vertical 1D SPT chain of the $v$ partons.  Similarly two of the four $h$ partons are coupled to adjacent bulk sites, forming a pair of horizontal 1D SPT chains of $h$ partons.  Away from the corner, the remaining pair of $v$ and $h$ partons may be paired up in an on-site fashion without violating subsystem symmetry.%(\ref{hhh}).
 %Additionally on vertical (horizontal) edges there is a \emph{single} unpaired non-trivial 1D SPT chain of the $v$ ($h$) partons. 
Similar considerations apply to the horizontal edges, with the roles of $v$ and $h$ interchanged.
Thus at the corners, where the vertical and horizontal edges meet, both parton species contribute only one unpaired zero mode.

At mean-field level, therefore, our system has the following response theory that encodes the non-trivial edge polarization in terms of the auxiliary gauge field $b$, and the external U(1) gauge field $A$:
\begin{align} \label{Eq:Partons}
&\mathcal{L}_{pa}=\frac{\theta_v}{2\pi} (\partial_x  (b_0 +A_0/2)-\partial_t  (b_x +A_x/2))_{y=0}\nonumber\\
&-\frac{\theta_v}{2\pi}(\partial_x  (b_0 +A_0/2)-\partial_t  (b_x +A_x/2))_{y=L}\nonumber\\
&+\frac{\theta_h}{2\pi}(\partial_y  (-b_0 +A_0/2)-\partial_t  (-b_y +A_y/2))_{x=0}\nonumber\\
&-\frac{\theta_h}{2\pi}(\partial_y  (-b_0 +A_0/2)-\partial_t  (-b_y +A_y/2))_{x=L},
\end{align}
where we require $\theta_v = \theta_h$ to ensure $C_4$ symmetry.
Equivalently, in bulk form:
\begin{align}  \label{Eq:PartonResponse}
&\mathcal{L}_{pa}=\frac{\theta_v}{2\pi} \partial_y(\partial_x  (b_0 +A_0/2)-\partial_t  (b_x +A_x/2))\nonumber\\
&+\frac{\theta_h}{2\pi}\partial_x (\partial_y  (-b_0 +A_0/2)-\partial_t  (-b_y +A_y/2)).
\end{align}
Here $\theta_h,\theta_v=\pi$ in the bulk and are zero elsewhere.  The first two lines of Eq. (\ref{Eq:Partons}) denote the charge polarization of the $v$ parton on the two $x$-boundaries, and the last two lines denote the charge polarization of the $h$ parton on the y-boundaries. Since $v,h$ carry $\pm 1$ gauge charge for $b,$ and half a U(1) charge for $A$, this description provides a simple interpretation of the $1/4$ polarization on the edge of the SSPT HOTI phase. Namely, focusing on the electromagnetic response to $A_\mu$, we have a usual 1D SPT phase on the edge, but it is made of half-charged partons. Thus the usual electromagnetic boundary  polarization of 1/2, combined with the half-charge of each parton, leads to a net polarization of $1/4.$ Additionally, at the corners $\theta_h,\theta_v$ each contribute a  charge of $1/4$,  yielding a physical   charge of $1/2$ on each corner, as expected.

Going beyond mean field, we expect that including dynamics of the compact U(1) gauge field $b$ will lead to confinement due to instanton proliferation.  In this event, the two partons $v,h$ can appear only as the bound state $a = h v$. 
Correspondingly, the two charge $1/4$ corner modes from $v$ and $h$ individually are projected to a single charge-$1/2$ corner mode of $a$. 

Since $C_4$ symmetry guarantees
\begin{align}
 \theta_v=\theta_h  \equiv \theta \ ,
\end{align}
the effective electromagnetic response is:
\begin{align} 
&\mathcal{L}=\frac{\theta}{2\pi} (\partial_y\partial_x A_0 - \partial_t  (\partial_x A_y+\partial_y A_x)/2).
\end{align}
Since confinement effectively imposes subsystem charge conservation, there is no vector gauge field, and we can make the replacement $(\partial_x A_y+\partial_y A_x)/2=A_{xy}$, which exactly reproduces the quadrupole response from Eq.~\ref{qua}.

\subsection{Generating bosonic HOTI via fermionic topological quadrupole insulator}

To complete our discussion of bosonic quadrupolar response theories in 2D, we show how the strongly interacting bosonic HOTI with quantized quadrupole moment\cite{you2018higher,dubinkin2018higher} is connected to the non-interacting fermionic topological quadrupole insulator from Refs.~\onlinecite{benalcazar2017quantized,benalcazar2017electric}. More precisely, we will show how the bosonic HOTI Hamiltonian (\ref{bosona}) arises by coupling a pair of fermionic topological quadrupole insulators to a layer of local bosonic degrees of freedom.

\begin{figure}[h]
  \centering
      \includegraphics[width=0.25\textwidth]{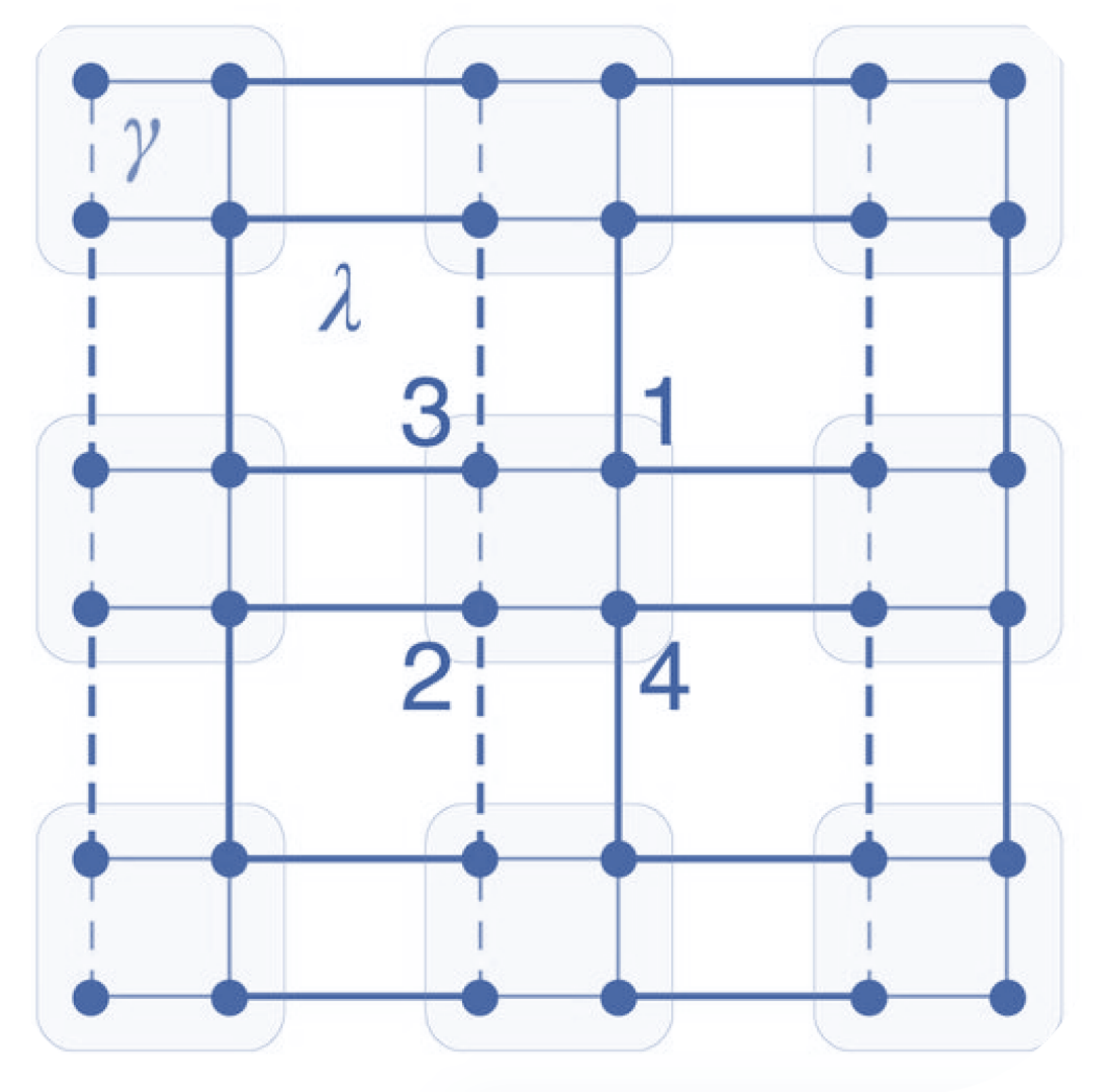}
      \caption{2D Topological quadrupole insulator with four spinless fermion orbitals per-unit cell. Each intra- and inter-cell plaquette contains a $\pi$ flux. The fermion only hops along the blue bonds within each plaquette. The dotted lines represent hopping terms with a relative minus sign compared to the solid lines. The edges resemble gapped SSH chains and the corners carry a fermion zero mode.}
  \label{f3} 
\end{figure}

Let us begin with a bilayer of the topological quadrupole insulator model proposed in Ref.~\cite{benalcazar2017quantized,benalcazar2017electric}. Each layer is a square lattice and consists of four spinless fermion orbitals per unit cell as shown in Fig. \ref{f3}. The tight binding lattice model for the quadrupole insulator has dimerized couplings in the $x$ and $y$ directions with an intra-cell coupling $\gamma,$ and an inter-cell coupling $\lambda$, where we take $|\lambda |> |\gamma|$ to put each layer in a HOTI phase.  Another key feature of the model is a $\pi$-flux in every inter- and intra-cell plaquette. In Fig. \ref{f3} we have shown a gauge choice were the dotted line represents a relative minus sign compared to the other hopping terms. This model has mirror symmetries $M_x, M_y,$ and $C_4$ symmetries up to a gauge transformation.

Let us take the fermions in the first (second) layers to have spin up  (spin down) respectively; we denote the corresponding  fermion operators as $f^{\dagger}_\uparrow$ $ (f^{\dagger}_\downarrow)$. Apart from the usual EM charge that couples to $A$, the bilayer system contains an additional U(1) symmetry  associated with conservation of $S^z$; the corresponding charge is the difference in fermion number of the up and down spins.   We will associate  a spin- $U(1)$ gauge field $A^s$ with this second $U(1)$ symmetry.  Provided both layers are identical, the bilayer system is also invariant under the usual time-reversal transformation that interchanges spin-up and spin-down fermions having $\mathcal{T}^2=-1$, as well as under a second anti-unitary symmetry that acts trivially on the spins.  Though the topological quadrupole insulator has a $\mathbb{Z}_2$ classification, with this additional interlayer U(1) symmetry, the interlayer coupling between two corner modes either breaks U(1) or $\mathcal{T}$ symmetry.

Next, we couple the bilayer of fermions with a layer of external Kondo spin degrees of freedom arranged on a square lattice. We choose a setup like that illustrated in Fig. \ref{f2} with four spins per unit cell, but without any inter-spin coupling. The four onsite spins are labeled as 1,2,3,4 in the same way as the fermion labeling in Fig.~\ref{f3}. If we express the spin operators in terms of hardcore bosons $a^{\dagger},$  then the we can represent the coupling between these Kondo spins and the  itinerant electrons via the interaction:
\begin{align} 
g(f^{\dagger}_{\downarrow,1}  f_{\uparrow,1} a^{\dagger}_1+f^{\dagger}_{\downarrow,2}  f_{\uparrow,2} a^{\dagger}_2+f^{\dagger}_{\downarrow,3}  f_{\uparrow,3} a^{\dagger}_3+f^{\dagger}_{\downarrow,4}  f_{\uparrow,4} a^{\dagger}_4+h.c),
\label{bbb}
\end{align}
\noindent in each unit cell. 
Since $a^\dag$ is electrically neutral, but raises $S^z$ by $1$, this induces an effective tunneling between the two quadrupole insulator layers that conserves $S^z$.

If we assume $g$ is small and implement a perturbative expansion from a topological quadrupole insulator Hamiltonian with $\gamma=0$ (which represents the HOTI state with zero-correlation length), the effective interaction between Kondo spins is, 
\begin{align} 
H=&g^2  \sum_{{\bf{R}}}(a^{\dagger}_{{\bf{R}},1} a^{\phantom{\dagger}}_{{\bf{R}}+e_x,3}+a^{\dagger}_{{\bf{R}},2} a^{\phantom{\dagger}}_{{\bf{R}}+e_x,4}\nonumber\\&+a^{\dagger}_{{\bf{R}},1} a^{\phantom{\dagger}}_{{\bf{R}}+e_y,4}+a^{\dagger}_{{\bf{R}},2} a^{\phantom{\dagger}}_{{\bf{R}}+e_y,4}+h.c).
\label{boson}
\end{align}
This exactly reproduces the bosonic HOTI model in Eq.~\ref{bosona} which has a topological phase protected by $U(1)\times \mathcal{T}\times C_4$ symmetry, where here $U(1)$ is conservation of $S^z$. 
Based on our previous discussions, this system will exhibit a quadrupolarization response
\begin{align} 
&\mathcal{L}_{a}=\frac{1}{2} (\partial_x \partial_y A^{s}_0-\partial_t (\partial_x A^{s}_y+\partial_y A^{s}_x)/2).
\end{align}

\section{3D HOTI with hinge modes: A dipolar Chern-Simons theory}

In this section we extend the scope of our concept of a topological dipole response to 3D HOTIs with gapless hinge modes. The first  2nd-order TI in 3D with protected, gapless hinge states was proposed in non-interacting, fermionic band theory\cite{benalcazar2017electric,langbehn2017reflection,song2017d,schindler2017higher}, and was later extended to strongly interacting bosons and fermions\cite{you2018higher,tiwari2019unhinging}. While the phenomenological understanding and mathematical classification of HOTIs in 3D is rather complete, a topological field theory description, and the associated quantized response properties, is less explored \cite{you2018higher}. %{\fb But there is some work -- can we stick the refs here too?} 

In first order topological phases, such topological response theories have provided a wealth of new theoretical avenues for discovery.
Notably, in many cases our theoretical understanding of these phases can be connected with their characteristic experimental signatures through a topological field theory, from which both bulk topology and the experimentally relevant topological responses can be derived.  Among the most striking examples are those where the bulk topology leads to quantized observable properties, such as a quantized bulk or surface Hall conductance. In this Section, our aim is to develop a topological response theory describing for 3D, 2nd-order HOTIs that are  protected by $C^{\mathcal{T}}_4$ symmetry (i.e., the product of $C_4$ and $\mathcal{T}$). 

Specifically, we describe a topological field theory that can be viewed as a Chern-Simons-like topological response of {\it dipoles}; we therefore refer to it as a dipolar Chern-Simons theory.  Similar to the situation in 2D discussed above for the quadrupolarization response, the HOTI response can be viewed as emerging from the topological response of a particular higher-rank gauge theory, upon relaxing subsystem symmetry to global charge conservation, thus extending the close relationship between HOTI phases and subsystem symmetric systems to three dimensions. In Section \ref{sec:higherdcs} we show that our theory is one member of a hierarchy of topological multipole responses, of which the conventional 2D Chern-Simons response for a Chern insulator is the lowest member. To this end we provide a family of multipole Chern-Simons terms, one for each spatial dimension, that describe transport of charge and multipole moments.  

Our dipolar Chern-Simons theory predicts a number of experimentally measurable responses that are  characteristic of a second order 3D HOTI.  In analogy with 2D Chern-Simons theory, our 3D dipolar Chern Simons theory predicts a current anomaly, but this time at the hinges. It also predicts transverse charge and dipole currents in the presence of electric field gradients and uniform electric fields respectively. Finally, it predicts a magnetic-quadrupole response to an applied scalar potential. While the derivation of our 3D dipolar Chern Simons theory is proposed at a field theory level, its physical consequences, including the quantized dipole current response, can be sharply demonstrated at a microscopic level in terms of the transport generated by the low-energy modes of a free-fermion 3D, 2nd-order HOTI.   Thus our response theory provides a set of straightforward experimental signatures to verify and measure the salient features of the 2nd order HOTI. Our discussion also highlights the relation between the 3D chiral hinge HOTI and the 2D quadrupole insulator via a dimensional reduction and charge pumping process\cite{benalcazar2017electric}. More generally, these arguments can be extended to higher dimensions, and thus can be used to predict the topological responses of higher dimensional HOTIs and to connect to a multipolar version of axion electrodynamics that we breifly discuss below in Section \ref{sec:higherdcs}.

\subsection{Review of 2D Chern-Simons response}

\begin{figure}[h]
  \centering
      \includegraphics[width=0.5\textwidth]{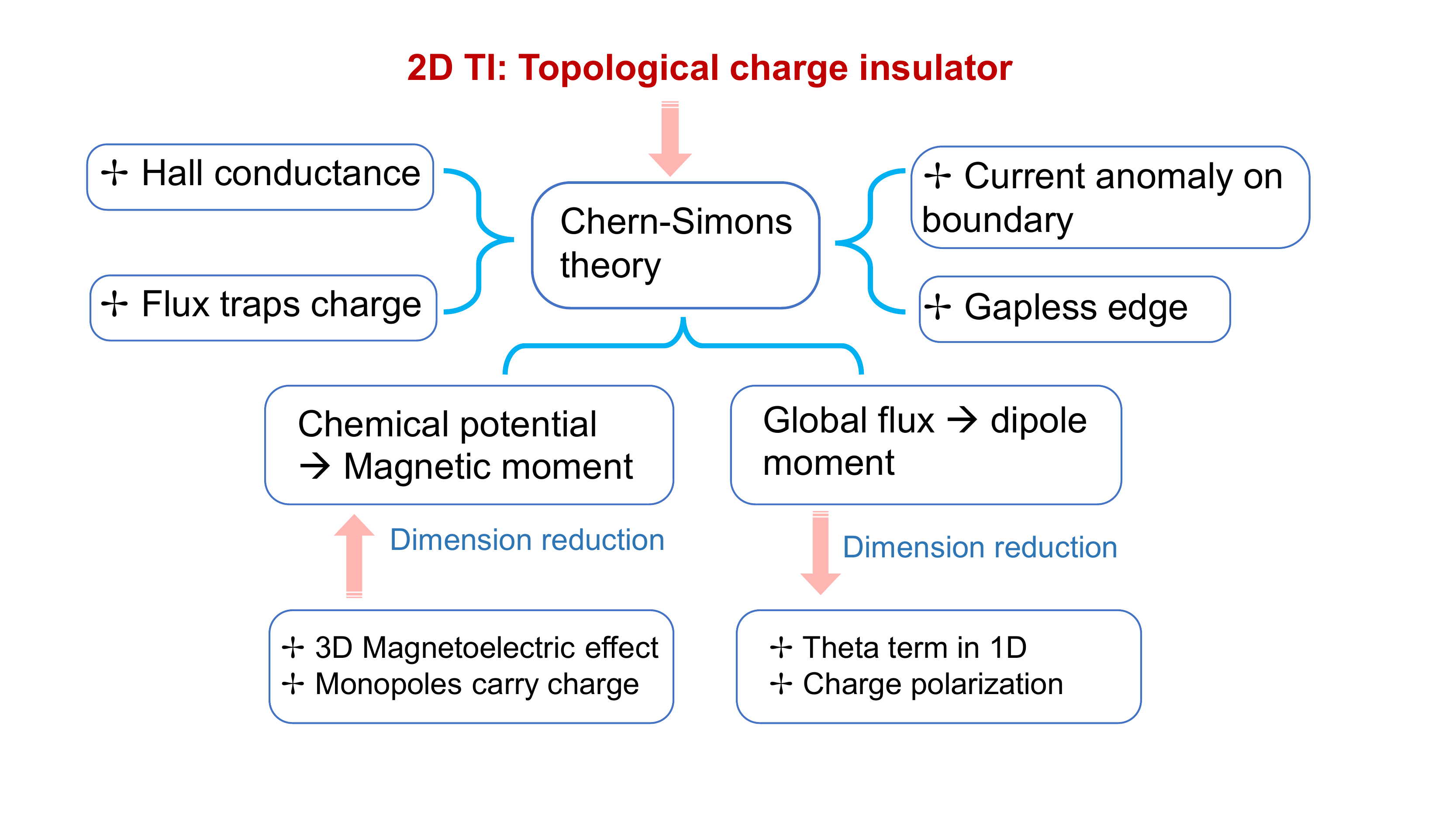}
      \caption{Summary of 2D Chern-Simons response and related effects in 1D and 3D.}\label{fig:cs1}
\end{figure}

To set the stage for the hierarchy of Chern-Simons responses, let us begin with the first-order Chern-Simons term that describes the quantum Hall response of a Chern insulator.  In 2D Chern insulators, the Chern-Simons response theory reveals both the nontrivial topological structure in the bulk, and the corresponding gapless edge. The Chern-Simons response theory is also at the crux of a variety of important topological phenomena, not just in 2D, but also 1D and 3D as diagrammed in Fig. \ref{fig:cs1}.

\begin{figure}[h]
  \centering
      \includegraphics[width=0.5\textwidth]{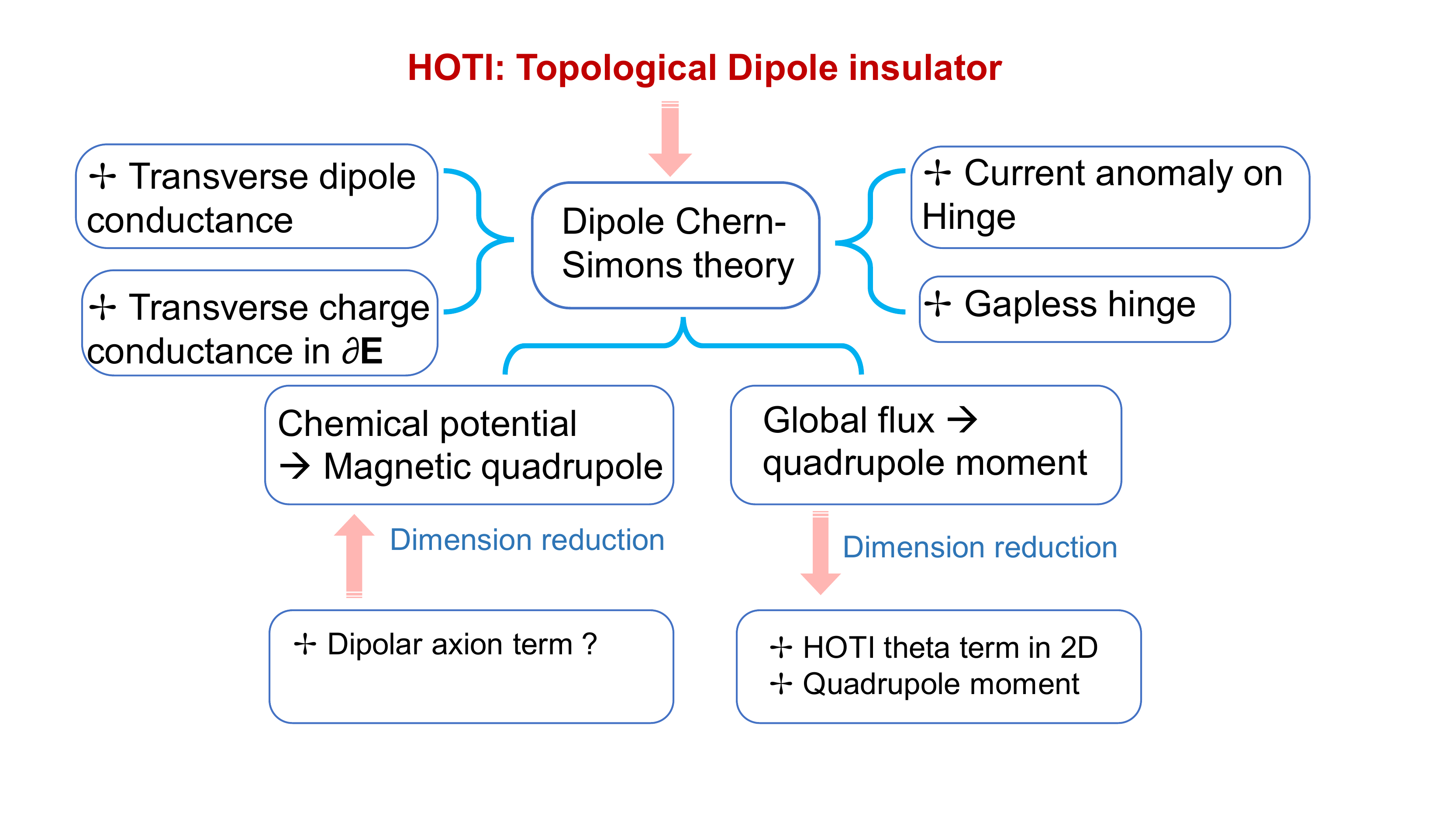}
      \caption{Summary of 3D dipole Chern-Simons response and related effects in 2D and 4D.}
  \label{fig:dipsummary} 
\end{figure}

The 2D Chern-Simons response to a background electromagnetic field is given by
\begin{align} 
&\mathcal{L}_{cs}=\frac{K}{4\pi}\epsilon^{\mu\nu\rho}A_\mu \partial_\nu A_\rho , 
\end{align}\noindent where $K$ is an integer (as we will confirm below). Let us discuss some of the consequences of the Chern-Simons response that we will find analogies for in the dipole version.
First and foremost, the Chern-Simons term indicates a charge Hall response where a current is generated  transverse to an applied electric field (see Fig. \ref{fig:fullresponse}a),
\begin{align} 
&j^i =\frac{K}{2\pi} \epsilon^{ij} E_j. \label{eq:hallcurrent}
\end{align}
Furthermore, it is well-known that the Chern-Simons term lacks gauge invariance for a system with boundary and requires a chiral charge mode circulating along the edge to restore invariance (see Appendices \ref{app:dimred},\ref{app:bosonization}).
In addition, if we change the background scalar potential $\Delta A_0=U$, a finite change in magnetic dipole moment is induced according to (see Fig. \ref{fig:fullresponse}c):
\begin{align} 
&\Delta M=\frac{K}{2\pi} U,
\end{align}\noindent where the magnetization $M$ is calculated by varying $\mathcal{L}_{cs}$ with respect to the magnetic field $B.$
Such a magnetic dipole moment is exemplified through the (bound) chiral edge current circulating along the boundary. The latter effect can also be connected to the Streda formula\cite{streda1982}
\begin{align}
    &\rho=\frac{K}{2\pi}B,\nonumber\\
    \implies &\frac{\partial\rho}{\partial B}=\frac{K}{2\pi}.\label{eq:streda}
    \end{align}

Next, let us make a connection between the Chern-Simons response and electric polarization. By expanding out all of the terms, we can write
\begin{equation}
    \mathcal{L}_{cs}=\frac{K}{4\pi}\left(A_x E_y - A_y E_x+A_0 B\right).
    \end{equation} If we take variations with respect to the electric fields we find the polarization
    \begin{equation}
        P^{i}=-\frac{K}{2\pi}\epsilon^{ij}A_j .\label{eq:anomalouspolarization}
        \end{equation}\noindent While these equations are not gauge invariant, we can identify some physical predictions. First, if we take a time-derivative of $P^{i}$ then we recover the expression for the Hall current in Eq. \ref{eq:hallcurrent}, and if we take the spatial divergence of $P_i$ we should find a bound charge density, and indeed we recover Eq. \ref{eq:streda}. 
        
        Furthermore, we can consider a 2D Chern insulator on a torus and insert a flux $\phi_y=2\pi$ into a cycle of the torus. In terms of the gauge fields, the flux insertion is described by a shift of the gauge field $A_y\rightarrow A_y + 2\pi/L_y$.  Thus after such a process, the bulk action is changed by
\begin{align} 
&\delta [ \int dxdydt \mathcal{L}_{cs}]=K \int dx dt E_x,
\end{align}
which represents a shift of the $x$-component of the 2D charge polarization on the torus by $K/L_y$ (since the dipole couples to the electric field in the action).
If the Chern insulator contains a unique gapped ground state on the torus, then the large gauge transformation does not change the ground state manifold, so the shift of the polarization $K$ can only be an integer multiple of $1/L_y.$ This establishes the level quantization of the Hall conductivity, and provides a robust signature for experimental probes. 

As a comparison, we could perform this experiment on a thin cylinder. The effect would be to create a dipole moment of $K L_x$ (hence a polarization $P^x = K/L_y$) on the cylinder, that will lead to a bound charge of $\pm K$ on each end of the cylinder (see Fig. \ref{fig:fullresponse}b). If we only insert $\pi$-flux (or an odd-multiple of $\pi$) and shrink the radius of the cylinder in a dimensional reduction procedure, then this system can represent a 1D polarized SPT if we require additional symmetries to fix the quantization of the polarization in the $x$-direction\cite{axion2}. This will produce charges of $\pm K/2$ on the boundaries as expected for a 1D SPT. This  effect is manifest in the Chern-Simons response, which precisely reduces to the 1D $\theta$ term $\mathcal{L}_P$ where the Wilson loop of the vector potential in the compactified direction plays the role of $\theta$ after the dimensional reduction\cite{axion2} (see Appendix \ref{app:dimred}).  

Finally, one can make other connections between the Chern-Simons response and 3D and 1D phenomena. Beyond the dimensional reduction mentioned above, we can associate the Chern-Simons response to the 3D axion response from a dimensional extension point of view. Indeed, the 2D Chern-Simons term appears at the boundary of a 3D topological insulator with axion electrodynamics\cite{axion2}. The interesting feature is that the level of the surface Chern-Simons theory is half that of which can be generated in an ordinary Chern insulator in 2D. Additionally, starting from 1D we can also imagine a dimensional extension where we take a 1D polarized topological insulator and generate an analogy to the Chern Simons response via an adiabatic charge pumping process\cite{thouless1983quantization}, which makes an additional dimensional extension connection from 1D to 2D. We will discuss analogies to some of these properties in the rest of this section and in Section \ref{sec:higherdcs}.

\subsection{3D Rank-2 Dipolar Chern-Simons response}
\label{sec:rank2}
\begin{figure}[t!]
  \centering
      \includegraphics[width=0.4\textwidth]{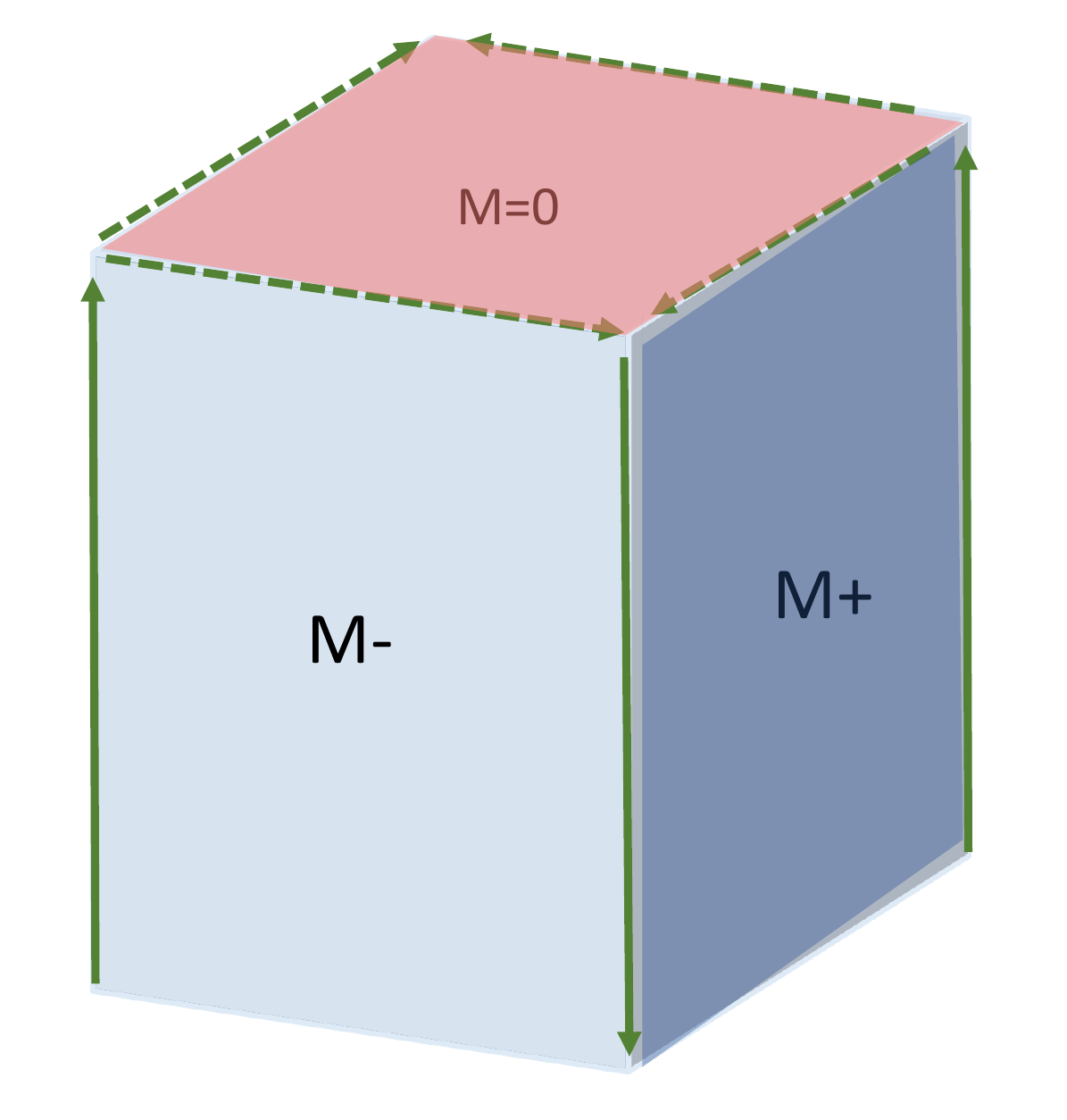}
      \caption{Surface $\mathcal{T}$ breaking patterns on a free-fermion chiral hinge insulator. We show an example where the red-shaded area is $\mathcal{T}$-invariant and $C_4$-invariant, and thus it harbors at gapless surface Dirac cone. The solid green lines indicate a chiral charge mode with a single quantum of conductance, the dashed lines effectively have half of a conductance quantum. Different choices for the top surface termination that are consistent with $C_4\mathcal{T}$-symmetry yield the same topological response.}
  \label{shrink} 
\end{figure}

We will now proceed to formulate a dipolar Chern-Simons term and describing the analogous features to many of the 2D Chern-Simons response phenomena mentioned above.
We expect this response theory to describe a 2$^{nd}$-order HOTI in 3D that harbors chiral modes along the hinges parallel to, say, the $z$-direction. The HOTI we will consider is protected by $U(1) \times C^{\mathcal{T}}_4$ symmetry, where $U(1)$ is global charge conservation, and $C^{\mathcal{T}}_4$ symmetry is the combination of a $\pi/2$ rotation around the $z$-axis, together with a time-reversal  transformation $\mathcal{T}$. A free-fermion version of such a HOTI can be generated starting from a strong, $\mathcal{T}$-invariant TI in 3D, and then breaking $\mathcal{T}$ while keeping $C^{\mathcal{T}}_4.$ The resulting system will generically have gapped surface states on the side-facing $xz$- and $yz$-planes. To preserve the symmetry, these surfaces are gapped by a time-reversal breaking mass that is odd under $C_4$. Such a pattern implies that the magnetic mass switches sign on neighboring surfaces, hence generating $\mathcal{T}$-breaking mass domain walls on the $z$-hinges between different side faces. These domain walls naturally  bind 1D chiral modes (see Fig. \ref{shrink}). The top and bottom surfaces ($xy$-planes) can remain gapless, but at the very least, they must be able to compensate the chiral hinge currents that flow between the top and bottom surfaces. We will now develop a response theory that describes these phenomena. 

%{\fb Here or earlier: insert summary of previous descriptions of response theories.  Explain why the theory we obtain is different (no bulk domain walls in $\theta$, etc.)}{\bf{TLH: Do you mean the discussion of the 3+1-d theta-term? If so I added it below, but above where it was before.}}

\begin{figure*}[t!]
  \centering
      \includegraphics[width=0.90\textwidth]{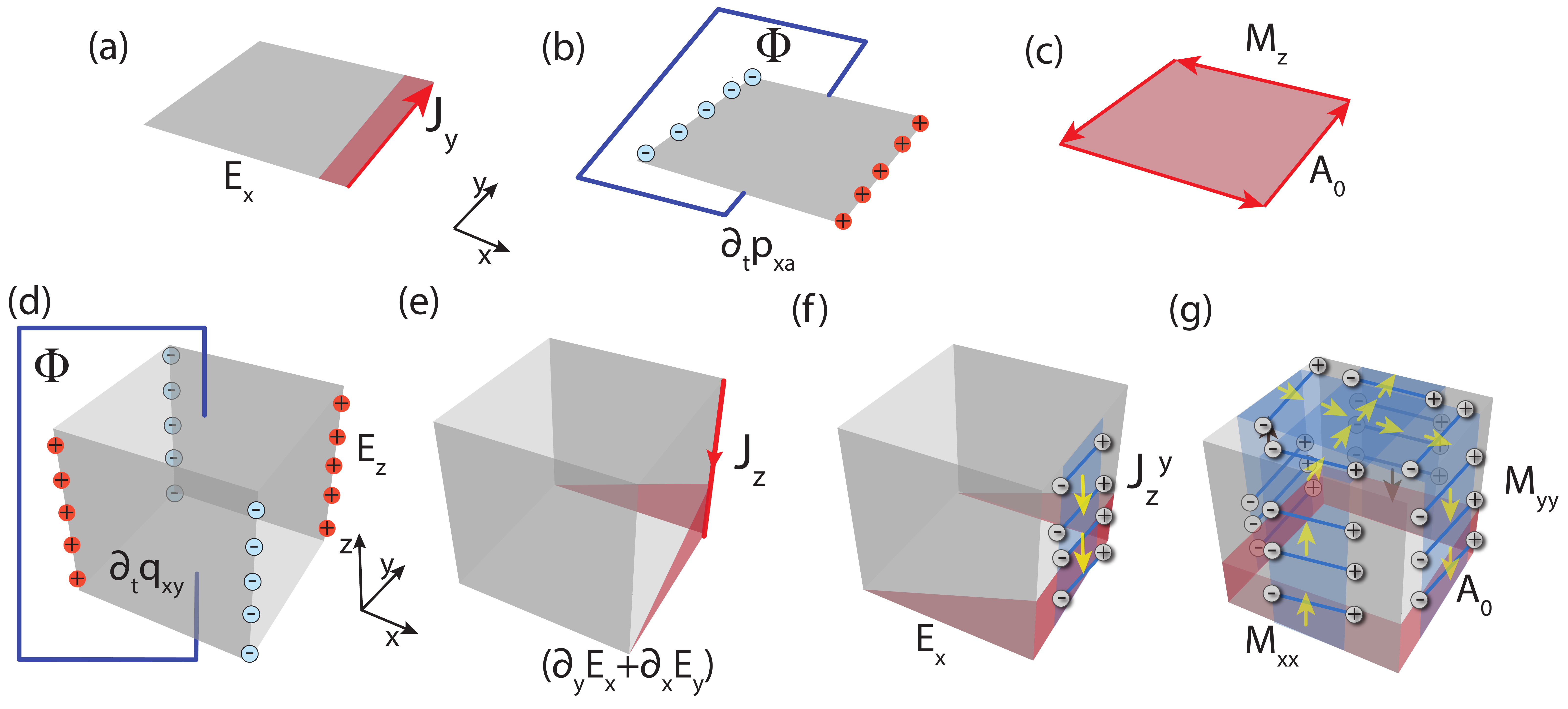}
      \caption{Illustration of Chern-Simons response phenomena for a 2D Chern insulator and a chiral hinge insulator. In subfigures (a),(c),(e),(f),(g) the shaded red regions indicate the application of a scalar potential $A_0.$ For the Chern insulator in (a),(b),(c) the first two subfigures have periodic boundary conditions in $y$ and open in $x$ while (c) is open in both. (a) is charge (Hall) current transverse to an applied electric field, (b) is the generation of dipole moment when flux is threaded in the periodic $y$-direction, and (c) is the change in magnetic moment when the scalar potential is changed indicated by the bound chiral currents. For the hinge Chern insulator subfigures (d),(e),(f) are periodic in $z$ and open in $x$ and $y$ while (g) is open in all three directions. (d) generation of quadrupole moment as flux is threaded in periodic $z$-direction (e) $J_z$ current transverse to an electric field gradient ($\partial_x E_y+\partial_y E_x)$ (f) dipole current transverse to an applied electric field (g) change in magnetic quadrupole as scalar potential is changed as indicated by the two pairs of circulating dipole currents. }
  \label{fig:fullresponse} 
\end{figure*}

To establish a topological response theory for this type of 3D HOTI, we adopt a phenomenological approach and propose a dipolar Chern-Simons theory. Instead of deriving a Chern-Simons response through the diagrammatic calculations of a microscopic model, we propose a response theory based on a connection between the 3D SSPT (or alternatively HOTI) and the 2D quadrupolar SSPT (or HOTI). We expect that such an approach will be successful since topological properties are essentially insensitive to details, and indeed we show that the resulting theory phenomenologically matches the microscopic behavior of the 3D, 2nd order phases of interest. We illustrate a variety of consequences of our response theory that are summarized in Fig. \ref{fig:dipsummary} and Fig. \ref{fig:fullresponse}d,e,f,g, and show how it fits into a natural dimensional hierarchy in Section \ref{Sec:Generalizations}.

%Historically, the Chern-Simons type response in 2D QAHE can be obtained via one-loop diagrammatic calculations of a band insulator with non-zero Chern number or via a phenomenological field theory argument. In 2D parity odd electromagnetic theory, the leading order gauge invariant coupling is the Chern-Simons term, which provides transverse Hall response and gapless boundary modes. As a result, the linear response theory for Chern insulators should contain a Chern-Simons term and such term survives even in the presence of strong interaction. 

%\subsubsection{3D Mixed-Rank Subsystem Symmetry Chern-Simons Term}
We begin by taking the same route to the HOTI that we did for the 2D case, i.e., we first look into a higher rank gauge theory response in a 3D subsystem symmetric system, and then consider breaking the subsystem symmetry down to a global symmetry. Let us consider a natural 3D generalization of our 2D SSPT to obtain a mixed-rank theory with $A_{xy}$ associated with ring-exchange (i.e., dipole hopping) terms in the $xy$ planes, and $A_z$ associated with charge hopping in the $z$-direction. This describes a system in which charge is conserved in each $xz$- and $yz$-plane, but not in each $xy$ plane. Consequently, a charge can hop only along the $z$-axis, while an $x$($y$)-dipole can hop along the $y$($x$)-direction.

For such a theory, we can write down a Chern-Simons-like response to the mixed-rank gauge fields,
\ba
& \mathcal{L}=\frac{1}{4\pi}[A_z E_{xy} + A_{xy} E_z - A_0 B]\nonumber\\
& A_z\rightarrow A_z+\partial_z \alpha,  A_0\rightarrow A_0+\partial_t \alpha,\nonumber\\
&A_{xy}\rightarrow A_{xy}+\partial_x \partial_y \alpha,\nonumber\\
&\mathcal{T}: A_0 \rightarrow A_0, A_{z},A_{xy} \rightarrow -A_{z},-A_{xy},\nonumber\\
&\mathcal{C}_4: A_0, A_z \rightarrow A_0, A_z , \ A_{xy} \rightarrow -A_{xy}
\label{rank}
\ea
where $B$ is a gauge-invariant operator involving only spatial derivatives of the gauge field,
\be
B =   \partial_x \partial_y A_z-\partial_z A_{xy}.
\ee
Such a Chern-Simons coupling breaks $C_4$ and $\mathcal{T}$ symmetry, but keeps the product $C_4 \mathcal{T}$ invariant.  Despite not deriving this action from a microscopic model, we claim that this is the only set of terms that is gauge invariant in the bulk and, like the usual Chern-Simons term, is bilinear in the gauge fields $A_0, A_{xy}, A_z$ and the physical fields $E_{xy}, E_z$, and $B$.  As such, it is the most relevant physical term that we can write down for this theory.  %We will see that like the usual Chern-Simons term, our action is gauge invariant only up to boundary contributions, which 

%We need to qualify this last statement.  $B$ by itself is gauge-invariant and odd under $C_4$ and $T$; $E_Z$ by itself is gauge invariant and even under $C_4$ and $T$.  Why do we exclude these terms?  We also are excluding total derivative terms like $A_{xy} \partial_\mu A_{xy}$, which are gauge invariant up to a boundary term.  Similarly for e.g. $\partial_t (A_{xy} A_z)$.  What seems to me to be special about this term is that the gauge variation is linear in $\alpha$ on the boundaries.}

Interestingly, when the gauge fields are non-singular this term can be reduced to a total derivative in spacetime. The theory thus describes a boundary response, and does not affect the bulk action (see Appendix \ref{app:no-go} for a general discussion of why this must be the case). This is in sharp contrast to both the usual 2D Chern-Simons term, and also to %Chern-Simons term should be distinguished from 
the fracton Chern-Simons and BF theories of Refs.~\onlinecite{you2019fractonic,Slagle2017-la}, which do describe a bulk response and nontrivial braiding statistics. It should also be contrasted with a conventional $\theta$-term in 3D which is a total derivative, but is also fully gauge invariant.  %In particular, we would like to emphasize that the rank-2 Dipolar Chern-Simons theory we proposed here is different from the Fracton Chern-Simons theory or BF theory in Ref.~\onlinecite{you2019fractonic,Slagle2017-la} which renders a bulk statistical effect and long range entanglement.

Specifically,  Eq.~\ref{rank} can be reduced to boundary actions on the $t, x,y,z,$ and boundaries, and the $xy$-hinges: 
\begin{eqnarray} \label{Eq:LbdyHO}
 \mathcal{L}_t&=&-\frac{1}{4\pi}[A_z A_{xy} |_{t=T}-A_z A_{xy}|_{t=0}],\nonumber\\
 \mathcal{L}_x&=&\frac{1}{4\pi}[A_z\partial_y A_{0}|_{x=L}-A_z\partial_y A_{0}|_{x=0}],\nonumber\\
 \mathcal{L}_y&=&\frac{1}{4\pi}[A_z \partial_x A_{0}|_{y=L}-A_z\partial_x A_{0}|_{y=0}], \nonumber\\
 \mathcal{L}_z&=&\frac{1}{4\pi}[A_0 A_{xy}|_{z=L}-A_0 A_{xy}|_{z=0}]\nonumber\\
\mathcal{L}_{xy}&=&-\frac{1}{4\pi}[A_0 A_z|_{(x=L,y=L)}+A_0 A_z|_{(x=0,y=0)}\nonumber\\&-&A_0 A_z|_{(x=0,y=L)}-A_0 A_z|_{(x=L,y=0)}].
\end{eqnarray}  %Heuristically, we can interpret these responses as follows. For $\mathcal{L}_x$ we see that applying a potential difference in the $y$-direction on $\hat{x}$ surfaces generates a charge current in the $z$-direction. This is a transverse current response with opposite response on opposite surfaces (note that we do not get a full Hall effect since charges are frozen in the $y$-direction). There is a similar transverse current response on the $\hat{y}$ surfaces. For $\mathcal{L}_z$ we find that changing the potential $A_0$ gives a change in the $xy$ dipole current localized on the $\hat{z}$ surfaces, and thus will change the quadrupole density $q_{xy}$ on those surfaces. Finally, on the hinges parallel to the $z$-direction we see that if we change the potential we generate a current in the $z$-direction. 
The responses in Eq. \ref{Eq:LbdyHO} do not appear manifestly gauge invariant, and instead can be summarized through a set of anomalous conservation laws on the surfaces and vertical hinges: 
\begin{eqnarray}
\partial^t j^{(x\pm)}_{0}+\partial^z j^{(x\pm)}_z&=&\pm\frac{1}{4\pi}\partial_y E_z\nonumber\\
\partial^t j^{(y\pm)}_{0}+\partial^z j^{(y\pm)}_z&=&\pm\frac{1}{4\pi}\partial_x E_z\nonumber\\
\partial^t j^{(z\pm)}_{0}+\partial^{x}\partial^{y} j^{(z\pm)}_{xy}&=&\mp\frac{1}{4\pi}E_{xy}\nonumber\\
\partial^z j^{(t\pm)}_{z}+\partial^{x}\partial^{y} j^{(t\pm)}_{xy}&=&\pm\frac{1}{4\pi}B\nonumber\\
\partial^t j^{(xy)}_{0}+\partial^{z} j^{(xy)}_{z}&=&-\frac{1}{4\pi}E_{z},\label{eq:rank2anomalies}
\end{eqnarray}\noindent where the superscript on the currents represents the surface on which the current is localized, and the $\pm$ indicates the direction of the normal vector on that surface (for the hinge term we have chosen to list only the response for the hinge at $(x=L, y=L)$ to avoid clutter, and the other hinges can be obtained by adding a $-$ sign for each 90-degree rotation). From the form of these equations we see that there is an anomalous dipole current on the top and bottom surfaces in the presence of a rank 2 electric field, an anomalous current in the $z$-direction on the side surfaces in the presence of an electric field gradient, and a conventional chiral anomaly-like response on the hinges. 

%{\fb Alternative phrasing to the following.  Unlike in CS theory, pure gauge is not a constraint here. If this way of motivating it is overly complicated, at the very least, I think we need an alternative way of justifying looking at $A_0 = 0$, $A_{xy}$ and $A_z$ pure gauge.   

In fact, the anomalies in Eq. (\ref{eq:rank2anomalies}) are not the full story: the action (\ref{rank}) is gauge invariant only up to boundary terms.  Thus to recover a fully gauge invariant theory, we must add boundary degrees of freedom that couple to gauge fields in such a way as to restore gauge invariance. 
%This can be accomplished by adding scalar fields at the boundary that couple to the gauge fields via $A_0 \partial_t \phi - A_{xy} \partial_x \partial_y \phi$ on the top and bottom surfaces, and $A_0 \partial_t \phi - A_z \partial_i \partial_z \phi$ on $i-z$ surfaces, with $i=x,y$.  These scalar fields are associated with the boundary action ... 
%
%Interestingly, these equations are not the full story as we must also include the effects of any consistent anomalies arising from boundary degrees of freedom. Indeed, what we have calculated so far is the contribution of the dipolar Chern-Simons response to boundary/hinge currents. To determine the form of possible boundary degrees of freedom we can use the standard arguments\cite{wensomething} and, for a moment, treat the gauge fields as dynamical fields. In particular, if we 
This is equivalent to letting $A_0$ be a Lagrange multiplier imposing the constraint that $B=(\partial_x\partial_y A_z-\partial_z A_{xy})=0.$ One possible solution for this constraint is $A_z=\partial_z \phi,A_{xy}=\partial_x \partial_y\phi.$
Plugging these forms into the dipolar Chern-Simons action we find
\begin{eqnarray}
S&=&\frac{1}{4\pi}\int d^4 x (-\partial_z \phi \partial_t \partial_x\partial_y \phi- \partial_x\partial_y\phi \partial_t\partial_z \phi)\nonumber\\
&=&\frac{1}{4\pi}\int d^4 x[\partial_x (\partial_y\partial_z\phi\partial_t \phi)+\partial_y (\partial_x\partial_z\phi\partial_t \phi)\nonumber\\&-&\partial_z (\partial_x\partial_y\phi\partial_t \phi)-\partial_x \partial_y(\partial_z\phi\partial_t \phi)].
\end{eqnarray} This will lead to a hinge action at $(x,y)=(L,L)$ of the form
\begin{equation}
    \mathcal{S}_{hinge}^{(L,L)}=-\frac{1}{4\pi}\int dz dt \partial_t\phi\partial_{z}\phi
\end{equation}\noindent which is exactly the form for a chiral mode on the hinge. Thus, we see that the full anomalous current on the hinge itself is contributed partly by the Chern-Simons response, and partly by the consistent anomaly of the hinge degrees of freedom yielding a total (covariant) hinge anomaly 
\begin{equation}
    \partial^t j^{(xy)}_{0}+\partial^{z} j^{(xy)}_{z}=-\frac{1}{2\pi}E_{z}
\end{equation} that alternates around the four hinges in a $C_4\mathcal{T}$-symmetric pattern.

Interestingly, this action also generates terms on the surfaces. For example, on a $z$-surface we find
\begin{equation}
    \mathcal{S}_{surf}^{z=L}=-\frac{1}{4\pi}\int dx dy dt \partial_t\phi\partial_y\partial_{x}\phi.
\end{equation}\noindent We show in Appendix \ref{app:bosonization}   that such a bosonic theory has a consistent anomaly given by exactly the form of the $z$-surface anomaly in Eq. \ref{eq:rank2anomalies} bringing the full (covariant) anomalous current response to
\begin{equation}
    \partial^t j_0^{(z+)}+\partial^x\partial^y j_{xy}^{(z+)}=-\frac{1}{2\pi}E_{xy}.
\end{equation}
 Similarly, the other anomalous conservation laws are modified by the consistent anomalies of the boundary degrees of freedom such that they all have a coefficient of $\frac{1}{2\pi}.$

To provide further intuition about this response action, let us now connect this response to a dimensionally reduced response action. We can perform dimensional reduction on Eq. \ref{Eq:LbdyHO} by compactifying the $z$-direction and threading $\pi$-flux through the compact direction by taking $A_z=\pi/L_z.$ As we take $L_z\to 0$ all fields  get projected onto the modes with no variation in $z$. In this limit we find
\begin{eqnarray}
S&=&\frac{1}{4\pi}\int d^4 x (A_z E_{xy}-A_{xy}\partial_t A_z-A_0\partial_x\partial_y A_z)\nonumber\\
&=&\frac{1}{4\pi}\int dx dy dt  [\partial_x (\theta\partial_y A_0)+\partial_y(\theta\partial_x A_0)\nonumber\\&-&\partial_x\partial_y(A_0 \theta)-\partial_t (A_{xy}\theta)],
\end{eqnarray}\noindent where in the last line  we defined $\theta (x,y,t)\equiv \int dz A_z(x,y, t).$ 

The ordinary 2D Chern-Simons action dimensionally reduces to the $\theta$-term encoding charge polarization\cite{axion2} (see App. \ref{app:dimred}), so let us compare this compactified dipole Chern-Simons action with the quadrupole response
\begin{equation}
S_{Q}=\frac{1}{2\pi}\int dx dy dt\, \theta(\partial_x \partial_y A_0 -\partial_t A_{xy}).\nonumber 
\end{equation} We can rewrite this as
\begin{eqnarray}
S_{Q}&=&\frac{1}{2\pi}\int dx dy dt [\partial_x (\theta\partial_y A_0)+\partial_y( \theta \partial_x A_0)-\partial_x\partial_y (\theta A_0)\nonumber\\&-&\partial_t(\theta A_{xy})+A_0\partial_x \partial_y \theta+A_{xy}\partial_t \theta ]
\end{eqnarray}\noindent where we have carefully separated bulk (last two terms) and boundary (first four terms) pieces. In comparison, we find that there are two ways in which this response does not seem to match the compactified dipolar Chern-Simons term: (i) there are two terms that depend on derivatives of $\theta,$ and (ii) the overall coefficient of $S_{Q}$ is twice as big.  For (i), since the associated SPT is described by a constant value of $\theta$ in the bulk, and we have carefully treated all of the boundary response terms, then we can drop the terms depending on derivatives of $\theta$ and treat the response as a purely boundary effect. Thus, $S_Q$ and the dimensionally reduced dipolar Chern-Simons have the same form up to a factor of two. This second discrepancy is illusory, but subtle, and arises because we are trying to obtain a lower dimensional response action by dimensionally reducing the \emph{boundary} response of a Chern-Simons response theory. We show in App. \ref{app:dimred} that dimensionally reducing the boundary currents of a Chern-Simons response will always produce a lower dimensional response that is too small by a factor of two because it does not include the consistent anomaly of the degrees of freedom on the boundary (as we discussed above). To recap, the underlying issue is that the boundary response derived from Chern-Simons is only part of the story, and boundary degrees of freedom provide an additional contribution (the consistent anomaly) that combines with the Chern-Simons piece to exactly compensate for the missing factor of two (i.e., the combined form generates the covariant anomaly). Thus, our dipolar Chern-Simons theory, when dimensionally reduced, should represent a lower-dimensional quadrupole SPT response theory with, e.g., charges of $\pm 1/2$ on the corners of a sample as expected from $S_Q$ when the anomalies from the boundary degrees of freedom are accounted for (see Appendix \ref{app:bosonization}).

\subsection{3D HOTI (Rank-1) Dipolar Chern-Simons Response}
As in the 2-dimensional case discussed above, we can connect our SSPT response theory to a related HOTI response %if we use the discussion in the previous sections where we made a connection between an SSPT and a HOTI 
by breaking the subsystem symmetry down to a global symmetry. Breaking subsystem symmetry demotes the rank-2 gauge field $A_{xy}$ to the rank-1 combination $\frac{1}{2} ( \partial_x A_y + \partial_y A_x ).$ Here,  in order to study the HOTI version of this response we can make this same substitution to find the action:
\begin{align} 
\mathcal{S}_{dcs}=\int d^4 x\frac{1}{8\pi}[&A_z (\partial_x E_y+\partial_y E_x)+ (\partial_x A_y +  \partial_yA_x ) E_z
\nonumber\\
&- A_0 (\partial_x  B_x-\partial_y  B_y) ],
\label{fff}\end{align}  which we call the HOTI dipolar Chern-Simons response.
 Let us now evaluate the symmetry properties of this term. The $C_4^{\mathcal{T}}$ symmetry acts on the gauge field as,
\begin{align} 
&\mathcal{T}\colon t \rightarrow -t, A_0 \rightarrow A_0, A_i \rightarrow -A_i,\nonumber\\
&C_4\colon (x,y) \rightarrow (y,-x), (A_x, A_y, A_z )\rightarrow (-A_y,A_x, A_z ).
\end{align}
 Hence the dipolar Chern-Simons term in Eq.~\ref{fff} is invariant under $C_4^{\mathcal{T}}.$ It is also invariant under charge-conjugation ($\mathcal{C}\colon A_\mu\to -A_\mu$) and $C_2$ rotation symmetry around the $z$ axis. Additionally it is odd under $C_4$, $\mathcal{T}$, reflection of the $z$-coordinate, and full coordinate inversion, so these last four symmetries must be broken for the response to be activated.

We will explore this 3D response further below where we  find that the physical consequences of the response theory are anomalous dipole and charge currents in response to electric fields and electric field gradients respectively, and additionally a potential-induced magnetic quadrupole moment. Explicitly we will find
\begin{eqnarray}
q_{xy}&=&\frac{A_z}{2\pi}\implies j_{xy}=-\frac{E_z}{2\pi}\label{eq:anomalousquad}\\
j_z &=& \frac{1}{4\pi}(\partial_x E_y +\partial_y E_x)\\
M_{xx}&=&-M_{yy}=\frac{A_0}{4\pi}
\end{eqnarray} where $q_{xy}$ is the 3D quadrupolarization, $j_{xy}$ is the combined dipole current for $x$-dipoles moving in $y$ and vice-versa, and $M_{ij}$ are the magnetic quadrupole moments with $j$-th magnetic moments separated in the $i$-th direction and vice versa. We will see below that these features result from the $C_{4}^{\mathcal{T}}$-symmetric pattern of chiral hinge currents. Thus, we will demonstrate that our effective response theory generates the characteristic response properties of a class of 3D chiral hinge insulators. 

To gain some initial intuition about this response we can again perform dimensional reduction by compactifying the $z$-direction and defining $\theta \equiv \int dz A_z (x,y,t).$ After doing so, this theory exactly reproduces the quadrupole moment response in Eq.~\ref{quad} up to the same caveats mentioned above for the SSPT version (i.e., that we take $\theta$ to be constant in the bulk of our sample, and account for the consistent anomaly contributions arising from boundary degrees of freedom to correct for a factor of two). The connection between the 3D dipolar Chern-Simons response  and the 2D quadrupole response,  succinctly captured by Eq. \ref{eq:anomalousquad}, turns out to be essential for the level quantization of the dipolar Chern-Simons theory, as we will return to in Section \ref{sec:laughlindipole}.

Finally, before we move on to a more detailed discussion, we would like to comment on another field theory description of HOTIs with $C_4^{\mathcal{T}}$ symmetry. In Ref.~\onlinecite{you2018higher}, a description of interacting HOTIs was introduced in terms of a non-linear $\sigma$ model with Wess-Zumino-Witten terms. In particular, for the HOTI protected by $C_4^{\mathcal{T}}$ symmetry, such a theory can be reduced to an axion electrodynamic response with a spatially dependent $\theta$ term (such a theory was first introduced in the context of non-interacting fermionic HOTIs in Ref. \onlinecite{schindler2017higher}):
\begin{align} 
& \mathcal{L}_{\theta}=\frac{\theta(\vec{r})}{16\pi^2}\epsilon^{\mu\nu\rho\sigma} \partial_\mu A_\nu  \partial_\rho A_\sigma.
\label{axionvor}
\end{align}
Here $\theta(\vec{r})$ is spatial-dependent, and transforms under $C_4^{\mathcal{T}}$ as,
\begin{align} 
& C_4^{\mathcal{T}}: \theta(x,y,z)\rightarrow -\theta(y,-x,z).
\end{align}
Subsequently, $2\pi$ vortex lines of $\theta(\vec{r})$ mark the locations of hinges and produce a current anomaly\cite{callan1985anomalies} that is compensated by chiral hinge modes. Despite the distinct forms of the two field theory descriptions, i.e., the dipolar Chern-Simons in Eq.~\ref{fff}, and the axion electrodynamic response in  Eq.~\ref{axionvor}, they predict similar underlying physics generated by chiral hinge modes. However, as  $\theta(\vec{r})$ is a function of space, and its vortex structure is essentially added by hand (under the appropriate symmetry constraints), it can be affected by  microscopic interactions. On the other hand, the dipolar Chern-Simons response has a constant, quantized (as we show below) coefficient and is not sensitive to such continuous deformations.

%Additionally, by symmetry and obstruction analysis, one can show that the dipole Chern-Simons theory is the only `topological' term, i.e., response term having a quantized coefficient, that is allowed in this class of HOTIs. 

%\begin{align} 
%&\mathcal{L}_{dcs}=\frac{1}{4\pi}[( \partial_x A_y +\partial_y A_x) (\partial_z A_0- \partial_t A_z)\nonumber\\
%&-A_z\partial_t ( \partial_x A_y +\partial_y A_x)+ 2 A_z\partial_x \partial_y A_{0}+\partial_x A_z \partial_y A_{0} \nonumber\\
%&+A_0\partial_z ( \partial_x A_y +\partial_y A_x)+\partial_x A_0  \partial_y A_z].
%\label{fff}
%\end{align} 

\subsubsection{Surface and Hinge Responses}
Let us now begin our detailed analysis of the physical phenomena associated to the dipolar Chern-Simons response.
The dipolar Chern-Simons theory is a total derivative on a closed manifold. As a result, it does not change the bulk constituent relations, but it does affect the boundary physics, and yields an interesting bulk-boundary correspondence as we indicated in the SSPT case above. %We will find that the response theory describes the response properties of a system having chiral hinge states on hinges parallel to $z,$ and chiral currents on the $x$ and $y$-hinges that carry half of the current on a $z$-hinge. 
To simplify our discussion let us assume the time direction is periodic so we can ignore total derivatives in the time direction. Then, one can decompose the theory into pieces that live on the spatial boundaries,
\begin{align} 
&\mathcal{S}_{dcs}=\mathcal{S}_{x}+\mathcal{S}_y+\mathcal{S}_z+\mathcal{S}_{xy}.
\end{align}
For side surfaces normal to the $x$ and $y$-directions, the dipolar Chern-Simons term will generate: \begin{align} &\mathcal{S}_{x}=\frac{1}{4\pi}\int dydzdt \left[
A_z \partial_y A_0|_{x=L}-A_z \partial_y A_0|_{x=0}\right]\nonumber\\
&\mathcal{S}_{y}=\frac{1}{4\pi}\int dxdzdt \left[A_z \partial_x A_0|_{y=L}-A_z \partial_x A_0|_{y=0}\right].\label{eq:sidesurfaceresponse}
\end{align} As we will show below, these terms imply that the side surfaces exhibit dipole currents transverse to an applied electric field, in addition to a charge response to a non-uniform electric field. The action $\mathcal{S}_{xy}$ appears at the intersection of surfaces normal to $x$ and $y$, i.e., at hinges parallel to the $z$ direction. On the hinges this action yields
\begin{eqnarray}
&&\mathcal{S}_{xy}=-\frac{1}{4\pi}\int dz dt[A_zA_0|_{(x,y)=(L,L)}+A_zA_0|_{(x,y)=(0,0)}\nonumber\\&-&A_zA_0|_{(x,y)=(0,L)}-A_zA_0|_{(x,y)=(L,0)}].\label{eq:hingeaction}
\end{eqnarray} This action predicts exactly the type of current anomaly that one finds at the boundary of a system described by a 2D Chern-Simons theory, and thus indicates the presence of chiral hinge modes.

For top and bottom surfaces normal to the $z$-direction we have the action 
\begin{align} 
\mathcal{S}_{z}=\frac{1}{8\pi}\int dx dy dt&\left[A_0\left(\partial_y A_x + \partial_x A_y   \right)|_{z=L}\right.\nonumber\\
&\left.-A_0\left(\partial_y A_x + \partial_x A_y   \right)|_{z=0}\right].
\label{eq:zsurface}
\end{align}
These terms represent the absorption/emission of the chiral currents that flow along the $z$-hinges and then split when they hit the top and bottom surfaces. We will return to this issue in more detail below. Interestingly, unlike the SSPT case we can integrate the terms in $\mathcal{S}_z$ by parts to put them in the same form as $\mathcal{S}_x, \mathcal{S}_y.$ This generates hinge actions on $xz$ and $yz$ hinges:
\begin{eqnarray}
&&\mathcal{S}_{xz}=\frac{1}{8\pi}\int dy dt[A_yA_0|_{(x,z)=(L,L)}+A_yA_0|_{(x,z)=(0,0)}\nonumber\\&-&A_yA_0|_{(x,z)=(0,L)}-A_zA_0|_{(x,z)=(L,0)}],\nonumber\\
&&\mathcal{S}_{yz}=\frac{1}{8\pi}\int dx dt[A_xA_0|_{(y,z)=(L,L)}+A_xA_0|_{(y,z)=(0,0)}\nonumber\\&-&A_xA_0|_{(y,z)=(0,L)}-A_xA_0|_{(y,z)=(L,0)}].
\end{eqnarray} These also appear to have a chiral anomaly, but with half the size of the vertical hinges. As we will see below, this has to do with the two-in-one-out chiral current flow on the hinges of an open cube sample where the side surfaces meet the top and bottom surfaces. Equivalently the hinge contribution can appear because in the rank-1 case the $x$-oriented dipoles are not equivalent to $y$-oriented dipoles, as would be the case for rank-2 case, and the conversion between the two inequivalent dipoles, which is necessary to satisfy $C_4\mathcal{T}$ symmetry, generates currents localized between the side and top surfaces.

%\begin{figure}[h]
  %\centering
  %    \includegraphics[width=0.3\textwidth]{hoti.png}
  %    \caption{}
 % \label{hoti} 
%\end{figure}

\subsubsection{Anomalous Quadrupole Moment}\label{sec:dipoleHallresponse}

In this section we will show that the anomalous conservation laws on the side surfaces are a manifestation of a bulk quadrupolarization $q_{xy}$ that is anomalous. Since anomalous does not have a necessarily have a unique meaning in this context, let us clarify that we are calling it anomalous in the following sense: a monopole of chiral currents, i.e., a single chiral current, has an anomalous current, and in particular $\rho\sim A_x.$ For a dipole of chiral currents, as one would find at the edges of a Chern insulator, one finds anomalous dipole currents $P_x\sim A_y.$ Finally, in our case, where we have a quadrupole of chiral currents and we might expect to find $q_{xy}\sim A_z,$ which is exactly what we will now show.

Let us start from Eq. \ref{eq:sidesurfaceresponse} to analyze this result in more detail. Similar to the 2D Chern-Simons action discussed above, we can calculate two related responses by taking variations with respect to the fields $A_z, A_0$ (to find current, charge) or the gradients $\partial_i A_z, \partial_i A_0$ (to find dipole currents and densities). From this data we show in Appendix \ref{app:anomalyderivation} how we can derive the following set of anomalous conservation laws for the charge currents:
\begin{eqnarray}
\partial^t j^{(x\pm)}_{0}+\partial^z j^{(x\pm)}_z&=&\pm\frac{1}{2\pi}\partial_y E_z\nonumber\\
\partial^t j^{(y\pm)}_{0}+\partial^z j^{(y\pm)}_z&=&\pm\frac{1}{2\pi}\partial_x E_z\nonumber\\
\partial^t j^{(xy)}_{0}+\partial^{z} j^{(xy)}_{z}&=&-\frac{1}{2\pi}E_{z},\label{eq:rank1xyanomalies}
\end{eqnarray}\noindent where again we have only included the result in the last equation for the hinge at $(x,y)=(L,L).$ For the dipole currents we find
\begin{eqnarray}\partial^t j^{(x\pm),y}_{0}+\partial^z j^{(x\pm),y}_z&=&\mp\frac{1}{2\pi} E_z\nonumber\\
\partial^t j^{(y\pm),x}_{0}+\partial^z j^{(y\pm),x}_z&=&\mp\frac{1}{2\pi} E_z.\label{eq:rank1xydipoleanomalies}
\end{eqnarray}

Now let us try to interpret these equations in a simple square-cylinder geometry where the $z$-direction is periodic and the $x$ and $y$-directions are open (see Fig. \ref{fig:fullresponse}d). Consider the application of a uniform electric field $E_z=-\partial_t A_z.$ We find the equations
\begin{eqnarray}
\partial_t \rho^{(x\pm)}&=&\mp\frac{1}{2\pi}\partial_t \partial_y A_z\nonumber\\
\partial_t \rho^{(y\pm)}&=&\mp\frac{1}{2\pi}\partial_t \partial_x A_z\nonumber\\
\partial_t \rho^{(xy)}&=&\frac{1}{2\pi}\partial_t A_z.\label{eq:surfpolarization}
\end{eqnarray} From these equations we can determine an interesting relationship. Let us focus on the surfaces $x=L$ and $y=L$ that intersect at the hinge where $(x,y)=(L,L).$ From the first equation we find that the anomalous surface charge density is just arising from a spatial variation of a surface polarization defined through $\rho=-\partial_y P^{y}_{surf}$ where $P^{y}_{surf}\equiv\frac{A_z}{2\pi}\vert_{x=L}$ (up to a non-anomalous constant of integration). From the second equation we find an analogous result with a surface polarization $P^x_{surf}=\frac{A_z}{2\pi}\vert_{y=L}$ on the $y$-surface. Finally, we find that the corner charge density right at the hinge is $\rho_{corner}=\frac{A_z}{2\pi}.$ Now we can use the formula from Ref. \onlinecite{benalcazar2017quantized} for the definition of the quadrupole moment density \begin{equation}
    q_{xy}\equiv P^{x}_{surf}+P^{y}_{surf}-\rho_{corner}=\frac{A_z}{2\pi}.\label{eq:anomalousqxy}\end{equation}
This result is remarkable as it establishes that, in analogy to the 2D Chern-Simons term (see Eq. \ref{eq:anomalouspolarization}), where threading a flux quantum generates a polarization (manifest through opposite integer charges on the opposing boundaries of a cylinder), here the flux quantum threading generates a quadrupole moment $q_{xy}$ (manifest through alternating integer charges localized on the hinges). Indeed this is exactly what one would expect in a chiral hinge insulator since some free-fermion, microscopic models of the 3D chiral hinge insulator can be viewed as 2D quadrupole insulators undergoing an adiabatic, topological pumping process as a function of $k_z,$ where $q_{xy}$ changes by an integer during the process\cite{benalcazar2017electric} (N.B. $k_z\to k_z+A_z$ in minimal coupling). Furthermore, we will argue that this response can be used to prove the quantization of the dipolar Chern-Simons coefficient in Section \ref{sec:laughlindipole}.

\subsubsection{Transverse charge current in response to electric field gradient}
In a related effect we can also see see that there is a transverse \emph{charge} current response if we impose a \emph{non-uniform} electric field that induces a potential imbalance between the four $z$-hinges (see Fig. \ref{fig:fullresponse}e).  Let us assume all of the fields are static, and we apply an electrostatic potential $A_0(x,y).$ From our response equations, on the $xz$-surfaces we find
\begin{align} 
& j_z|_{y=L}= \frac{1}{2\pi}\partial_x A_0\implies\nonumber\\
&J_z|_{y=L} = \int dx~j_z|_{y=L}= \frac{1}{2\pi}(A_0|_{x=L,y=L}-A_0|_{x=0,y=L}),\end{align}
and similarly
\begin{align}
J_z|_{y=0}= -\frac{1}{2\pi}(A_0|_{x=L,y=0}-A_0|_{x=0,y=0}).
\end{align}
Likewise, the response on the  $yz$ boundaries has a similar form,
\begin{align} 
&J_z|_{x=L}= \frac{1}{2\pi}(A_0|_{x=L,y=L}-A_0|_{x=L,y=0}),\nonumber\\
&J_z|_{x=0}= -\frac{1}{2\pi}(A_0|_{x=0,y=L}-A_0|_{x=0,y=0}).
\end{align} Finally we can take into account the current coming directly from the hinges
\begin{eqnarray}
J_z|_{x=L,y=L}&=&-\frac{1}{2\pi}A_0|_{x=L,y=L}\nonumber\\
J_z|_{x=0,y=L}&=&\frac{1}{2\pi}A_0|_{x=0,y=L}\nonumber\\
J_z|_{x=L,y=0}&=&\frac{1}{2\pi}A_0|_{x=L,y=0}\nonumber\\
J_z|_{x=0,y=0}&=&-\frac{1}{2\pi}A_0|_{x=0,y=0}.\nonumber\\
\end{eqnarray} 

Now we can add up all the contributions and we find that the total current in the $z$-direction is given by
\begin{eqnarray}
    J_z&=&\frac{1}{2\pi}(A_0|_{x=L,y=L}-A_0|_{x=0,y=L}\nonumber\\&-&A_0|_{x=L,y=0}+A_0|_{x=0,y=0})\nonumber\\
    &\equiv & \frac{1}{2\pi} \Delta_x \Delta_y A_0.
\end{eqnarray}
Thus, by applying an electric field gradient $(\partial_x E_y+\partial_y E_x)$ a transverse charge current is generated in the $z$-direction\begin{equation}
    j_z=\frac{1}{2\pi}(\partial_x E_y+\partial_y E_x).
\end{equation} If we apply a non-uniform potential as illustrated in Fig. \ref{fig:fullresponse}e, that, for example, takes a value of $\alpha$ on the $(x=L, y=L)$ hinge, and is vanishing on the other three hinges we find \begin{align} 
& J_z=\frac{\alpha}{2\pi}.
\label{surhall}
\end{align} 

Now consider the same setup but with fully open boundary conditions. To realize this potential we can take $A_0|_{x=L}=\alpha y/L, A_0|_{y=L}=\alpha x /L, A_0|_{z=L}=A_0|_{z=0}=\alpha x y/L^2$ on the surfaces. We have already calculated the the full current flowing in the $z$-direction so we just  need to know the responses on the top and bottom surfaces. Starting from Eq. \ref{eq:zsurface} we can derive (see Appendix \ref{app:anomalyderivation}) the anomalous conservation laws:
\begin{align} 
&\partial^\mu j^{(z\pm)}_\mu=\mp\frac{1}{4\pi}(\partial_x E_y +\partial_y E_x)\nonumber\\
&\partial^{\mu}j^{x,(z\pm)}_{\mu}=\pm\frac{1}{4\pi}E_y\nonumber\\
&\partial^{\mu}j^{y,(z\pm)}_{\mu}=\pm\frac{1}{4\pi}E_x,\label{eq:topsurfaceanomalies}
\end{align}\noindent where the signs are correlated with the surface normal being $\pm \hat{z}$  respectively. We note that these anomalies have a coefficient which is half that of the side surfaces; we will comment on this further below.

For our field configuration we can calculate $j_x=-\frac{\alpha x}{4\pi L^2}$ and $j_y=-\frac{\alpha y}{4\pi L^2}$ from the anomalous responses on the top surface. Then we can take the divergence of the surface currents $\partial_x j_x+\partial_y j_y=-\frac{\alpha}{2\pi L^2}.$ Thus, if we integrate this over the top surface we find that the total charge entering/exiting the surface is $\partial_t Q|_{z=L}=\frac{\alpha}{2\pi}.$ If we repeat the calculation for $z=0$ we find $\partial_t Q|_{z=L}=-\frac{\alpha}{2\pi}$ which matches the total current flow from the side surfaces. Thus, for the entire system the total charge is conserved, and in the presence of a non-uniform electric field, charge flows between the top and bottom surfaces along the side surfaces and hinges. 

For another helpful illustration let us revist the free fermion HOTI which harbors chiral hinge currents. A typical $xy$ surface state is illustrated in Fig.~\ref{shrink}, where the top $xy$ surface contains a gapless Dirac cone. The four side faces have a sign-alternating, $\mathcal{T}$-breaking mass that gaps out the Dirac cones on the side faces. This configuration has domain walls at the hinges that bind gapless chiral fermion modes. Due to the gapless nature of the top surface, there are no generic, localized chiral fermion \emph{modes} on the hinges parallel to the x or y directions. However, Ref. \onlinecite{chu2011surface} showed that for this configuration of surface masses the gapless Dirac cone on the top surface just acts as a ``wide" domain wall, and half of the chiral current itself will still be localized at each of the hinges parallel to the x and y-directions. When the chiral mode traveling along the z hinge impinges on the top or bottom surface, the current it carries splits in half to flow evenly along the x and y-hinges. Thus, each x or y-hinge effectively exhibits \emph{half} of the conductance as the z-hinges, which, in this case, is half of a conductance quantum. This phenomenology is exactly what is described by some pieces of the dipolar Chern-Simons theory.

Before moving on let us also remark that the half-integer value ($1/4\pi$ in our units) for the coefficients of the anomalous conservation laws for the $z$-surfaces are a salient property of the HOTI with $C_4^{\mathcal{T}}$ symmetry. If one merely decorates the four side surfaces of a trivial bulk with a Chern insulator in a $C_4^{\mathcal{T}}$ symmetric way to generate chiral hinge currents, the decoration always attaches full chiral fermion modes on the hinges parallel to the $x$ and $y$-directions. Such a surface decoration changes the top surface anomaly by
\begin{align} 
&\partial^{\mu} j_{\mu}=\frac{1}{2\pi}(\partial_x E_y+\partial_y E_x),
\end{align} which is twice the value found in Eq. \ref{eq:topsurfaceanomalies}.
 Subsequently, this surface decoration can change the dipole anomaly in Eq.~\ref{eq:topsurfaceanomalies} by an even number and therefore only an odd anomaly is an intrinsic feature coming from the HOTI bulk.

\subsubsection{Transverse dipole current in an electric field}
Now consider a simple experiment for a system that is periodic in the $z$-direction, but open in the $x$ and $y$ directions. We want to apply an electric field, say $E_x=\partial_x A_0=\alpha$ and consider the response. We find a dipole current $j_{z}^y=-\frac{\alpha}{2\pi}$ at $x=L,$ and the dipole currents at $y=0,L$ $j_{z}^x\vert_{y=L}=-j_{z}^y\vert_{y=0}=-\frac{\alpha x}{2\pi}.$ The spatially dependent dipole currents generate a charge current $j_z=\partial_x j_{z}^{x}$ on the $y$-surfaces.  The net current on each such surface is canceled by the $J_z$ currents at the hinges $(x=L,y=L)$ and $(x=L,y=0).$ Thus the remaining response is just the surface dipole current (see Fig. \ref{fig:fullresponse}f) \begin{equation}
    j_{z}^y=-\frac{E_x}{2\pi}.
    \end{equation} One can find the related response \begin{equation}
    j_{z}^x=-\frac{E_y}{2\pi},
    \end{equation} by symmetry.

If we keep the same field configuration, but now have open boundary conditions in all three directions, we find the dipole currents
$j_{y}^x=j_{x}^y=\frac{\alpha x}{4\pi L_x}$ on the top surface, and in particular $j_{y}^x=j_{x}^y=\frac{\alpha}{4\pi}$ where the $x$-surface hits the $z$-surface at $x=L$. From this we see that the dipole current traveling up the $x$-surface splits into two equal pieces on the top surface which represents a quadrupolar-like charge flow on the top surface where charges enter the $z$ surface then split into the $x$ and $y$ directions symmetrically. This provides one interpretation for why the anomalous conservation laws on the top/bottom surface have an extra factor of $1/2$ when compared with the side surfaces. Indeed, this matches the one-in,two-out pattern expected for the chiral current flow from the side surfaces to the top surface of the HOTI. An additional point of interest is that as the dipole current flows up the side surface (say $yz$ surface) and then hits the top surface half of the flowing dipoles that were pointing in $y$ are converted to dipoles pointing along $x.$ This non-conservation of the dipole leaves a charge current localized at the hinge with a magnitude equal to the amount of converted dipole moment, i.e.,  $\left|J_y\vert_{hinge}\right|=\left|\frac{A_0}{4\pi}|_{hinge}\right|.$

\subsubsection{Magnetic Quadrupole Moment}
 In a 2D Chern insulator, changing the  electrostatic potential creates a non-zero magnetic moment which can couple with an external magnetic field (see Fig. \ref{fig:fullresponse}c). As the potential is increased, the gapless chiral boundary modes acquire momentum and generate circulating currents that are bound to the edges; a signature that is indicative of a nonzero orbital magnetic moment in the bulk: 
\begin{align} 
& \frac{\delta S_{cs}}{\delta B}=M=\frac{A_0}{2\pi}.
\end{align} Despite the right hand side not being gauge invariant we can use this response to predict the \emph{change} in orbital magnetization when a change in the scalar potential $A_0$ is made.

 For a 3D HOTI, the application of a chemical potential gives rise to an analogous effect (see Fig. \ref{fig:fullresponse}g). In this case a finite potential does not  generate a non-zero magnetic moment in the $xy$ plane since there is $C^{\mathcal{T}}_4$ symmetry; instead, it induces a magnetic quadrupole moment. Such a quadrupole moment will naturally couple to a non-uniform magnetic field, e.g., a magnetic field gradient, and it is essentially tied to the spatially separated circulating hinge currents. One can calculate the change in magnetic quadrupole moment in response to a change in $A_0$ by calculating the hinge current response and relating them to a change in spatially separated magnetic dipole moments. When we apply $\Delta A_0$ we find
 \begin{eqnarray}
 \Delta M_{x}\vert_{x=L_x}=-\frac{\Delta A_0}{4\pi}L_y L_z=-\Delta M_x\vert_{x=0}\nonumber\\
 \Delta M_{y}\vert_{y=L_y}=\frac{\Delta A_0}{4\pi}L_x L_z=-\Delta M_y\vert_{y=0}.
 \end{eqnarray} To calculate the magnetic quadrupolarization we can simply weight these magnetic moments by their positions and divide by the volume to find
\begin{equation}
\Delta M_{xx}=-\Delta M_{yy}=-\frac{\Delta A_0}{4\pi},\label{eq:magquad}
\end{equation}\noindent where $M_{ij}$ represents opposite $i$ magnetic moments separated in the $j$ direction. This result is analogous to the connection between the scalar potential and magnetic dipole moment in the 2D Chern-Simons theory.

Although we presented the discussion in terms of loops of charge currents, we could have just as well explained it is current loops of dipoles. Indeed, when $A_0$ is changed for a system with fully open boundary conditions there are dipole currents circulating in the $xz$-planes with dipoles pointing in $y$ and dipole currents circulating in the $yz$-plane with dipoles pointing in $x$ as shown in Fig. \ref{fig:fullresponse}g. The magnitude of each of the two circulating dipole current is $A_0/4\pi$ and they clearly generate the two magnetic quadrupole components.

 We can also try to work out an analogy to the Streda formula. Let us assume periodic boundary conditions along the $z$-direction and open boundary conditions in $x$ and $y.$ We can calculate the density response from Eq. \ref{eq:sidesurfaceresponse} and we find
\begin{align}
 \rho(x=0)+\rho(x=L_x)=-\Delta_x \frac{1}{2\pi} (\partial_y A_z), \nonumber\\ \rho(y=0)+\rho(y=L_y)=-\Delta_y \frac{1}{2\pi} (\partial_x A_z).
\end{align}
If we add these contributions we find
\begin{align}
 \rho=-\Delta_x \frac{1}{2\pi} (\partial_y A_z)-\Delta_y \frac{1}{2\pi} (\partial_x A_z)\nonumber\\
 =\frac{1}{2\pi}\left(\Delta_y B_y-\Delta_x B_x\right),
\end{align}
where we have used the fact that terms containing a $\partial_z$ are assumed to vanish because of periodic boundary conditions. We can interpret this result using a Streda formula
\begin{equation}
    \frac{\partial \rho}{\partial (\partial_y B_y)}=-\frac{\partial \rho}{\partial (\partial_x B_x)}=\frac{1}{2\pi},
\end{equation} where the right hand side is the twice the coefficient of the dipolar Chern-Simons term in analogy with the usual Streda formula.

\subsection{Dipole Laughlin Argument, Quantization, and Dipole Pumping}\label{sec:laughlindipole}

At this stage, we have developed a dipolar Chern-Simons theory description for the electromagnetic response of a class of 3D HOTIs. This theory correctly captures the chiral hinge currents separating side surfaces, and an anomalous quadrupole moment $q_{xy}.$ While we have indicated that the coefficient of the dipolar Chern-Simons term is dimensionless and universal, we now answer the explicit question of whether it must be quantized in a HOTI, independent of microscopic details in the Hamiltonian. 

To elucidate the quantization we borrow the Laughlin argument, and the charge pumping picture for the usual quantum Hall quantization in 2D. One way to understand the quantization of the Hall conductance in an ordinary 2D Chern-Simons theory is to consider inserting $2\pi$ flux through a cylinder.  This flux insertion changes the  charge polarization of the ground state along the cylinder's length by $\sigma_{xy}$ (see Fig. \ref{fig:fullresponse}a for a cylinder geometry). If the Chern insulator has a unique, gapped ground state in the bulk, then this change in polarization just results from a net transfer of charge from one boundary to the other.  The change in total charge at each boundary is an integer only if $\sigma_{xy}$ is an integer.  Thus for weakly interacting Chern insulators, $K \in \mathbb{Z}$.  %, the charge polarization is invariant only under an integer shift, so the Hall conductance $\sigma_{xy}$ must be an integer. We also discussed that on a cylinder geometry such a flux insertion ultimately amounts to changing the charges on the edges by an integer.

This argument can be extended to a 3D HOTI with a dipolar Chern-Simons response. We illustrate the basic idea in Fig. \ref{fig:fullresponse}d where we have shown that threading flux through the periodic cycle of a 3D HOTI can generate a shift in the quadrupole moment analogous to the shift of the dipole moment mentioned above. Let us calculate the change in quadrupole moment  in terms of the dipolar Chern-Simons response. First, let us parameterize the flux insertion by the vector potential
\begin{equation}
A_z=\begin{cases} 0 & t < 0 \\
 \frac{2\pi t}{L_zT} & 0 \leq t \leq T \\ 
 \frac{2\pi t}{L_zT} & t > T \\
 \end{cases}
 \end{equation}
for some large time $T$. We can obtain the change of quadrupolarization  based on Eq. \ref{eq:anomalousqxy}, i.e., $q_{xy}=K\frac{A_z}{2\pi}$ where we have added an arbitrary coefficient $K.$ During the flux threading process $\Delta q_{xy}= K,$ which also indicates that $\Delta P^x_{surf}=\Delta P^{y}_{surf}=\Delta Q_{corner}=K.$ Thus by inserting $2\pi$ flux into the hole of the cylinder, the local charge on each hinge is shifted by $\pm 1$ depending on its chirality.

%\begin{figure}[h]
%  \centering
%      \includegraphics[width=0.3\textwidth]{cyl.pdf}
 %     \caption{Place the HOTI on thin annulus with PBC along z. The blue lines illustrated the chiral hinge modes. Inserting a $2\pi$ gauge flux in the hollow hole shifts the total quadrupole moment.}
 % \label{cyl} 
%\end{figure}

From this result, we can argue for the quantization of $K$. For any HOTI with a unique gapped ground state, any global $2\pi$ flux insertion should leave the infrared theory invariant. We find that if $K$ is an integer then the quadrupolarization changes by an integer quantum, the surface polarizations change by an integer quantum, and the amount of charge on the hinges changes by an integer, in exact analogy with the polarization and edge charges changing by a quantum in the 2D Chern-Simons response. Such a quantization of the coefficient of the dipolar Chern-Simons term confirms a robust transport signature for experiments, and represents a universal property that is insensitive to any microscopic details, and remains valid in the strong interacting limit.

Finally, let us comment on the topological classification of HOTIs described by our response theory. While we have found that the coefficient of the dipolar Chern-Simons term is quantized to an integer, this does not necessarily imply a $\mathbb{Z}$ classification of the topological phases. In general, one can always decorate the four side faces of a cubic sample with integer quantum Hall states in a $C_4^{\mathcal{T}}$ symmetric way. Such a surface decoration changes the integer coefficient by an even number. Since  bulk properties should remain unaffected by a change to the surface, only odd coefficients of the dipolar Chern-Simons theory in Eq.~\ref{fff} represent nontrivial bulk topological classes, so the classification is  $\mathbb{Z}_2$.

\section{Generalizations and Extensions of Multipole Response Theories} \label{Sec:Generalizations}
 %Although the HOTI itself does not manifest any subsystem symmetry or fracton behavior, the topological response characterized by the dipole Chern-Simons term implies the system carries a quantized dipole Hall conductivity in the absence of charge Hall conductance. Based on this observation, the HOTI is a trivial charge insulator but a topological dipole insulator. This phenomenon, although with different origin, is akin to the Fracton phase of matter where the charge degree of freedom is immoble so the low energy hydrodynamics is manifested at the dipole level. In particular, one can apply the Laughlin argument and dipole pumping process to related such 3D dipole Chern-Simons term to 2D quadrupole moment. This connections the relation between HOTI in 2D and 3D via momentum polarization and dimension reduction.
So far, we have developed response theories for a quarupolarization response and a dipolar Chern-Simons theory for 2D and 3D HOTIs. From a fundamental point of view, such theories can provide a phenomenological field theory description for HOTIs at the interacting level. We now wrap up the paper with a brief discussion of the extensions of our topological response theories to other contexts and other dimensions. We will leave a full discussion of these new directions to future research. 
\subsection{Higher Multipole Polarization Responses}
In this subsection we introduce a multipole polarization response theory in higher dimensions, with a primary focus on the topological octupole insulator. In Refs.~\onlinecite{benalcazar2017quantized,benalcazar2017electric}, the authors introduced a topological octupole insulator in 3D with fractional corner charge and midgap fermion bound states localized at the corners of a cubic sample. This model also exhibits gapped bulk, surfaces, and hinges.  Such a band insulator with corner zero modes in 3D, termed as a `third order TI', can be generalized to strongly interacting boson or fermion systems whose corners contain zero modes forming projective representations  under certain symmetries\cite{you2018higher,dubinkin2018higher,rasmussen2018classification,rasmussen2018intrinsically}. 

As we have done throughout the draft, to motivate the response of the topological octupole insulator, we can first consider a fracton model with subsystem symmetry. We consider a class of  symmetry enriched fracton models on a 3D cubic lattice where charge is conserved on each line. Due to this 1D subsystem symmetry, the charges and dipoles are immobile objects, and quadrupoles have their motion restricted to move only  along a direction $k$ perpendicular to its quadrupole component $q_{ij}$. The resultant system couples to a rank-3 gauge field and can support an octupole response,
\begin{align} 
& \mathcal{L}_{O}= \frac{\theta}{2\pi}( \partial_x\partial_y\partial_z A_0-\partial_t A_{xyz})=\frac{\theta}{2\pi}E_{xyz},\label{eq:octresponse}\\
& A_0 \rightarrow A_0+\partial_t \alpha,\;\; A_{xyz} \rightarrow A_{xyz}+\partial_x \partial_y \partial_z \alpha,\nonumber
\end{align}
where $A_{xyz}$ is the rank-3 gauge field that couples to the quadrupole current $J_{xyz}$, and in the second line we have indicated the relevant gauge transformation properties. Just as in the dipole and quadrupole cases, one can argue that $\theta$ has a $2\pi$ ambiguity. Thus, if we enforce a symmetry under which $\theta \rightarrow -\theta,$ then the response has a quantized value for $\theta$ (the octupole moment density) $\theta=0,\pi$ ($o_{xyz}=\theta/2\pi=0,1/2$). For the case when $o_{xyz}=1/2,$ the response theory predicts half-charges on the corners. Additionally, the subsystem symmetry protects the corner charges/modes from hybridizing with each other via hinge/surface transitions.

\begin{figure}[h]
  \centering
      \includegraphics[width=0.45\textwidth]{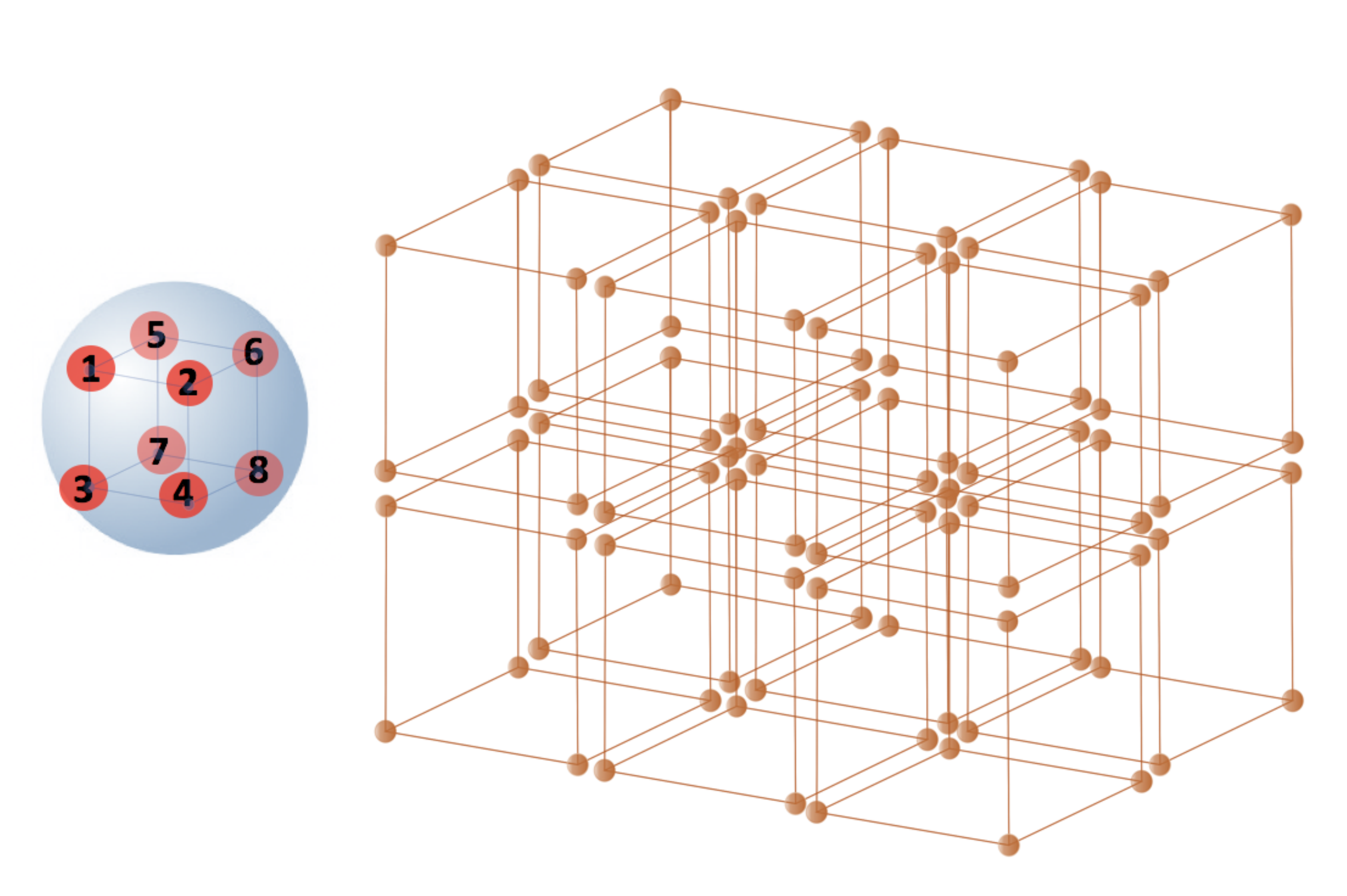}
      \caption{3D symmetry enriched Fracton on cubic lattice. Each site has eight spin-1/2 and every eight spins in the cube cluster is projected into a unique entangled state. The hinge and surface free modes can be gapped out locally leaving the corner with additional spin-1/2 zero mode.}
  \label{fig:oct} 
\end{figure}

A typical lattice model that produces a fracton octupole response is illustrated in Fig.~\ref{fig:oct}. This model is defined on a cubic lattice where each unit cell of the lattice contains eight spin-1/2 degrees of freedom. Each spin interacts with a spin in each of the eight cubes adjacent to it with an eight-spin cluster interaction,
\begin{align}
&H=-\sum_{{\bf{R}}}(  S^{+}_{{\bf{R}},1} S^{-}_{{\bf{R}}+e_x,3} S^{+}_{{\bf{R}}+e_x+e_y,4} S^{-}_{{\bf{R}}+e_y,2}\nonumber\\
&\times S^{-}_{{\bf{R+e_z}},5} S^{+}_{{\bf{R}}+e_z+e_x,7} S^{-}_{{\bf{R}}+e_z+e_x+e_y,6} S^{+}_{{\bf{R}}+e_z+e_y,8}+h.c ).
\end{align}
This Hamiltonian preserves global $\mathcal{T}$-symmetry and subsystem U(1) symmetry, the latter of which preserves $S_z$ on each line. The cluster interaction projects the eight interacting spins in each bulk cluster into a unique state $\frac{1}{\sqrt{2}}(|\downarrow\uparrow\downarrow\uparrow\downarrow\uparrow\downarrow\uparrow\rangle+|\uparrow\downarrow\uparrow\downarrow\uparrow\downarrow\uparrow\downarrow\rangle).$  
The surface unit cells and hinge unit cells contain an even number of dangling spins (four and two respectively) that can be projected into a symmetry preserving singlet state via intra-cell interactions. The $S_z$ number conservation on each line prohibits spin interactions between unit cells, or among plaquettes that span multiple unit cells, on the hinges or surfaces respectively.
The corners have an additional free spin-1/2 protected by global $\mathcal{T}$ and subsystem U(1) symmetries. If we cast the spins into the hardcore boson representation then we end up with a bosonic octupole insulator with half charges localized on the corners of an open cubic sample.

In analogy with the quadrupole case, we can now imagine a scenario where the subsystem $U(1)$ symmetry is broken down to either a planar subsystem $U(1)$ symmetry, or a global $U(1)$ symmetry. In the former case, the system will exhibit mobile dipoles and quadrupoles, but the charges will still be frozen. In the latter case, the system will represent a HOTI having a response to a conventional $U(1)$ gauge field. We will leave a thorough discussion of these symmetry breaking possibilities to future work, and for now we will just consider the HOTI octupole response that descends from the subsymmetry protected response when \emph{all} subsystem symmetry is broken.

The octupole response of a HOTI can thus be written\cite{kang2018many,wheeler2018many}:
\begin{align} 
&\mathcal{L}_{O*}= \frac{\theta}{2\pi} [\partial_t V_{xyz}- \partial_x \partial_y \partial_z A_0]\nonumber\\
& V_{xyz}=\frac{1}{3}(\partial_x \partial_z A_y +\partial_y \partial_x A_z+\partial_y \partial_z A_x).
\end{align}\noindent This can be rewritten in terms of electric fields as
\begin{equation}
 \mathcal{L}_{O*}=\frac{\theta}{6\pi}\left[\partial_x\partial_y E_z+\partial_y\partial_z E_x+\partial_z\partial_x E_y \right],  
    \end{equation} which represents the phenomenology of an octupole moment $o_{xyz}.$ One can argue for the $2\pi$ periodicity of $\theta$, under conditions for which the octupole moment is well-defined (i.e., vanishing charge, dipole, and quadrupole moments) in analogy with the discussion in Appendix \ref{app:ambiguity}. Hence, any symmetry under which the octupole moment is odd, e.g. charge-conjugation or cubic symmetry, enforces the quantization condition $\theta=N\pi$. 
    
    Interestingly, even if the octupole moment takes  nontrivial values, i.e., $\theta=(2N+1)\pi,$ it does not guarantee protected corner modes. To illustrate this subtlety we note that since there are \emph{three} hinges that terminate at a corner, then, in principle, one can always decorate the hinges with nontrivial 1D SPT chains on all hinges to cancel the corner modes. Consequently, there is no `intrinsic  topological octupole insulator' protected by cubic symmetry with robust corner modes. However, it is worth mentioning that there exist several 3D bosonic HOSPT states whose protected corner mode is akin to the edge of 1D SPT chain with a $\mathbb{Z}_3$ classification\cite{rasmussen2018intrinsically,rasmussen2018classification}, though these examples are not likely described by our response theory.

\subsection{Generalized Multipole Chern-Simons and Axion Responses}\label{sec:higherdcs}
In addition to higher polarization responses, our results suggest a possible hierarchical generalization of Chern-Simons theories to multipole Chern-Simons theories. Suppose that in spatial dimension $d$ we have a system with subsystem symmetry that couples to the background fields $A_0, A_{x_1x_2\ldots x_{d-1}},$ and $A_{x_d}.$ Then if $d>1$ is even we can write the Lagrangian 
\begin{eqnarray}
 \mathcal{L}^{CS}_d&=&\frac{K}{4\pi}(A_{x_1x_2\ldots x_{d-1}}E_{x_d}-A_{x_{d}}E_{x_1x_2\ldots x_{d-1}}+A_0 B),\nonumber\\
 \end{eqnarray}\noindent while if $d$ is odd we can write
 \begin{eqnarray}
 \mathcal{L}^{CS}_d&=&\frac{K}{4\pi}(A_{x_1x_2\ldots x_{d-1}}E_{x_d}+A_{x_{d}}E_{x_1x_2\ldots x_{d-1}}-A_0 B),\nonumber\\
 \end{eqnarray} where
 \begin{eqnarray}
 E_{x_1x_2\ldots x_{d-1}}&=&\partial_{x_1}\partial_{x_2}\ldots\partial_{x_{d-1}}A_0-\partial_t A_{x_1x_2\ldots x_{d-1}}\nonumber\\
 B&=&\partial_{x_1}\partial_{x_2}\ldots\partial_{x_{d-1}}A_{x_d}-\partial_{x_d}A_{x_1x_2\ldots x_{d-1}}.\nonumber
\end{eqnarray}\noindent The alternating signs in each spatial dimension are necessary for these actions to be gauge invariant (up to boundary terms) using the gauge transformation $A_{x_1x_2\ldots x_{d-1}}\to A_{x_1x_2\ldots x_{d-1}}+\partial_{x_1}\partial_{x_2}\ldots\partial_{x_{d-1}}\alpha$. 

When $d$ is odd the action is a total derivative and generates boundary responses. The first non-trivial action appears when $d=3,$ and we have already discussed this theory at length in this article so let us consider even values of $d.$ When $d$ is even the system has the bulk responses
\begin{eqnarray}
\rho&=&\frac{K}{2\pi}B\nonumber\\
j_{x_d}&=&-\frac{K}{2\pi}E_{x_1x_2\ldots x_{d-1}}\nonumber\\
j_{x_1x_2\ldots x_{d-1}}&=&\frac{K}{2\pi}E_{x_d}.
\end{eqnarray} Additionally, one can perform dimensional reduction by compactifying the $x_d$ direction. When $\pi$-flux is inserted in the compactified direction, i.e., when $A_{x_d}=\frac{\pi}{L_d},$ then the multipole Chern-Simons action reduces
\begin{eqnarray}
\mathcal{L}^{\theta}_{d-1}&=&\frac{\theta}{2\pi}E_{x_1x_2\ldots x_{d-1}}\\
\theta&\equiv& \int dx_{d}A_{x_d}.\nonumber
\end{eqnarray}

As an example we can consider $d=4$ having the Lagrangian
\begin{eqnarray}
\mathcal{L}^{CS}_4=\frac{K}{4\pi}(A_{xyz}E_u-A_uE_{xyz}+A_0B)
\end{eqnarray}\noindent where the spatial coordinates are $(x,y,z,u).$ The bulk response equations are
\begin{eqnarray}
\rho&=&\frac{K}{2\pi}B\nonumber\\
j_{u}&=&-\frac{K}{2\pi}E_{xyz}\nonumber\\
j_{xyz}&=&\frac{K}{2\pi}E_{u},
\end{eqnarray} which represents a transverse charge current in the presence of a higher-rank electric field, and a transverse quadrupole current in the presence of rank-1 electric field. This is an example of a mixed rank Hall effect. If we dimensionally reduce this action we find
\begin{equation}
    \mathcal{L}^{\theta}_{3}=\frac{\theta}{2\pi}E_{xyz}\nonumber
\end{equation}\noindent which is exactly the octupole response $\mathcal{L}_{O}$ in Eq. \ref{eq:octresponse}. To find the response of the HOTI system we can replace the higher-rank field by a suitable symmetrized combination of rank-1 fields $A_{x_1x_2\ldots x_{d-1}}= \frac{1}{d}(\partial_{x_1} \partial_{x_2}\ldots\partial_{x_{d-2}}A_{x_{d-1}}+{\rm{permutations}}).$ After doing this we would find $\mathcal{L}_{O*}$ in this example.

If we consider even values of $d$ we can make a clear connection to a mixed-rank axion electrodynamics in $d+1$ dimensions. We augment our theory by an additional rank-1 field $A_{d+1}$ and we can consider a Lagrangian
\begin{eqnarray}
\mathcal{L}^{\phi}_{d+1}&=&\frac{\phi}{4\pi^2} (E_{x_{d+1}}B_{x_1 x_2\ldots x_d}+E_{x_1x_2\ldots x_{d-1}}B_{x_d x_{d+1}}\nonumber\\&+&E_{x_d}B_{x_{d+1} x_1 x_2\ldots x_{d-1}})
\end{eqnarray}\noindent where
\begin{eqnarray}
B_{x_d x_{d+1}}&=&\partial_{x_d}A_{x_{d+1}}-\partial_{x_{d+1}}A_{x_d}\nonumber\\
B_{x_1 x_2\ldots x_d}&=&\partial_{x_1}\partial_{x_2}\ldots\partial_{x_{d-1}}A_{x_d}-\partial_{x_d}A_{x_1x_2\ldots x_{d-1}}\nonumber\\
B_{x_{d+1} x_1 x_2\ldots x_{d-1}}&=&\partial_{x_{d+1}}A_{x_1x_2\ldots x_{d-1}}-\partial_{x_1}\partial_{x_2}\ldots\partial_{x_{d-1}}A_{x_{d+1}},\nonumber\\
\end{eqnarray}\noindent and $\phi$ is the scalar axion field (that will be quantized to $\pi$ in the presence of, e.g., inversion or time-reversal symmetry). This action is a total derivative and on a $w$ or $u$ surface one finds an action similar in form to $\mathcal{L}_{4}^{CS},$ whereas on the two-dimensional, higher order surface intersection of $x_1, x_2, x_3,\ldots x_{d-1}$ surfaces one finds an ordinary $d=2$ Chern-Simons theory $\mathcal{L}^{CS}_2$ that is a functional of $A_0, A_{x_d}, A_{x_{d+1}}.$ 

Let us conclude with a few important notes about this generalized mixed-rank axion electrodynamics. First, we have chosen the normalization constant such that the surface Chern-Simons coefficients have half of the value they normally take in a lower-dimensional bulk system. We note that, while this is natural, we have not proven this result, nor do we have a microscopic model for which this is the response action. Second, if a similar construction is formed for cases where $d$ is odd the resulting action vanishes, i.e., there is not an analogous \emph{bulk} action that reduces to $\mathcal{L}^{CS}_{2n+1}$ on its surfaces. This is perhaps not unexpected since we know that each $\mathcal{L}^{CS}_{2n+1}$ is already a total derivative. Thus, finding the axion extension of $\mathcal{L}^{CS}_{2n+1}$ is an open problem. Third, we have only treated the response to one flavor of gauge field, and in principle we could generalize our response actions to hydrodynamic field theories where the coefficient $K$ could become a matrix to allow for more exotic SPTs as well as possible fractional states. Fourth, we have primarily discussed the higher dimensional cases for subsystem symmetric systems. We expect that lower-rank field substitutions will provide new types of responses for systems with less subsystem symmetry, all the way down to the case of a HOTI with no subsystem symmetry. We leave full discussions of these research directions for future work.

\section{Conclusion and outlook}\label{sec: summary}

In this work we have proposed  topological multipole field theories for higher order topological insulators in 2D and 3D. %interacting bosons and fermions. 
Our theory predicts various topological dipole responses in HOTI with measurable experiment signatures. Notably, our dipole Chern-Simons description of the 3D HOTI creates a complete connection between gapless hinge modes, dipole Hall response and surface anomalies. It also yields the 2D quadrupole insulator upon dimensional reduction.

While interest in HOTIs thus far has mainly focused on the gapless hinge or corner modes, both our 2D and 3D field theories highlight the previously overlooked fact that the gapped boundaries in these systems are also anomalous.   These anomalies are reflected in important topological features of the bulk, such as the quantized quadrupole moment and magnetic quadrupole response in 3D.  In 2D, our topological response theory engenders a $\mathcal{T}$ and $C_4$ invariant fractional dipole moment on the edge which does not exist in pure 1D systems. %The dipolar Chern-Simons theory has surface anomalies for dipole currents, which are a characteristic feature of the 3D HOTI.

An interesting corrolary of our analysis is that some subsystem-symmetry protected phases can essentially be viewed as examples of HOTI for which spatial symmetries are not required to protect the gapless corner modes. 
Indeed, the two are described by closely related quantized dipole responses, with a straightforward substitution being largely sufficient to pass from the subsystem symmetry protected phase to a more conventional model with spatial symmetries and global charge conservation.  This correspondence is rooted in the fact that a HOTI by definition must have a trivial charge response, so that in both cases the topological field theory describes dipolar response.

Our results suggest various directions and open questions for future study. (1) We have written down a multipolar axion-type term in odd spatial dimension, but what about even spatial dimensions? (2) Similar to the dipolar Chern-Simons description of the HOTI, we expect there to be similar gravitational responses for higher order topological superconductrors that will capture the chiral Majorana mode at the hinge. (3) For $C_n$ symmetric crystalline insulators with vanishing polarization, there exist corner-localized charges quantized as $e/n$. The field theory and topological response of such fractional corner charge still remains unclear and thus worth pursuing.

\begin{acknowledgments}
We are grateful to Titus Neupert, Mike Stone, and Chong Wang for insightful comments and discussions. YY is supported by PCTS Fellowship at Princeton University. FJB is grateful for the financial support of NSF-DMR 1352271 and the Sloan Foundation FG-2015-65927. YY, TLH are supported in part by the National Science Foundation under Grant No.NSF PHY-1748958(KITP) during the Topological Quantum Matter program.
TLH thanks the US National Science Foundation under grant DMR 1351895-CAR, and the MRSEC program under NSF Award Number DMR-1720633 (SuperSEED) for support.
This work(YY,FJB,TLH) was initiated at Aspen Center for Physics, which is supported by National Science Foundation grant PHY-1607611.

\end{acknowledgments}

\appendix

\section{Ambiguity of $\theta \sim 2\pi$ in path integral}\label{app:ambiguity}
In this section, we focus on the $2\pi$ shift equivalency of $\theta$ in the quadrupole response term. Let us first review the $2\pi$ equivalency of $\theta$ for the dipole response in 1D. The 1D polarization term 
\begin{equation}
\frac{\theta}{2\pi}\int dx dt \epsilon^{\mu\nu}\partial_\mu A_\nu,    
\end{equation}
is a total derivative in the action so naively one would expect it to vanish if we impose PBCs in spacetime. However, we need to be careful when applying PBCs for gauge fields. If we implement a large gauge transformation by inserting a global flux $\int A_x dx=2\pi$ at time $t=T$, the vector potential shifts so the total path integral of the polarization term is changed by $\theta$,
\begin{align} 
&\int dx \int_0^T dt \frac{\theta}{2\pi} \partial_t A_{x}= \theta.
\end{align}
Since any $2\pi$ contribution in the path integral does not affect the IR physics, $\theta$ (the polarization) has a $2\pi$ (integer) ambiguity.

A physical interpretation of this LGT is a quench process where we turn an electric field on and off by inserting global flux during a time period $T:$
\begin{align} 
&\int  dx \int_0^T E_x dt = \Delta \Phi=2\pi.
\end{align}
The electric field couples directly with the dipole moment $P_x E_x$ so the quenched procedure changes the total action as,
\begin{align} 
&\Delta \mathcal{S}=\int dx \int_0^T E_x P_x dt =2\pi P_x. 
\end{align}
A key constraint is that the system has vanishing charge density. The dipole moment is only well-defined when the system is neutral. Furthermore, any charge density would couple with the gauge potential $A_0$ to generate an additional change in the action during the quench process, and thus make it unable to uniquely identify the contribution from the polarization. When $P_x\in \mathbb{Z}$, the process only changes the path integral by $2\pi \mathbb{Z}$, which has no effect. Thus, the dipole moment has an integer ambiguity, which implies $\theta$ has a $2\pi$ ambiguity. Finally, we comment that the minimal value of the electric field quench is imposed by maintaining PBCs after flux insertion.

We argued in the main text that the same $2\pi$ ambiguity applies for quadrupole moments in symmetry enriched fracton phases with the response action:
\begin{align} 
\frac{\theta}{2\pi}\int dx dy dt (\partial_x \partial_y A_0-\partial_t A_{xy}).
\end{align}
Since the charge is conserved on each line of the system, let us apply a large gauge transformation for the rank-2 gauge field by implementing a $2\pi$ global flux insertion during a time-period T for the charges on a specific row:
\begin{align} 
\int A_{xy}(y=y_i,t=T) dx=2\pi \delta(y_i). 
\end{align}
Under this LGT, the path integral is changes by
\begin{align} 
\int dx dy  \int_0^T  dt \frac{\theta}{2\pi} (\partial_x \partial_y A_0-\partial_t A_{xy})=\theta.
\end{align}
Thus, any shift of $\theta$ by $2\pi$ does not affect the path integral and the quadrupole moment $q_{xy}=\tfrac{\theta}{2\pi}$ has an integer ambiguity.

Now we turn to the quadrupole moment in a conventional HOTI with a global U(1) charge symmetry, but without subsystem symmetry. For a quadrupole moment to be well-defined, it is essential to require that the ground state contains no net charge density or dipole moment, otherwise the quadrupole moment can change under arbitrary coordinate shifts, and is ill-defined.
To measure the total quadrupole moment of a neutral, unpolarized many-body system, one can add a non-uniform electric field, with a non-vanishing gradient. The electric field gradient couples with the total quadrupole moment as $q_{xy}(\partial_x E_y+\partial_y E_x)/2$.   Thus, we can imagine a process where we turn on and off a uniform electric field gradient $(\partial_x E_y+\partial_y E_x)/2$ over a time period $T,$ and identify the change of the action during this process:
\begin{align} 
&\Delta \mathcal{S}=\frac{1}{2}\int dx dy \int_0^T q_{xy}(\partial_x E_y+\partial_y E_x) dt. 
\end{align}

As the system must respect PBCs after the quench, the flux insertion for each row or column on the lattice must be an integer multiple of $2\pi$. Based on these criteria, we can have a gauge field configuration,
\begin{equation}
    A_i= \begin{cases} 0 & t \leq T \\
    \frac{2\pi\sigma_{ij} x^j t}{aLT} &  0<t<T\\
    \frac{2\pi\sigma_{ij} x^j }{aL } & t \geq T \\
    \end{cases}
    \label{tr}
\end{equation}
where $\sigma_{xy}=\sigma_{yx}=1,$ $L$ is the spatial length in the $x$ and $y$ directions, and $a$ is the lattice spacing. Since the theory is defined on a square lattice, the coordinates $x,y$ only take discrete values, and thus, after a time $T$ the system still respects PBCs. This gauge field configuration includes both large and small gauge transformations and creates a spatially non-uniform electric field during the quench process.  It is chosen to introduce a uniform electric field gradient in the system. %Despite the change of the microscopic Hamiltonian, the IR theory remains unaffected because, as long as the theory is an insulator with no net dipole moment, there nothing will couple to the uniform electric field. 
Such a process changes the action as:
\begin{align} 
&\Delta \mathcal{S}=\int dx dy \int_0^T \frac{2\pi}{aLT}q_{xy} dt =2\pi q_{xy} L/a .
\end{align}
The quantity $L/a$ is the number of rows (or columns) which, without loss of generality, can be chosen to be an odd number. Thus, if $q_{xy}$ is an integer, the change of the action due to the process does not affect the path integral, and such a $2\pi$ ambiguity cannot be tracked. 

Finally, we comment the validity of this quenched measurement. The gauge field configuration in Eq.~\ref{tr} is not equivalent to a large gauge transformation as it produces a flux for both poloidal and
toroidal directions after the process. However, as we are trying to extract the quadrupole moment in a gapped insulator with no net charge density, and no net dipole moment, the spatially non-uniform electric field only couples with the quadrupole moment so the change of the path integral is attributed to only the quadrupolarization. During the quench procedure, a non-uniform electric field is generated that could potentially trigger a charge or dipole excitation from the ground state. However, as long as the charge and dipole gap are large compared to the electric field ${\bf{E}}$, the IR theory is unaffected. Indeed, our argument relies on the fact that the ground state is fully gapped with no charge or dipole moment. As long as the charge degree of freedom has a large gap, the background potential $A_0$ appearing during the quench process will not trigger any free charge excitations. Likewise, the ${\bf{E}}$ field appearing during the quench process can potentially activate a free dipole excitation and generate extra contributions to $\Delta S$ that would spoil the argument. Based on these observations, the
validity of this quenched measurement requires that the system does not  generate any free charges or dipoles during the process. Subsequently, the non-uniform ${\bf{E}}$ field during the quench only activates the quadrupole degree of freedom, and thus changes the action by $2\pi q_{xy}$. These criteria must be met to have a well-defined quadrupolar value of $\theta$ that is ambiguous under a $2\pi$ shift. We note that subsystem symmetry automatically forbids any net dipole moment, so the only requirement for the SSPT HOTI would be that the system remain gapped. In the absence of subsystem symmetry, we can have a fixed  dipole moment if spatial symmetries, e.g., $C_4, C_2$, or mirror, remain unbroken during the quench process.

\section{Dimensional Reduction of Chern-Simons Response}\label{app:dimred}
In Ref. \onlinecite{axion2} the authors showed a connection between topological insulators in a series of spatial dimensions. One important example was a connection between the 2D Chern-Simons response of a Chern insulator and a 1D topological insulator with quantized charge polarization. The connection between these two systems was illustrated using a dimensional reduction procedure where one spatial direction is compactified and shrunk down to form a quasi-1D system. Importantly the compactified direction allows for flux threaded through it, and the space-time profile of that flux becomes a background field $\theta$ in the 1D insulator. 

Let us explicitly show how this works beginning with the 2D Chern-Simons response:
\begin{eqnarray}
S_{cs}&=&\frac{1}{4\pi}\int d^3 x [A_x (\partial_y A_0-\partial_t A_y)-A_y(\partial_x A_0 -\partial_t A_x)\nonumber\\&+&A_0(\partial_x A_y -\partial_y A_x)],
\end{eqnarray}
which we will compactify in the $y$-direction. As the compactified direction is shrunk all fields become independent of $y$ and we have
\begin{eqnarray}
    S_{cs}&=&\frac{1}{4\pi}\int dx dt \int dy[-A_x\partial_t A_y-A_y E_x+A_0\partial_x A_y]\nonumber\\
    &=&\frac{1}{4\pi}\int dx dt\int dy[2A_y E_x-\partial_t(A_x A_y)+\partial_x(A_0 A_y)]\nonumber\\&=&\frac{1}{4\pi}\int dx dt [2\theta E_x-\partial_t(\theta A_x)+\partial_x (\theta A_0)]\nonumber\\&=&\frac{1}{2\pi}\int dx dt [\theta E_x]+\frac{1}{4\pi}\int dx dt (\partial_x (\theta A_0)-\partial_t (\theta A_x)].\nonumber\\ \label{eq:csdimred}
    \end{eqnarray}
In the last line we see we have arrived at the 1D $\theta$-term response for the polarization. If we consider the system to be a topological insulator then $\theta=\pi$ is a constant background field in the bulk of the system, and $\theta=0$ outside of the system. 

Let us consider the resulting action carefully. The first term is a ``bulk"-response piece and it only generates something non-vanishing when $\theta$ depends on space or time. Indeed we find
\begin{equation}
    j_{bulk}^{\mu}=\frac{1}{2\pi}\epsilon^{\mu\nu}\partial_\nu \theta.
\end{equation} However, for a TI we are treating $\theta$ as a constant in the bulk, and the only response arises at a physical boundary, which we have carefully kept track of in Eq. \ref{eq:csdimred}. Thus, if we assume that $\theta$ only changes at a boundary in space or time we find the response action
\begin{eqnarray}
S=\frac{1}{4\pi}\int dt \theta A_0\vert_{x=0}^{x=L}-\frac{1}{4\pi}\int dx \theta A_x\vert_{t=0}^{t=T}.
\end{eqnarray}  Hence, on an $x$-boundary, say $x=L,$ we find a charge of $\frac{\theta}{4\pi}$, which is actually half of the expected value, i.e., it predicts a charge of $1/4$ instead of $1/2$ when $\theta=\pi$ in the bulk. 

While this factor of two may initially seem puzzling, it has a well-known resolution. The 2D bulk Chern-Simons action has a bulk current that flows into the edge, and a current localized on the boundary. Indeed, terms representing both of these processes appear in Eq. \ref{eq:csdimred}. However, in order for the boundary response to properly match the bulk inflow we need to add additional degrees of freedom on the boundary that make up the difference. The boundary degrees of freedom contribute the \emph{consistent} anomaly, and when combined with the boundary current of the reduced Chern-Simons response, they form the \emph{covariant} anomaly, which properly matches the bulk inflow\cite{naculich1988,stone2012}. Thus, if we carefully perform dimensional reduction starting from a Chern-Simons response, and we treat $\theta$ as a constant to yield a purely boundary response, then we find the wrong answer by a factor of two as we have just shown. There is an additional response contributed by boundary degrees of freedom, and the sum of the Chern-Simons boundary current, and the current from the auxiliary edge degrees of freedom, gives the full response, which in this case would be
\begin{eqnarray}
S_{cov}=\frac{1}{2\pi}\int dt \theta A_0\vert_{x=0}^{x=L}-\frac{1}{2\pi}\int dx \theta A_x\vert_{t=0}^{t=T}.
\end{eqnarray} To summarize, when dimensional reducing from a Chern-Simons response, the boundary currents only indicates half of the overall response, while any bulk currents would have the correct, full coefficient. The same arguments apply to the dimensional reduction of the dipolar Chern-Simons term.

\section{Anomalies from a bosonic perspective}  \label{app:bosonization}
\subsection{Bosonization Picture for Chiral Anomaly}
In this section we review the well-known theory of the 1D chiral anomalies from a bosonic theory closely following the calculations in Ref. \onlinecite{fradkin2013field}, and then extend the ideas to the new type of bosonic theory we find on the surfaces of the dipolar Chern-Simons. In order to calculate the chiral anomaly we can perform the calculation for a non-chiral theory and use those results to determine the \emph{consistent} anomaly for a chiral theory\cite{naculich1988}. The consistent anomaly can be added to the boundary current response derived from our dipolar Chern-Simons theory, and the combination will generate the full response\cite{naculich1988,stone2012}.

Let us begin with a free 1D bosonic field $\phi$ with the Lagrangian
\begin{equation}
\mathcal{L}_B=\frac{1}{2}\int dx dt [(\partial_t \phi)^2-(\partial_x \phi)^2],   
\end{equation}
and satisfying the equal-time commutation relations
\begin{equation}
    [\phi(x,t),\Pi (x',t)]=i\delta(x-x'),
    \end{equation}\noindent with its canonical momentum $\Pi(x,t)=\partial_t \phi(x,t).$ We assume that this scalar field has a Goldstone-Wilczek type current response\cite{goldstone1981fractional}:
    \begin{equation}
        j_0=\frac{1}{\sqrt{\pi}}\partial_x \phi,\,\,\, j_x=\frac{1}{\sqrt{\pi}}\partial_t \phi=\frac{1}{\sqrt{\pi}}\Pi
        \end{equation} which is exactly conserved
        \begin{equation}\partial^\mu j_\mu=\partial_t j_0-\partial_x j_x=\frac{1}{\sqrt{\pi}}(\partial_t\partial_x \phi-\partial_x\partial_t\phi)=0\end{equation} (note our convention in this appendix is that the spatial components of the diagonal Lorentz metric are negative). By applying the canonical commutation relations we can generate the $U(1)$ Kac-Moody current algebra
        \begin{equation}
            [j_0(x,t), j_x(x',t)]=\frac{i}{\pi}\partial_x \delta(x-x').
        \end{equation}

Now let us consider the chiral current $j_{\mu}^{5}=\epsilon_{\mu\nu}j^{\mu},$ i.e.,  $j_{0}^{5}=-\frac{1}{\sqrt{\pi}}\partial_t \phi,  j_{x}^{5}=\frac{1}{\sqrt{\pi}}\partial_x \phi.$ The conservation law for chiral current is then
\begin{equation}
\sqrt{\pi}\partial^{\mu}j_{\mu}^5=-\partial_{t}^2\phi +\partial_{x}^2\phi=-\partial^{\mu}\partial_{\mu}\phi.    
\end{equation} Thus, since we have specified a free boson theory, the right hand side vanishes (at least on-shell), and the chiral current is conserved. 

However, this analysis does not capture all of the important physics, for which we need to couple the scalar field to an electromagnetic field $A_{\mu}$. The coupling term we add is
\begin{equation}
    \mathcal{L}_{EM}=j_{\mu}A^{\mu}=\frac{1}{\sqrt{\pi}}\partial_x\phi A^{0}+\frac{1}{\sqrt{\pi}}\partial_t \phi A^{x}.
    \end{equation} In the bosonic action, this term acts like a source for $\phi$
    \begin{equation}
        S_{EM}=-\frac{1}{\sqrt{\pi}}\int dx dt \phi (\partial_x A^0 +\partial_t A^x)=-\frac{1}{\sqrt{\pi}}\int dx dt \phi E_x .
        \end{equation} This modifies the equations of motion to $-\partial^\mu \partial_\mu \phi=\frac{1}{\sqrt{\pi}}E_x.$ Thus we arrive at the anomalous conservation law
        $\partial^{\mu}j^{5}_\mu=\frac{1}{\pi}E_x$
            We derived this result for a non-chiral boson theory (typically derived from bosonizing a Dirac fermion). To find the result for the chiral theory we need to divide by a factor of four (not two!\cite{naculich1988}) to arrive at
            \begin{equation}
          \partial^{\mu}j^{5}_\mu=\frac{1}{4\pi}E_x.
                \end{equation}

\subsection{Bosonization Picture for Chiral Dipole Anomaly} Now we will repeat the arguments above to argue for the form of the anomaly for the bosonic theory we find at the surface of the dipolar Chern-Simons theory. We will again start with a non-chiral version of the 2D surface boson theory 
\begin{equation}
    \mathcal{L}_B=\frac{1}{2}\int dx dy dt [(\partial_t \phi)^2-(\partial_x \partial_y \phi)^2]
\end{equation} which has an equation of motion
\begin{equation}
\partial_t^2 \phi +\partial_x\partial_y\partial_x\partial_y\phi=0.   
\end{equation} The field $\phi$ satisfies the equal-time commutation relations
\begin{equation}
    [\phi(x,y,t),\Pi (x',y',t)]=i\delta(x-x')\delta(y-y'),
    \end{equation}\noindent with its canonical momentum $\Pi(x,y,t)=\partial_t \phi(x,y,t).$

In the previous subsection we assumed that the boson exhibited a Goldstone-Wilczek response. That type of current response is identical to that of charge polarization $P_x$ if we identify $P_x=\tfrac{\phi}{2\pi}.$ In analogy, we can treat the 2D $\phi$ in this case as a quadrupole moment $q_{xy}$ and write an analogous response
\begin{equation}
    j_0=\frac{1}{\sqrt{\pi}}\partial_x\partial_y\phi,\;\;\; j_{xy}=\frac{1}{\sqrt{\pi}}\partial_t\phi.
\end{equation} This current minimally couples to the gauge fields via $\mathcal{L}_{gauge}=j_0 A^0+j_{xy} A^{xy},$ and from the gauge transformation properties of $A_0, A_{xy}$ we can derive the conservation law
\begin{equation}
    -\partial_t j_0+\partial_x \partial_y j_{xy}=0,
    \end{equation} which is indeed satisfied by the current above. Using the canonical commutation relations we can write down a generalization of the $U(1)$ Kac-Moody algebra as well \begin{equation}
            [j_0(x,y,t), j_{xy}(x',y',t)]=\frac{i}{\pi}\partial_x \delta(x-x')\partial_y \delta (y-y').
        \end{equation}
    
    Now let us consider the chiral current\begin{equation}
        j^{5}_0=\frac{1}{\sqrt{\pi}}\partial_t\phi, \;\;\; j^{5}_{xy}=-\frac{1}{\sqrt{\pi}}\partial_x\partial_y \phi.
    \end{equation}  It satisfies
    \begin{equation}
        -\partial_{t} j^{5}_0+\partial_x\partial_y j^{5}_{xy}=\frac{1}{\sqrt{\pi}}(-\partial^2_{t}\phi-\partial_x\partial_y\partial_x\partial_y\phi),
        \end{equation} which vanishes if we apply the equation of motion. Just as in the prevous case we can turn on the background gauge fields $A_0, A_{xy}$ via $\mathcal{L}_{gauge}$ and they will act like a source for $\phi$ in the action
        \begin{eqnarray}
            S_{gauge}&=&\frac{1}{\sqrt{\pi}}\int dx dy dt \phi(\partial_x\partial_y A^0-\partial_t A^{xy})\nonumber\\&=&\frac{1}{\sqrt{\pi}}\int dx dy dt\; \phi E_{xy}.
        \end{eqnarray} This serves to modify the equation of motion $\partial_t^2 \phi +\partial_x\partial_y\partial_x\partial_y\phi=\frac{1}{\sqrt{\pi}}E_{xy}.$ Thus, we now have an anomalous conservation law. We have argued for this anomaly from the non-chiral theory, and if we take the chiral theory alone we expect the anomaly to be reduced by a factor of four\cite{naculich1988}
        \begin{equation}
            \partial_t j_0^5-\partial_x \partial_y j_{xy}^5=\frac{1}{4\pi}E_{xy}.
        \end{equation}
\section{Deriving the anomalous conservation laws for the HOTI dipolar Chern-Simons response}\label{app:anomalyderivation}
\subsection{Side surface conservation laws} Let us start from  Eq. \ref{eq:sidesurfaceresponse} to analyze this result in more detail. Similar to the 2D Chern-Simons action discussed above, we can calculate two related responses by taking variations with respect to the fields $A_z, A_0$ (to find current, charge) or the gradients $\partial_i A_z, \partial_i A_0$ (to find dipole currents and densities).

For the first case we find
\begin{eqnarray}
j_z\vert_{y=L}&=&-\frac{1}{4\pi}\partial_x A_0=-j_z\vert_{y=0}\nonumber\\
\rho\vert_{y=L}&=&-\frac{1}{4\pi}\partial_x A_z=-\rho\vert_{y=0},\end{eqnarray} for surfaces normal to $y,$ and 
\begin{eqnarray}
j_z\vert_{x=L}&=&-\frac{1}{4\pi}\partial_y A_0=-j_z\vert_{x=0},\nonumber\\
\rho\vert_{x=L}&=&-\frac{1}{4\pi}\partial_y A_z=-\rho\vert_{x=0},
\end{eqnarray} for surfaces normal to $x.$ For a $y$-surface, if we apply a potential difference $\partial_x A_0$ in the $x$-direction to create a uniform electric field $E_x$, then a current will flow in the $z$-direction. However, the two surfaces $(y=0, L)$ have opposite current flows, and thus the total current is vanishing. We also find currents on the hinges
\begin{eqnarray}
j_z|_{(x=L, y=L)}&=&j_z|_{(x=0, y=0)}=-\frac{A_0}{4\pi}\nonumber\\
\rho|_{(x=L,y=L)}&=&\rho|_{(x=0,y=0)}=-\frac{A_z}{4\pi}\nonumber\\
j_z|_{(x=L, y=0)}&=&j_z|_{(x=L, y=0)}=\frac{A_0}{4\pi}\nonumber\\
\rho|_{(x=L,y=0)}&=&\rho|_{(x=0,y=L)}=\frac{A_z}{4\pi}.
\end{eqnarray}

For the dipole currents and densities we find 
\begin{eqnarray}
j_0^x\vert_{y=L}&=&\frac{1}{4\pi}A_z=-j_0^x\vert_{y=0}\nonumber\\
j_z^x\vert_{y=L}&=&-\frac{1}{4\pi}A_0=-j_z^x\vert_{y=0}\nonumber\\
j_0^y\vert_{x=L}&=&\frac{1}{4\pi}A_z=-j_0^y\vert_{x=0}\nonumber\\
j_z^y\vert_{x=L}&=&-\frac{1}{4\pi}A_0=-j_z^y\vert_{x=0},
\end{eqnarray}\noindent where the notation $j_{\mu}^{a}$ is the current in the $\mu$-direction carrying dipole moment pointing in the $a$-direction. If we change the potential $A_0$ then dipole currents will flow in the $z$-direction, however we again find that if we account for each pair of surfaces then the total dipole current flowing in the $z$-direction generated by a shift of the scalar potential vanishes.

These current/density responses for the charge and dipole responses are not manifestly gauge invariant; indeed they are anomalous. We find
\begin{eqnarray}
(\partial^z j_z+\partial^t \rho)\vert_{x=L,0}&=&\pm \frac{1}{4\pi}\partial_y E_z\vert_{x=L,0}\nonumber\\
(\partial^z j_z+\partial^t \rho)\vert_{y=L,0}&=&\pm \frac{1}{4\pi} \partial_x E_z\vert_{y=L,0},
\end{eqnarray} for the charge response, and
\begin{eqnarray}
(\partial^z j_z^y+\partial^t j_0^y)\vert_{x=L,0}&=&\mp  \frac{1}{4\pi} E_z\vert_{x=L,0}\nonumber\\
(\partial^z j_z^x+\partial^t j_0^x)\vert_{y=L,0}&=&\mp \frac{1}{4\pi} E_z\vert_{y=L,0},
\end{eqnarray}\noindent for the dipole response, where the signs are coordinated with the choice of surface at $x,y =0, L.$ As an example, on the surface $x=L$ we have the equations
\begin{eqnarray}
\partial^\mu j_{\mu}&=&\frac{1}{4\pi}\partial_y E_z\\
\partial^\mu j_{\mu}^{y}&=&-\frac{1}{4\pi}E_z,
\end{eqnarray}\noindent where $\mu=0,z.$ 
Interestingly, on the $x$ or $y$ surfaces we find a dipole current with an anomalous conservation law that resembles the form of the usual chiral anomaly, i.e., it is proportional to the electric field. We also find that the charge has an anomalous conservation law in the presence of an electric field gradient. Finally, on the hinges we find an anomalous conservation law: 
\begin{equation}
    \partial^t j_0+\partial^z j_z =-\frac{1}{4\pi}E_z
    \end{equation}\noindent 
for the hinge at $(x,y)=(L,L)$ and the others can be obtained by adding extra - signs for each 90-degree rotation. 
    
As discussed in the previous section for the SSPT, and informed by the discussions in Appendix \ref{app:bosonization}, we need to add the consistent anomaly coming from boundary degrees of freedom to these anomalous conservation laws. In doing so we finally arrive at
\begin{eqnarray}
\partial^t j^{(x\pm)}_{0}+\partial^z j^{(x\pm)}_z&=&\pm\frac{1}{2\pi}\partial_y E_z\nonumber\\
\partial^t j^{(y\pm)}_{0}+\partial^z j^{(y\pm)}_z&=&\pm\frac{1}{2\pi}\partial_x E_z\nonumber\\
\partial^t j^{(xy)}_{0}+\partial^{z} j^{(xy)}_{z}&=&-\frac{1}{2\pi}E_{z},
\end{eqnarray}\noindent for the charge currents where again we have only included the result in the last equation for the hinge at $(x,y)=(L,L).$ For the dipole currents we find
\begin{eqnarray}
\partial^t j^{(x\pm),y}_{0}+\partial^z j^{(x\pm),y}_z&=&\mp\frac{1}{2\pi} E_z\nonumber\\
\partial^t j^{(y\pm),x}_{0}+\partial^z j^{(y\pm),x}_z&=&\mp\frac{1}{2\pi} E_z.
\end{eqnarray}
\subsection{Top and bottom surface conservation laws}
 From Eq. \ref{eq:zsurface} we find that the action localized on, say the $z=L$ surface, generated by the dipolar Chern-Simons term is 
\begin{align} 
&S|_{z=L}=\frac{1}{8\pi}\int dx dy dt[ (\partial_x A_y +\partial_y A_x)  A_0].
\label{zsur}
\end{align} We can extract the collection of charge and dipole densities and currents. For the charge we have
\begin{align}
    &\rho=\frac{1}{8\pi}(\partial_y A_x+\partial_x A_y)\nonumber\\
    &j_x=-\frac{1}{8\pi}\partial_y A_0\nonumber\\
    &j_y=-\frac{1}{8\pi}\partial_x A_0.\label{eq:chargetopresponse}
    \end{align} In addition, the density and currents of x-dipoles on the surface can be extracted as,
\begin{align} 
&j_0^x=\frac{\delta S}{\delta (\partial_x A_0)}=-\frac{A_y}{8\pi}\nonumber\\
&j_y^x= \frac{\delta S}{\delta (\partial_x A_y)}=\frac{A_0}{8\pi}\nonumber\\
&j_x^x=\frac{\delta S}{\delta (\partial_x A_x)}=0.
\end{align}
Similar definitions apply for the density and currents of dipoles  oriented in the y-direction, and  we find
\begin{align}
&j_0^y=-\frac{A_x}{8\pi}\nonumber\\
&j_x^y=\frac{A_0}{8\pi}\nonumber\\
&j_y^y=0.
\end{align}
Similar to the $x$ and $y$ surfaces we find that these currents are all anomalous
\begin{align} 
&\partial^\mu j_\mu=-\frac{1}{8\pi}(\partial_x E_y +\partial_y E_x)\nonumber\\
&\partial^{\mu}j^x_{\mu}=\frac{1}{8\pi}E_y\nonumber\\
&\partial^{\mu}j^y_{\mu}=\frac{1}{8\pi}E_x,\nonumber
\end{align}\noindent where $\mu=x,y,t.$ 

Using the same arguments we have applied earlier, we know that we need to augment these anomalies by the consistent anomalies of the boundary degrees of freedom, and we ultimately find \begin{align} 
&\partial^\mu j_\mu=\mp\frac{1}{4\pi}(\partial_x E_y +\partial_y E_x)\nonumber\\
&\partial^{\mu}j^x_{\mu}=\pm\frac{1}{4\pi}E_y\nonumber\\
&\partial^{\mu}j^y_{\mu}=\pm\frac{1}{4\pi}E_x.
\end{align}

\section{Absence of dipole Chern-Simons effect in the bulk}
\label{app:no-go}

In Sec.~\ref{sec:rank2}, we conclude that the dipole Chern-Simons action is zero in the bulk so it does not produce any bulk effects like  the usual 2D Chern-Simons theory does. In this appendix, we provide a no-go theorem to demonstrate that the mixed rank Chern-Simons cannot induce any bulk statistical effects provided the gauge structure of the mixed rank Chern-Simons theory is fixed as in Eq.~\ref{rank}. As discussed in Section ~\ref{sec:rank2}, the fracton charge current couples with the mixed rank gauge field as,
\begin{align}
    \mathcal{L}=J_0 A_0+J_{xy}A_{xy}+J_z A_z.
\end{align}
This current obeys a special conservation law,
\begin{align}
\partial^0 J_0-\partial^x \partial^y J_{xy}+\partial^z J_z=0,
\label{aaa}
\end{align}\noindent where the sign on the second term arises from the two derivative gauge transformation of $A_{xy}.$

We can further re-interpret this conserved current in terms of derivatives of a dual gauge field,
\begin{align}
&J_0=-\partial_x \partial_y \mathcal{B}_z-\partial_z \mathcal{B}_{xy}\nonumber\\
&J_{xy}=\partial_z \mathcal{B}_0-\partial_t \mathcal{B}_{z}\nonumber\\
&J_z=\partial_t \mathcal{B}_{xy}+\partial_x \partial_y \mathcal{B}_0.
\end{align}
The mixed rank gauge field $\mathcal{B}$ has the gauge transformations
\begin{align}
&\mathcal{B}_0\rightarrow \mathcal{B}_0+\partial_t \beta \nonumber\\
&\mathcal{B}_{xy}\rightarrow \mathcal{B}_{xy}-\partial_x \partial_y \beta \nonumber\\
&\mathcal{B}_z\rightarrow \mathcal{B}_z+\partial_z \beta.
\end{align}
Such a  gauge field minimally couples with a vortex current via
\begin{align}
&\mathcal{L}_v=J^v_0 \mathcal{B}^0+J^v_{xy} \mathcal{B}^{xy}+J^v_z \mathcal{B}^z, \end{align} with corresponding conservation law \begin{align}
&\partial^0 J^v_0+\partial^x \partial^y J^v_{xy}+\partial^z J^v_z=0.
\end{align}
Interestingly we find that the vortex current has a different conservation law compared to the original current in Eq.~\ref{aaa}.

If the mixed rank Chern-Simons term manifests a bulk statistical effect, the  mixed rank gauge flux should bind mixed-rank gauge charge and the statistical effects would arise from the flux-charge binding give rise to fractional statistics as a consequence of an Aharonov-Bohm effect. However, as the original matter and its dual vortex matter contain distinct current conservation laws, it is impossible to bind the charge current with the gauge flux in a consistent way, thus  a statistical effect is absent.

The same argument can be extended to the rank-1 dipolar Chern-Simons theory we explored in Eq.~\ref{fff}. As we emphasized, such a theory does not induce any bulk effect. Indeed, we can show that no consistent bulk statistical effects can be generated via a similar no-go theorem. As the magnetic flux is a loop-object in 3D, by symmetry and dimensional analysis, a $C_4 \mathcal{T}$ invariant theory can only have a flux-dipole binding effect along the lines of
\begin{align}
&\rho=\frac{1}{4\pi} (\partial_x B_x-\partial_y B_y)\nonumber\\
&P_x=\frac{1}{4\pi} B_x,~P_y=\frac{-1}{4\pi} B_y
\end{align}
These equations would indicate that flux loops bind with a `dipole charge' parallel to the flux loop. The binding can be alternatively viewed as a charge polarization of the flux line.
However, as the flux in the bulk must form closed loops, each bulk flux loop should be charge neutral so there is no statistical effect between two fluxes. When the flux line hits the boundary, the effective open flux end carries an isolated charge living at the end of the dipole and this induces statistics between charge and flux on the surfaces, thus indicating why surface responses, but not bulk responses are viable in the dipolar Chern-Simons response.


\begin{thebibliography}{92}%
\makeatletter
\providecommand \@ifxundefined [1]{%
 \@ifx{#1\undefined}
}%
\providecommand \@ifnum [1]{%
 \ifnum #1\expandafter \@firstoftwo
 \else \expandafter \@secondoftwo
 \fi
}%
\providecommand \@ifx [1]{%
 \ifx #1\expandafter \@firstoftwo
 \else \expandafter \@secondoftwo
 \fi
}%
\providecommand \natexlab [1]{#1}%
\providecommand \enquote  [1]{``#1''}%
\providecommand \bibnamefont  [1]{#1}%
\providecommand \bibfnamefont [1]{#1}%
\providecommand \citenamefont [1]{#1}%
\providecommand \href@noop [0]{\@secondoftwo}%
\providecommand \href [0]{\begingroup \@sanitize@url \@href}%
\providecommand \@href[1]{\@@startlink{#1}\@@href}%
\providecommand \@@href[1]{\endgroup#1\@@endlink}%
\providecommand \@sanitize@url [0]{\catcode `\\12\catcode `\$12\catcode
  `\&12\catcode `\#12\catcode `\^12\catcode `\_12\catcode `\%12\relax}%
\providecommand \@@startlink[1]{}%
\providecommand \@@endlink[0]{}%
\providecommand \url  [0]{\begingroup\@sanitize@url \@url }%
\providecommand \@url [1]{\endgroup\@href {#1}{\urlprefix }}%
\providecommand \urlprefix  [0]{URL }%
\providecommand \Eprint [0]{\href }%
\providecommand \doibase [0]{http://dx.doi.org/}%
\providecommand \selectlanguage [0]{\@gobble}%
\providecommand \bibinfo  [0]{\@secondoftwo}%
\providecommand \bibfield  [0]{\@secondoftwo}%
\providecommand \translation [1]{[#1]}%
\providecommand \BibitemOpen [0]{}%
\providecommand \bibitemStop [0]{}%
\providecommand \bibitemNoStop [0]{.\EOS\space}%
\providecommand \EOS [0]{\spacefactor3000\relax}%
\providecommand \BibitemShut  [1]{\csname bibitem#1\endcsname}%
\let\auto@bib@innerbib\@empty
%</preamble>
\bibitem [{\citenamefont {Fu}(2011)}]{fu2011topological}%
  \BibitemOpen
  \bibfield  {author} {\bibinfo {author} {\bibfnamefont {L.}~\bibnamefont
  {Fu}},\ }\href@noop {} {\bibfield  {journal} {\bibinfo  {journal} {Physical
  Review Letters}\ }\textbf {\bibinfo {volume} {106}},\ \bibinfo {pages}
  {106802} (\bibinfo {year} {2011})}\BibitemShut {NoStop}%
\bibitem [{\citenamefont {Hughes}\ \emph {et~al.}(2011)\citenamefont {Hughes},
  \citenamefont {Prodan},\ and\ \citenamefont {Bernevig}}]{hughes2011}%
  \BibitemOpen
  \bibfield  {author} {\bibinfo {author} {\bibfnamefont {T.~L.}\ \bibnamefont
  {Hughes}}, \bibinfo {author} {\bibfnamefont {E.}~\bibnamefont {Prodan}}, \
  and\ \bibinfo {author} {\bibfnamefont {B.~A.}\ \bibnamefont {Bernevig}},\
  }\href@noop {} {\bibfield  {journal} {\bibinfo  {journal} {Phys. Rev. B}\
  }\textbf {\bibinfo {volume} {83}},\ \bibinfo {pages} {245132} (\bibinfo
  {year} {2011})}\BibitemShut {NoStop}%
\bibitem [{\citenamefont {Hsieh}\ \emph {et~al.}(2012)\citenamefont {Hsieh},
  \citenamefont {Lin}, \citenamefont {Liu}, \citenamefont {Duan}, \citenamefont
  {Bansil},\ and\ \citenamefont {Fu}}]{hsieh2012topological}%
  \BibitemOpen
  \bibfield  {author} {\bibinfo {author} {\bibfnamefont {T.~H.}\ \bibnamefont
  {Hsieh}}, \bibinfo {author} {\bibfnamefont {H.}~\bibnamefont {Lin}}, \bibinfo
  {author} {\bibfnamefont {J.}~\bibnamefont {Liu}}, \bibinfo {author}
  {\bibfnamefont {W.}~\bibnamefont {Duan}}, \bibinfo {author} {\bibfnamefont
  {A.}~\bibnamefont {Bansil}}, \ and\ \bibinfo {author} {\bibfnamefont
  {L.}~\bibnamefont {Fu}},\ }\href@noop {} {\bibfield  {journal} {\bibinfo
  {journal} {Nature communications}\ }\textbf {\bibinfo {volume} {3}},\
  \bibinfo {pages} {982} (\bibinfo {year} {2012})}\BibitemShut {NoStop}%
\bibitem [{\citenamefont {Cheng}\ \emph {et~al.}(2016)\citenamefont {Cheng},
  \citenamefont {Zaletel}, \citenamefont {Barkeshli}, \citenamefont
  {Vishwanath},\ and\ \citenamefont {Bonderson}}]{cheng2016translational}%
  \BibitemOpen
  \bibfield  {author} {\bibinfo {author} {\bibfnamefont {M.}~\bibnamefont
  {Cheng}}, \bibinfo {author} {\bibfnamefont {M.}~\bibnamefont {Zaletel}},
  \bibinfo {author} {\bibfnamefont {M.}~\bibnamefont {Barkeshli}}, \bibinfo
  {author} {\bibfnamefont {A.}~\bibnamefont {Vishwanath}}, \ and\ \bibinfo
  {author} {\bibfnamefont {P.}~\bibnamefont {Bonderson}},\ }\href@noop {}
  {\bibfield  {journal} {\bibinfo  {journal} {Physical Review X}\ }\textbf
  {\bibinfo {volume} {6}},\ \bibinfo {pages} {041068} (\bibinfo {year}
  {2016})}\BibitemShut {NoStop}%
\bibitem [{\citenamefont {Ando}\ and\ \citenamefont
  {Fu}(2015)}]{ando2015topological}%
  \BibitemOpen
  \bibfield  {author} {\bibinfo {author} {\bibfnamefont {Y.}~\bibnamefont
  {Ando}}\ and\ \bibinfo {author} {\bibfnamefont {L.}~\bibnamefont {Fu}},\
  }\href@noop {} {\bibfield  {journal} {\bibinfo  {journal} {Annu. Rev.
  Condens. Matter Phys.}\ }\textbf {\bibinfo {volume} {6}},\ \bibinfo {pages}
  {361} (\bibinfo {year} {2015})}\BibitemShut {NoStop}%
\bibitem [{\citenamefont {Slager}\ \emph {et~al.}(2013)\citenamefont {Slager},
  \citenamefont {Mesaros}, \citenamefont {Juri{\v{c}}i{\'c}},\ and\
  \citenamefont {Zaanen}}]{slager2013space}%
  \BibitemOpen
  \bibfield  {author} {\bibinfo {author} {\bibfnamefont {R.-J.}\ \bibnamefont
  {Slager}}, \bibinfo {author} {\bibfnamefont {A.}~\bibnamefont {Mesaros}},
  \bibinfo {author} {\bibfnamefont {V.}~\bibnamefont {Juri{\v{c}}i{\'c}}}, \
  and\ \bibinfo {author} {\bibfnamefont {J.}~\bibnamefont {Zaanen}},\
  }\href@noop {} {\bibfield  {journal} {\bibinfo  {journal} {Nature Physics}\
  }\textbf {\bibinfo {volume} {9}},\ \bibinfo {pages} {98} (\bibinfo {year}
  {2013})}\BibitemShut {NoStop}%
\bibitem [{\citenamefont {Hong}\ and\ \citenamefont
  {Fu}(2017)}]{hong2017topological}%
  \BibitemOpen
  \bibfield  {author} {\bibinfo {author} {\bibfnamefont {S.}~\bibnamefont
  {Hong}}\ and\ \bibinfo {author} {\bibfnamefont {L.}~\bibnamefont {Fu}},\
  }\href@noop {} {\bibfield  {journal} {\bibinfo  {journal} {arXiv preprint
  arXiv:1707.02594}\ } (\bibinfo {year} {2017})}\BibitemShut {NoStop}%
\bibitem [{\citenamefont {Qi}\ and\ \citenamefont
  {Fu}(2015)}]{qi2015anomalous}%
  \BibitemOpen
  \bibfield  {author} {\bibinfo {author} {\bibfnamefont {Y.}~\bibnamefont
  {Qi}}\ and\ \bibinfo {author} {\bibfnamefont {L.}~\bibnamefont {Fu}},\
  }\href@noop {} {\bibfield  {journal} {\bibinfo  {journal} {Physical review
  letters}\ }\textbf {\bibinfo {volume} {115}},\ \bibinfo {pages} {236801}
  (\bibinfo {year} {2015})}\BibitemShut {NoStop}%
\bibitem [{\citenamefont {Huang}\ \emph {et~al.}(2017)\citenamefont {Huang},
  \citenamefont {Song}, \citenamefont {Huang},\ and\ \citenamefont
  {Hermele}}]{huang2017building}%
  \BibitemOpen
  \bibfield  {author} {\bibinfo {author} {\bibfnamefont {S.-J.}\ \bibnamefont
  {Huang}}, \bibinfo {author} {\bibfnamefont {H.}~\bibnamefont {Song}},
  \bibinfo {author} {\bibfnamefont {Y.-P.}\ \bibnamefont {Huang}}, \ and\
  \bibinfo {author} {\bibfnamefont {M.}~\bibnamefont {Hermele}},\ }\href@noop
  {} {\bibfield  {journal} {\bibinfo  {journal} {Physical Review B}\ }\textbf
  {\bibinfo {volume} {96}},\ \bibinfo {pages} {205106} (\bibinfo {year}
  {2017})}\BibitemShut {NoStop}%
\bibitem [{\citenamefont {Teo}\ and\ \citenamefont
  {Hughes}(2013)}]{teo2013existence}%
  \BibitemOpen
  \bibfield  {author} {\bibinfo {author} {\bibfnamefont {J.~C.}\ \bibnamefont
  {Teo}}\ and\ \bibinfo {author} {\bibfnamefont {T.~L.}\ \bibnamefont
  {Hughes}},\ }\href@noop {} {\bibfield  {journal} {\bibinfo  {journal}
  {Physical review letters}\ }\textbf {\bibinfo {volume} {111}},\ \bibinfo
  {pages} {047006} (\bibinfo {year} {2013})}\BibitemShut {NoStop}%
\bibitem [{\citenamefont {Song}\ \emph
  {et~al.}(2017{\natexlab{a}})\citenamefont {Song}, \citenamefont {Huang},
  \citenamefont {Fu},\ and\ \citenamefont {Hermele}}]{song2017topological}%
  \BibitemOpen
  \bibfield  {author} {\bibinfo {author} {\bibfnamefont {H.}~\bibnamefont
  {Song}}, \bibinfo {author} {\bibfnamefont {S.-J.}\ \bibnamefont {Huang}},
  \bibinfo {author} {\bibfnamefont {L.}~\bibnamefont {Fu}}, \ and\ \bibinfo
  {author} {\bibfnamefont {M.}~\bibnamefont {Hermele}},\ }\href@noop {}
  {\bibfield  {journal} {\bibinfo  {journal} {Physical Review X}\ }\textbf
  {\bibinfo {volume} {7}},\ \bibinfo {pages} {011020} (\bibinfo {year}
  {2017}{\natexlab{a}})}\BibitemShut {NoStop}%
\bibitem [{\citenamefont {Watanabe}\ \emph {et~al.}(2017)\citenamefont
  {Watanabe}, \citenamefont {Po},\ and\ \citenamefont
  {Vishwanath}}]{watanabe2017structure}%
  \BibitemOpen
  \bibfield  {author} {\bibinfo {author} {\bibfnamefont {H.}~\bibnamefont
  {Watanabe}}, \bibinfo {author} {\bibfnamefont {H.~C.}\ \bibnamefont {Po}}, \
  and\ \bibinfo {author} {\bibfnamefont {A.}~\bibnamefont {Vishwanath}},\
  }\href@noop {} {\bibfield  {journal} {\bibinfo  {journal} {arXiv preprint
  arXiv:1707.01903}\ } (\bibinfo {year} {2017})}\BibitemShut {NoStop}%
\bibitem [{\citenamefont {Po}\ \emph {et~al.}(2017)\citenamefont {Po},
  \citenamefont {Vishwanath},\ and\ \citenamefont {Watanabe}}]{po2017symmetry}%
  \BibitemOpen
  \bibfield  {author} {\bibinfo {author} {\bibfnamefont {H.~C.}\ \bibnamefont
  {Po}}, \bibinfo {author} {\bibfnamefont {A.}~\bibnamefont {Vishwanath}}, \
  and\ \bibinfo {author} {\bibfnamefont {H.}~\bibnamefont {Watanabe}},\
  }\href@noop {} {\bibfield  {journal} {\bibinfo  {journal} {Nature
  Communications}\ }\textbf {\bibinfo {volume} {8}},\ \bibinfo {pages} {50}
  (\bibinfo {year} {2017})}\BibitemShut {NoStop}%
\bibitem [{\citenamefont {Isobe}\ and\ \citenamefont
  {Fu}(2015)}]{isobe2015theory}%
  \BibitemOpen
  \bibfield  {author} {\bibinfo {author} {\bibfnamefont {H.}~\bibnamefont
  {Isobe}}\ and\ \bibinfo {author} {\bibfnamefont {L.}~\bibnamefont {Fu}},\
  }\href@noop {} {\bibfield  {journal} {\bibinfo  {journal} {Physical Review
  B}\ }\textbf {\bibinfo {volume} {92}},\ \bibinfo {pages} {081304} (\bibinfo
  {year} {2015})}\BibitemShut {NoStop}%
\bibitem [{\citenamefont {Benalcazar}\ \emph
  {et~al.}(2017{\natexlab{a}})\citenamefont {Benalcazar}, \citenamefont
  {Bernevig},\ and\ \citenamefont {Hughes}}]{benalcazar2017quantized}%
  \BibitemOpen
  \bibfield  {author} {\bibinfo {author} {\bibfnamefont {W.~A.}\ \bibnamefont
  {Benalcazar}}, \bibinfo {author} {\bibfnamefont {B.~A.}\ \bibnamefont
  {Bernevig}}, \ and\ \bibinfo {author} {\bibfnamefont {T.~L.}\ \bibnamefont
  {Hughes}},\ }\href@noop {} {\bibfield  {journal} {\bibinfo  {journal}
  {Science}\ }\textbf {\bibinfo {volume} {357}},\ \bibinfo {pages} {61}
  (\bibinfo {year} {2017}{\natexlab{a}})}\BibitemShut {NoStop}%
\bibitem [{\citenamefont {Benalcazar}\ \emph
  {et~al.}(2017{\natexlab{b}})\citenamefont {Benalcazar}, \citenamefont
  {Bernevig},\ and\ \citenamefont {Hughes}}]{benalcazar2017electric}%
  \BibitemOpen
  \bibfield  {author} {\bibinfo {author} {\bibfnamefont {W.~A.}\ \bibnamefont
  {Benalcazar}}, \bibinfo {author} {\bibfnamefont {B.~A.}\ \bibnamefont
  {Bernevig}}, \ and\ \bibinfo {author} {\bibfnamefont {T.~L.}\ \bibnamefont
  {Hughes}},\ }\href@noop {} {\bibfield  {journal} {\bibinfo  {journal}
  {Physical Review B}\ }\textbf {\bibinfo {volume} {96}},\ \bibinfo {pages}
  {245115} (\bibinfo {year} {2017}{\natexlab{b}})}\BibitemShut {NoStop}%
\bibitem [{\citenamefont {Schindler}\ \emph {et~al.}(2017)\citenamefont
  {Schindler}, \citenamefont {Cook}, \citenamefont {Vergniory}, \citenamefont
  {Wang}, \citenamefont {Parkin}, \citenamefont {Bernevig},\ and\ \citenamefont
  {Neupert}}]{schindler2017higher}%
  \BibitemOpen
  \bibfield  {author} {\bibinfo {author} {\bibfnamefont {F.}~\bibnamefont
  {Schindler}}, \bibinfo {author} {\bibfnamefont {A.~M.}\ \bibnamefont {Cook}},
  \bibinfo {author} {\bibfnamefont {M.~G.}\ \bibnamefont {Vergniory}}, \bibinfo
  {author} {\bibfnamefont {Z.}~\bibnamefont {Wang}}, \bibinfo {author}
  {\bibfnamefont {S.~S.}\ \bibnamefont {Parkin}}, \bibinfo {author}
  {\bibfnamefont {B.~A.}\ \bibnamefont {Bernevig}}, \ and\ \bibinfo {author}
  {\bibfnamefont {T.}~\bibnamefont {Neupert}},\ }\href@noop {} {\bibfield
  {journal} {\bibinfo  {journal} {arXiv preprint arXiv:1708.03636}\ } (\bibinfo
  {year} {2017})}\BibitemShut {NoStop}%
\bibitem [{\citenamefont {Langbehn}\ \emph {et~al.}(2017)\citenamefont
  {Langbehn}, \citenamefont {Peng}, \citenamefont {Trifunovic}, \citenamefont
  {von Oppen},\ and\ \citenamefont {Brouwer}}]{langbehn2017reflection}%
  \BibitemOpen
  \bibfield  {author} {\bibinfo {author} {\bibfnamefont {J.}~\bibnamefont
  {Langbehn}}, \bibinfo {author} {\bibfnamefont {Y.}~\bibnamefont {Peng}},
  \bibinfo {author} {\bibfnamefont {L.}~\bibnamefont {Trifunovic}}, \bibinfo
  {author} {\bibfnamefont {F.}~\bibnamefont {von Oppen}}, \ and\ \bibinfo
  {author} {\bibfnamefont {P.~W.}\ \bibnamefont {Brouwer}},\ }\href@noop {}
  {\bibfield  {journal} {\bibinfo  {journal} {Physical review letters}\
  }\textbf {\bibinfo {volume} {119}},\ \bibinfo {pages} {246401} (\bibinfo
  {year} {2017})}\BibitemShut {NoStop}%
\bibitem [{\citenamefont {Song}\ \emph
  {et~al.}(2017{\natexlab{b}})\citenamefont {Song}, \citenamefont {Fang},\ and\
  \citenamefont {Fang}}]{song2017d}%
  \BibitemOpen
  \bibfield  {author} {\bibinfo {author} {\bibfnamefont {Z.}~\bibnamefont
  {Song}}, \bibinfo {author} {\bibfnamefont {Z.}~\bibnamefont {Fang}}, \ and\
  \bibinfo {author} {\bibfnamefont {C.}~\bibnamefont {Fang}},\ }\href@noop {}
  {\bibfield  {journal} {\bibinfo  {journal} {Physical review letters}\
  }\textbf {\bibinfo {volume} {119}},\ \bibinfo {pages} {246402} (\bibinfo
  {year} {2017}{\natexlab{b}})}\BibitemShut {NoStop}%
\bibitem [{\citenamefont {Serra-Garcia}\ \emph {et~al.}(2018)\citenamefont
  {Serra-Garcia}, \citenamefont {Peri}, \citenamefont {S{\"u}sstrunk},
  \citenamefont {Bilal}, \citenamefont {Larsen}, \citenamefont {Villanueva},\
  and\ \citenamefont {Huber}}]{serra2018}%
  \BibitemOpen
  \bibfield  {author} {\bibinfo {author} {\bibfnamefont {M.}~\bibnamefont
  {Serra-Garcia}}, \bibinfo {author} {\bibfnamefont {V.}~\bibnamefont {Peri}},
  \bibinfo {author} {\bibfnamefont {R.}~\bibnamefont {S{\"u}sstrunk}}, \bibinfo
  {author} {\bibfnamefont {O.~R.}\ \bibnamefont {Bilal}}, \bibinfo {author}
  {\bibfnamefont {T.}~\bibnamefont {Larsen}}, \bibinfo {author} {\bibfnamefont
  {L.~G.}\ \bibnamefont {Villanueva}}, \ and\ \bibinfo {author} {\bibfnamefont
  {S.~D.}\ \bibnamefont {Huber}},\ }\href@noop {} {\bibfield  {journal}
  {\bibinfo  {journal} {Nature}\ }\textbf {\bibinfo {volume} {555}},\ \bibinfo
  {pages} {342} (\bibinfo {year} {2018})}\BibitemShut {NoStop}%
\bibitem [{\citenamefont {Peterson}\ \emph {et~al.}(2018)\citenamefont
  {Peterson}, \citenamefont {Benalcazar}, \citenamefont {Hughes},\ and\
  \citenamefont {Bahl}}]{peterson2018quantized}%
  \BibitemOpen
  \bibfield  {author} {\bibinfo {author} {\bibfnamefont {C.~W.}\ \bibnamefont
  {Peterson}}, \bibinfo {author} {\bibfnamefont {W.~A.}\ \bibnamefont
  {Benalcazar}}, \bibinfo {author} {\bibfnamefont {T.~L.}\ \bibnamefont
  {Hughes}}, \ and\ \bibinfo {author} {\bibfnamefont {G.}~\bibnamefont
  {Bahl}},\ }\href@noop {} {\bibfield  {journal} {\bibinfo  {journal} {Nature}\
  }\textbf {\bibinfo {volume} {555}},\ \bibinfo {pages} {346} (\bibinfo {year}
  {2018})}\BibitemShut {NoStop}%
\bibitem [{\citenamefont {Imhof}\ \emph {et~al.}(2018)\citenamefont {Imhof},
  \citenamefont {Berger}, \citenamefont {Bayer}, \citenamefont {Brehm},
  \citenamefont {Molenkamp}, \citenamefont {Kiessling}, \citenamefont
  {Schindler}, \citenamefont {Lee}, \citenamefont {Greiter}, \citenamefont
  {Neupert} \emph {et~al.}}]{imhof2018}%
  \BibitemOpen
  \bibfield  {author} {\bibinfo {author} {\bibfnamefont {S.}~\bibnamefont
  {Imhof}}, \bibinfo {author} {\bibfnamefont {C.}~\bibnamefont {Berger}},
  \bibinfo {author} {\bibfnamefont {F.}~\bibnamefont {Bayer}}, \bibinfo
  {author} {\bibfnamefont {J.}~\bibnamefont {Brehm}}, \bibinfo {author}
  {\bibfnamefont {L.~W.}\ \bibnamefont {Molenkamp}}, \bibinfo {author}
  {\bibfnamefont {T.}~\bibnamefont {Kiessling}}, \bibinfo {author}
  {\bibfnamefont {F.}~\bibnamefont {Schindler}}, \bibinfo {author}
  {\bibfnamefont {C.~H.}\ \bibnamefont {Lee}}, \bibinfo {author} {\bibfnamefont
  {M.}~\bibnamefont {Greiter}}, \bibinfo {author} {\bibfnamefont
  {T.}~\bibnamefont {Neupert}},  \emph {et~al.},\ }\href@noop {} {\bibfield
  {journal} {\bibinfo  {journal} {Nature Physics}\ }\textbf {\bibinfo {volume}
  {14}},\ \bibinfo {pages} {925} (\bibinfo {year} {2018})}\BibitemShut
  {NoStop}%
\bibitem [{\citenamefont {Schindler}\ \emph {et~al.}(2018)\citenamefont
  {Schindler}, \citenamefont {Cook}, \citenamefont {Vergniory}, \citenamefont
  {Wang}, \citenamefont {Parkin}, \citenamefont {Bernevig},\ and\ \citenamefont
  {Neupert}}]{schindler2018higher}%
  \BibitemOpen
  \bibfield  {author} {\bibinfo {author} {\bibfnamefont {F.}~\bibnamefont
  {Schindler}}, \bibinfo {author} {\bibfnamefont {A.~M.}\ \bibnamefont {Cook}},
  \bibinfo {author} {\bibfnamefont {M.~G.}\ \bibnamefont {Vergniory}}, \bibinfo
  {author} {\bibfnamefont {Z.}~\bibnamefont {Wang}}, \bibinfo {author}
  {\bibfnamefont {S.~S.}\ \bibnamefont {Parkin}}, \bibinfo {author}
  {\bibfnamefont {B.~A.}\ \bibnamefont {Bernevig}}, \ and\ \bibinfo {author}
  {\bibfnamefont {T.}~\bibnamefont {Neupert}},\ }\href@noop {} {\bibfield
  {journal} {\bibinfo  {journal} {Science advances}\ }\textbf {\bibinfo
  {volume} {4}},\ \bibinfo {pages} {eaat0346} (\bibinfo {year}
  {2018})}\BibitemShut {NoStop}%
\bibitem [{\citenamefont {Song}\ and\ \citenamefont
  {Schnyder}(2017)}]{song2017interaction}%
  \BibitemOpen
  \bibfield  {author} {\bibinfo {author} {\bibfnamefont {X.-Y.}\ \bibnamefont
  {Song}}\ and\ \bibinfo {author} {\bibfnamefont {A.~P.}\ \bibnamefont
  {Schnyder}},\ }\href@noop {} {\bibfield  {journal} {\bibinfo  {journal}
  {Physical Review B}\ }\textbf {\bibinfo {volume} {95}},\ \bibinfo {pages}
  {195108} (\bibinfo {year} {2017})}\BibitemShut {NoStop}%
\bibitem [{\citenamefont {You}\ \emph {et~al.}(2018{\natexlab{a}})\citenamefont
  {You}, \citenamefont {Litinski},\ and\ \citenamefont {von
  Oppen}}]{you2018higher}%
  \BibitemOpen
  \bibfield  {author} {\bibinfo {author} {\bibfnamefont {Y.}~\bibnamefont
  {You}}, \bibinfo {author} {\bibfnamefont {D.}~\bibnamefont {Litinski}}, \
  and\ \bibinfo {author} {\bibfnamefont {F.}~\bibnamefont {von Oppen}},\
  }\href@noop {} {\bibfield  {journal} {\bibinfo  {journal} {arXiv preprint
  arXiv:1810.10556}\ } (\bibinfo {year} {2018}{\natexlab{a}})}\BibitemShut
  {NoStop}%
\bibitem [{\citenamefont {Rasmussen}\ and\ \citenamefont
  {Lu}(2018{\natexlab{a}})}]{rasmussen2018intrinsically}%
  \BibitemOpen
  \bibfield  {author} {\bibinfo {author} {\bibfnamefont {A.}~\bibnamefont
  {Rasmussen}}\ and\ \bibinfo {author} {\bibfnamefont {Y.-M.}\ \bibnamefont
  {Lu}},\ }\href@noop {} {\bibfield  {journal} {\bibinfo  {journal} {arXiv
  preprint arXiv:1810.12317}\ } (\bibinfo {year}
  {2018}{\natexlab{a}})}\BibitemShut {NoStop}%
\bibitem [{\citenamefont {Rasmussen}\ and\ \citenamefont
  {Lu}(2018{\natexlab{b}})}]{rasmussen2018classification}%
  \BibitemOpen
  \bibfield  {author} {\bibinfo {author} {\bibfnamefont {A.}~\bibnamefont
  {Rasmussen}}\ and\ \bibinfo {author} {\bibfnamefont {Y.-M.}\ \bibnamefont
  {Lu}},\ }\href@noop {} {\bibfield  {journal} {\bibinfo  {journal} {arXiv
  preprint arXiv:1809.07325}\ } (\bibinfo {year}
  {2018}{\natexlab{b}})}\BibitemShut {NoStop}%
\bibitem [{\citenamefont {Thorngren}\ and\ \citenamefont
  {Else}(2018)}]{thorngren2018gauging}%
  \BibitemOpen
  \bibfield  {author} {\bibinfo {author} {\bibfnamefont {R.}~\bibnamefont
  {Thorngren}}\ and\ \bibinfo {author} {\bibfnamefont {D.~V.}\ \bibnamefont
  {Else}},\ }\href@noop {} {\bibfield  {journal} {\bibinfo  {journal} {Physical
  Review X}\ }\textbf {\bibinfo {volume} {8}},\ \bibinfo {pages} {011040}
  (\bibinfo {year} {2018})}\BibitemShut {NoStop}%
\bibitem [{\citenamefont {Benalcazar}\ \emph {et~al.}(2018)\citenamefont
  {Benalcazar}, \citenamefont {Li},\ and\ \citenamefont
  {Hughes}}]{benalcazar2018quantization}%
  \BibitemOpen
  \bibfield  {author} {\bibinfo {author} {\bibfnamefont {W.~A.}\ \bibnamefont
  {Benalcazar}}, \bibinfo {author} {\bibfnamefont {T.}~\bibnamefont {Li}}, \
  and\ \bibinfo {author} {\bibfnamefont {T.~L.}\ \bibnamefont {Hughes}},\
  }\href@noop {} {\bibfield  {journal} {\bibinfo  {journal} {arXiv preprint
  arXiv:1809.02142}\ } (\bibinfo {year} {2018})}\BibitemShut {NoStop}%
\bibitem [{\citenamefont {Araki}\ \emph {et~al.}(2019)\citenamefont {Araki},
  \citenamefont {Mizoguchi},\ and\ \citenamefont {Hatsugai}}]{araki2019mathbb}%
  \BibitemOpen
  \bibfield  {author} {\bibinfo {author} {\bibfnamefont {H.}~\bibnamefont
  {Araki}}, \bibinfo {author} {\bibfnamefont {T.}~\bibnamefont {Mizoguchi}}, \
  and\ \bibinfo {author} {\bibfnamefont {Y.}~\bibnamefont {Hatsugai}},\
  }\href@noop {} {\bibfield  {journal} {\bibinfo  {journal} {arXiv preprint
  arXiv:1906.00218}\ } (\bibinfo {year} {2019})}\BibitemShut {NoStop}%
\bibitem [{\citenamefont {Tiwari}\ \emph {et~al.}(2019)\citenamefont {Tiwari},
  \citenamefont {Li}, \citenamefont {Bernevig}, \citenamefont {Neupert},\ and\
  \citenamefont {Parameswaran}}]{tiwari2019unhinging}%
  \BibitemOpen
  \bibfield  {author} {\bibinfo {author} {\bibfnamefont {A.}~\bibnamefont
  {Tiwari}}, \bibinfo {author} {\bibfnamefont {M.-H.}\ \bibnamefont {Li}},
  \bibinfo {author} {\bibfnamefont {B.}~\bibnamefont {Bernevig}}, \bibinfo
  {author} {\bibfnamefont {T.}~\bibnamefont {Neupert}}, \ and\ \bibinfo
  {author} {\bibfnamefont {S.}~\bibnamefont {Parameswaran}},\ }\href@noop {}
  {\bibfield  {journal} {\bibinfo  {journal} {arXiv preprint arXiv:1905.11421}\
  } (\bibinfo {year} {2019})}\BibitemShut {NoStop}%
\bibitem [{\citenamefont {Wieder}\ and\ \citenamefont
  {Bernevig}(2018)}]{wieder2018axion}%
  \BibitemOpen
  \bibfield  {author} {\bibinfo {author} {\bibfnamefont {B.~J.}\ \bibnamefont
  {Wieder}}\ and\ \bibinfo {author} {\bibfnamefont {B.~A.}\ \bibnamefont
  {Bernevig}},\ }\href@noop {} {\bibfield  {journal} {\bibinfo  {journal}
  {arXiv preprint arXiv:1810.02373}\ } (\bibinfo {year} {2018})}\BibitemShut
  {NoStop}%
\bibitem [{\citenamefont {Qi}\ \emph {et~al.}(2013)\citenamefont {Qi},
  \citenamefont {Witten},\ and\ \citenamefont {Zhang}}]{axion1}%
  \BibitemOpen
  \bibfield  {author} {\bibinfo {author} {\bibfnamefont {X.-L.}\ \bibnamefont
  {Qi}}, \bibinfo {author} {\bibfnamefont {E.}~\bibnamefont {Witten}}, \ and\
  \bibinfo {author} {\bibfnamefont {S.-C.}\ \bibnamefont {Zhang}},\ }\href
  {\doibase 10.1103/PhysRevB.87.134519} {\bibfield  {journal} {\bibinfo
  {journal} {Phys. Rev. B}\ }\textbf {\bibinfo {volume} {87}},\ \bibinfo
  {pages} {134519} (\bibinfo {year} {2013})}\BibitemShut {NoStop}%
\bibitem [{\citenamefont {Qi}\ \emph {et~al.}(2008)\citenamefont {Qi},
  \citenamefont {Hughes},\ and\ \citenamefont {Zhang}}]{axion2}%
  \BibitemOpen
  \bibfield  {author} {\bibinfo {author} {\bibfnamefont {X.-L.}\ \bibnamefont
  {Qi}}, \bibinfo {author} {\bibfnamefont {T.~L.}\ \bibnamefont {Hughes}}, \
  and\ \bibinfo {author} {\bibfnamefont {S.-C.}\ \bibnamefont {Zhang}},\ }\href
  {\doibase 10.1103/PhysRevB.78.195424} {\bibfield  {journal} {\bibinfo
  {journal} {Phys. Rev. B}\ }\textbf {\bibinfo {volume} {78}},\ \bibinfo
  {pages} {195424} (\bibinfo {year} {2008})}\BibitemShut {NoStop}%
\bibitem [{\citenamefont {Laughlin}(1981)}]{laughlin1981}%
  \BibitemOpen
  \bibfield  {author} {\bibinfo {author} {\bibfnamefont {R.~B.}\ \bibnamefont
  {Laughlin}},\ }\href@noop {} {\bibfield  {journal} {\bibinfo  {journal}
  {Phys. Rev. B}\ }\textbf {\bibinfo {volume} {23}},\ \bibinfo {pages} {5632}
  (\bibinfo {year} {1981})}\BibitemShut {NoStop}%
\bibitem [{\citenamefont {Thouless}(1983{\natexlab{a}})}]{thouless1983}%
  \BibitemOpen
  \bibfield  {author} {\bibinfo {author} {\bibfnamefont {D.}~\bibnamefont
  {Thouless}},\ }\href@noop {} {\bibfield  {journal} {\bibinfo  {journal}
  {Phys. Rev. B}\ }\textbf {\bibinfo {volume} {27}},\ \bibinfo {pages} {6083}
  (\bibinfo {year} {1983}{\natexlab{a}})}\BibitemShut {NoStop}%
\bibitem [{\citenamefont {Wheeler}\ \emph {et~al.}(2018)\citenamefont
  {Wheeler}, \citenamefont {Wagner},\ and\ \citenamefont
  {Hughes}}]{wheeler2018many}%
  \BibitemOpen
  \bibfield  {author} {\bibinfo {author} {\bibfnamefont {W.~A.}\ \bibnamefont
  {Wheeler}}, \bibinfo {author} {\bibfnamefont {L.~K.}\ \bibnamefont {Wagner}},
  \ and\ \bibinfo {author} {\bibfnamefont {T.~L.}\ \bibnamefont {Hughes}},\
  }\href@noop {} {\bibfield  {journal} {\bibinfo  {journal} {arXiv preprint
  arXiv:1812.06990}\ } (\bibinfo {year} {2018})}\BibitemShut {NoStop}%
\bibitem [{\citenamefont {Kang}\ \emph {et~al.}(2018)\citenamefont {Kang},
  \citenamefont {Shiozaki},\ and\ \citenamefont {Cho}}]{kang2018many}%
  \BibitemOpen
  \bibfield  {author} {\bibinfo {author} {\bibfnamefont {B.}~\bibnamefont
  {Kang}}, \bibinfo {author} {\bibfnamefont {K.}~\bibnamefont {Shiozaki}}, \
  and\ \bibinfo {author} {\bibfnamefont {G.~Y.}\ \bibnamefont {Cho}},\
  }\href@noop {} {\bibfield  {journal} {\bibinfo  {journal} {arXiv preprint
  arXiv:1812.06999}\ } (\bibinfo {year} {2018})}\BibitemShut {NoStop}%
\bibitem [{\citenamefont {Trifunovic}\ \emph {et~al.}(2019)\citenamefont
  {Trifunovic}, \citenamefont {Ono},\ and\ \citenamefont
  {Watanabe}}]{trifunovic2019geometric}%
  \BibitemOpen
  \bibfield  {author} {\bibinfo {author} {\bibfnamefont {L.}~\bibnamefont
  {Trifunovic}}, \bibinfo {author} {\bibfnamefont {S.}~\bibnamefont {Ono}}, \
  and\ \bibinfo {author} {\bibfnamefont {H.}~\bibnamefont {Watanabe}},\
  }\href@noop {} {\bibfield  {journal} {\bibinfo  {journal} {arXiv preprint
  arXiv:1904.11394}\ } (\bibinfo {year} {2019})}\BibitemShut {NoStop}%
\bibitem [{\citenamefont {Ono}\ \emph {et~al.}(2019)\citenamefont {Ono},
  \citenamefont {Trifunovic},\ and\ \citenamefont
  {Watanabe}}]{ono2019difficulties}%
  \BibitemOpen
  \bibfield  {author} {\bibinfo {author} {\bibfnamefont {S.}~\bibnamefont
  {Ono}}, \bibinfo {author} {\bibfnamefont {L.}~\bibnamefont {Trifunovic}}, \
  and\ \bibinfo {author} {\bibfnamefont {H.}~\bibnamefont {Watanabe}},\
  }\href@noop {} {\bibfield  {journal} {\bibinfo  {journal} {arXiv preprint
  arXiv:1902.07508}\ } (\bibinfo {year} {2019})}\BibitemShut {NoStop}%
\bibitem [{\citenamefont {Raab}\ \emph {et~al.}(2005)\citenamefont {Raab},
  \citenamefont {De~Lange}, \citenamefont {de~Lange} \emph
  {et~al.}}]{raab2005}%
  \BibitemOpen
  \bibfield  {author} {\bibinfo {author} {\bibfnamefont {R.~E.}\ \bibnamefont
  {Raab}}, \bibinfo {author} {\bibfnamefont {O.~L.}\ \bibnamefont {De~Lange}},
  \bibinfo {author} {\bibfnamefont {O.~L.}\ \bibnamefont {de~Lange}},  \emph
  {et~al.},\ }\href@noop {} {\emph {\bibinfo {title} {Multipole theory in
  electromagnetism: classical, quantum, and symmetry aspects, with
  applications}}},\ Vol.\ \bibinfo {volume} {128}\ (\bibinfo  {publisher}
  {Oxford University Press on Demand},\ \bibinfo {year} {2005})\BibitemShut
  {NoStop}%
\bibitem [{\citenamefont {Shitade}\ \emph {et~al.}(2018)\citenamefont
  {Shitade}, \citenamefont {Watanabe},\ and\ \citenamefont
  {Yanase}}]{shitade2018}%
  \BibitemOpen
  \bibfield  {author} {\bibinfo {author} {\bibfnamefont {A.}~\bibnamefont
  {Shitade}}, \bibinfo {author} {\bibfnamefont {H.}~\bibnamefont {Watanabe}}, \
  and\ \bibinfo {author} {\bibfnamefont {Y.}~\bibnamefont {Yanase}},\
  }\href@noop {} {\bibfield  {journal} {\bibinfo  {journal} {Phys. Rev. B}\
  }\textbf {\bibinfo {volume} {98}},\ \bibinfo {pages} {020407} (\bibinfo
  {year} {2018})}\BibitemShut {NoStop}%
\bibitem [{\citenamefont {Gao}\ and\ \citenamefont {Xiao}(2018)}]{gao2018}%
  \BibitemOpen
  \bibfield  {author} {\bibinfo {author} {\bibfnamefont {Y.}~\bibnamefont
  {Gao}}\ and\ \bibinfo {author} {\bibfnamefont {D.}~\bibnamefont {Xiao}},\
  }\href@noop {} {\bibfield  {journal} {\bibinfo  {journal} {Phys. Rev. B}\
  }\textbf {\bibinfo {volume} {98}},\ \bibinfo {pages} {060402} (\bibinfo
  {year} {2018})}\BibitemShut {NoStop}%
\bibitem [{\citenamefont {Haah}(2011)}]{Haah2011-ny}%
  \BibitemOpen
  \bibfield  {author} {\bibinfo {author} {\bibfnamefont {J.}~\bibnamefont
  {Haah}},\ }\href@noop {} {\bibfield  {journal} {\bibinfo  {journal} {Phys.
  Rev. A}\ }\textbf {\bibinfo {volume} {83}},\ \bibinfo {pages} {042330}
  (\bibinfo {year} {2011})}\BibitemShut {NoStop}%
\bibitem [{\citenamefont {Hal{\'a}sz}\ \emph {et~al.}(2017)\citenamefont
  {Hal{\'a}sz}, \citenamefont {Hsieh},\ and\ \citenamefont
  {Balents}}]{Halasz2017-ov}%
  \BibitemOpen
  \bibfield  {author} {\bibinfo {author} {\bibfnamefont {G.~B.}\ \bibnamefont
  {Hal{\'a}sz}}, \bibinfo {author} {\bibfnamefont {T.~H.}\ \bibnamefont
  {Hsieh}}, \ and\ \bibinfo {author} {\bibfnamefont {L.}~\bibnamefont
  {Balents}},\ }\href@noop {} {\  (\bibinfo {year} {2017})},\ \Eprint
  {http://arxiv.org/abs/1707.02308} {arXiv:1707.02308 [cond-mat.str-el]}
  \BibitemShut {NoStop}%
\bibitem [{\citenamefont {Vijay}\ \emph {et~al.}(2016)\citenamefont {Vijay},
  \citenamefont {Haah},\ and\ \citenamefont {Fu}}]{Vijay2016-dr}%
  \BibitemOpen
  \bibfield  {author} {\bibinfo {author} {\bibfnamefont {S.}~\bibnamefont
  {Vijay}}, \bibinfo {author} {\bibfnamefont {J.}~\bibnamefont {Haah}}, \ and\
  \bibinfo {author} {\bibfnamefont {L.}~\bibnamefont {Fu}},\ }\href@noop {}
  {\bibfield  {journal} {\bibinfo  {journal} {Phys. Rev. B Condens. Matter}\
  }\textbf {\bibinfo {volume} {94}},\ \bibinfo {pages} {235157} (\bibinfo
  {year} {2016})}\BibitemShut {NoStop}%
\bibitem [{\citenamefont {Vijay}\ \emph {et~al.}(2015)\citenamefont {Vijay},
  \citenamefont {Haah},\ and\ \citenamefont {Fu}}]{Vijay2015-jj}%
  \BibitemOpen
  \bibfield  {author} {\bibinfo {author} {\bibfnamefont {S.}~\bibnamefont
  {Vijay}}, \bibinfo {author} {\bibfnamefont {J.}~\bibnamefont {Haah}}, \ and\
  \bibinfo {author} {\bibfnamefont {L.}~\bibnamefont {Fu}},\ }\href@noop {}
  {\bibfield  {journal} {\bibinfo  {journal} {Phys. Rev. B Condens. Matter}\
  }\textbf {\bibinfo {volume} {92}},\ \bibinfo {pages} {235136} (\bibinfo
  {year} {2015})}\BibitemShut {NoStop}%
\bibitem [{\citenamefont {Chamon}(2005)}]{Chamon2005-fc}%
  \BibitemOpen
  \bibfield  {author} {\bibinfo {author} {\bibfnamefont {C.}~\bibnamefont
  {Chamon}},\ }\href@noop {} {\bibfield  {journal} {\bibinfo  {journal} {Phys.
  Rev. Lett.}\ }\textbf {\bibinfo {volume} {94}},\ \bibinfo {pages} {040402}
  (\bibinfo {year} {2005})}\BibitemShut {NoStop}%
\bibitem [{\citenamefont {Shirley}\ \emph
  {et~al.}(2018{\natexlab{a}})\citenamefont {Shirley}, \citenamefont {Slagle},\
  and\ \citenamefont {Chen}}]{shirley2018fractional}%
  \BibitemOpen
  \bibfield  {author} {\bibinfo {author} {\bibfnamefont {W.}~\bibnamefont
  {Shirley}}, \bibinfo {author} {\bibfnamefont {K.}~\bibnamefont {Slagle}}, \
  and\ \bibinfo {author} {\bibfnamefont {X.}~\bibnamefont {Chen}},\ }\href@noop
  {} {\bibfield  {journal} {\bibinfo  {journal} {arXiv preprint
  arXiv:1806.08625}\ } (\bibinfo {year} {2018}{\natexlab{a}})}\BibitemShut
  {NoStop}%
\bibitem [{\citenamefont {Slagle}\ and\ \citenamefont
  {Kim}(2017{\natexlab{a}})}]{Slagle2017-ne}%
  \BibitemOpen
  \bibfield  {author} {\bibinfo {author} {\bibfnamefont {K.}~\bibnamefont
  {Slagle}}\ and\ \bibinfo {author} {\bibfnamefont {Y.~B.}\ \bibnamefont
  {Kim}},\ }\href@noop {} {\  (\bibinfo {year} {2017}{\natexlab{a}})},\ \Eprint
  {http://arxiv.org/abs/1704.03870} {arXiv:1704.03870 [cond-mat.str-el]}
  \BibitemShut {NoStop}%
\bibitem [{\citenamefont {Ma}\ \emph {et~al.}(2017{\natexlab{a}})\citenamefont
  {Ma}, \citenamefont {Lake}, \citenamefont {Chen},\ and\ \citenamefont
  {Hermele}}]{Ma2017-qq}%
  \BibitemOpen
  \bibfield  {author} {\bibinfo {author} {\bibfnamefont {H.}~\bibnamefont
  {Ma}}, \bibinfo {author} {\bibfnamefont {E.}~\bibnamefont {Lake}}, \bibinfo
  {author} {\bibfnamefont {X.}~\bibnamefont {Chen}}, \ and\ \bibinfo {author}
  {\bibfnamefont {M.}~\bibnamefont {Hermele}},\ }\href@noop {} {\  (\bibinfo
  {year} {2017}{\natexlab{a}})},\ \Eprint {http://arxiv.org/abs/1701.00747}
  {arXiv:1701.00747 [cond-mat.str-el]} \BibitemShut {NoStop}%
\bibitem [{\citenamefont {Hsieh}\ and\ \citenamefont
  {Hal{\'a}sz}(2017)}]{Hsieh2017-sc}%
  \BibitemOpen
  \bibfield  {author} {\bibinfo {author} {\bibfnamefont {T.~H.}\ \bibnamefont
  {Hsieh}}\ and\ \bibinfo {author} {\bibfnamefont {G.~B.}\ \bibnamefont
  {Hal{\'a}sz}},\ }\href@noop {} {\  (\bibinfo {year} {2017})},\ \Eprint
  {http://arxiv.org/abs/1703.02973} {arXiv:1703.02973 [cond-mat.str-el]}
  \BibitemShut {NoStop}%
\bibitem [{\citenamefont {Vijay}(2017)}]{Vijay2017-ey}%
  \BibitemOpen
  \bibfield  {author} {\bibinfo {author} {\bibfnamefont {S.}~\bibnamefont
  {Vijay}},\ }\href@noop {} {\  (\bibinfo {year} {2017})},\ \Eprint
  {http://arxiv.org/abs/1701.00762} {arXiv:1701.00762 [cond-mat.str-el]}
  \BibitemShut {NoStop}%
\bibitem [{\citenamefont {Slagle}\ and\ \citenamefont
  {Kim}(2017{\natexlab{b}})}]{Slagle2017-gk}%
  \BibitemOpen
  \bibfield  {author} {\bibinfo {author} {\bibfnamefont {K.}~\bibnamefont
  {Slagle}}\ and\ \bibinfo {author} {\bibfnamefont {Y.~B.}\ \bibnamefont
  {Kim}},\ }\href@noop {} {\bibfield  {journal} {\bibinfo  {journal} {Phys.
  Rev. B Condens. Matter}\ }\textbf {\bibinfo {volume} {96}},\ \bibinfo {pages}
  {195139} (\bibinfo {year} {2017}{\natexlab{b}})}\BibitemShut {NoStop}%
\bibitem [{\citenamefont {Williamson}(2016)}]{Williamson2016-lv}%
  \BibitemOpen
  \bibfield  {author} {\bibinfo {author} {\bibfnamefont {D.~J.}\ \bibnamefont
  {Williamson}},\ }\href@noop {} {\bibfield  {journal} {\bibinfo  {journal}
  {Phys. Rev. B Condens. Matter}\ }\textbf {\bibinfo {volume} {94}},\ \bibinfo
  {pages} {155128} (\bibinfo {year} {2016})}\BibitemShut {NoStop}%
\bibitem [{\citenamefont {Ma}\ \emph {et~al.}(2017{\natexlab{b}})\citenamefont
  {Ma}, \citenamefont {Schmitz}, \citenamefont {Parameswaran}, \citenamefont
  {Hermele},\ and\ \citenamefont {Nandkishore}}]{Ma2017-cb}%
  \BibitemOpen
  \bibfield  {author} {\bibinfo {author} {\bibfnamefont {H.}~\bibnamefont
  {Ma}}, \bibinfo {author} {\bibfnamefont {A.~T.}\ \bibnamefont {Schmitz}},
  \bibinfo {author} {\bibfnamefont {S.~A.}\ \bibnamefont {Parameswaran}},
  \bibinfo {author} {\bibfnamefont {M.}~\bibnamefont {Hermele}}, \ and\
  \bibinfo {author} {\bibfnamefont {R.~M.}\ \bibnamefont {Nandkishore}},\
  }\href@noop {} {\  (\bibinfo {year} {2017}{\natexlab{b}})},\ \Eprint
  {http://arxiv.org/abs/1710.01744} {arXiv:1710.01744 [cond-mat.str-el]}
  \BibitemShut {NoStop}%
\bibitem [{\citenamefont {You}\ \emph {et~al.}(2019{\natexlab{a}})\citenamefont
  {You}, \citenamefont {Bi},\ and\ \citenamefont {Pretko}}]{you2019emergent}%
  \BibitemOpen
  \bibfield  {author} {\bibinfo {author} {\bibfnamefont {Y.}~\bibnamefont
  {You}}, \bibinfo {author} {\bibfnamefont {Z.}~\bibnamefont {Bi}}, \ and\
  \bibinfo {author} {\bibfnamefont {M.}~\bibnamefont {Pretko}},\ }\href@noop {}
  {\bibfield  {journal} {\bibinfo  {journal} {arXiv preprint arXiv:1908.08540}\
  } (\bibinfo {year} {2019}{\natexlab{a}})}\BibitemShut {NoStop}%
\bibitem [{\citenamefont {Pretko}\ and\ \citenamefont
  {Radzihovsky}(2017)}]{pretko2017fracton}%
  \BibitemOpen
  \bibfield  {author} {\bibinfo {author} {\bibfnamefont {M.}~\bibnamefont
  {Pretko}}\ and\ \bibinfo {author} {\bibfnamefont {L.}~\bibnamefont
  {Radzihovsky}},\ }\href@noop {} {\bibfield  {journal} {\bibinfo  {journal}
  {arXiv preprint arXiv:1711.11044}\ } (\bibinfo {year} {2017})}\BibitemShut
  {NoStop}%
\bibitem [{\citenamefont {Ma}\ \emph {et~al.}(2018)\citenamefont {Ma},
  \citenamefont {Hermele},\ and\ \citenamefont {Chen}}]{ma2018fracton}%
  \BibitemOpen
  \bibfield  {author} {\bibinfo {author} {\bibfnamefont {H.}~\bibnamefont
  {Ma}}, \bibinfo {author} {\bibfnamefont {M.}~\bibnamefont {Hermele}}, \ and\
  \bibinfo {author} {\bibfnamefont {X.}~\bibnamefont {Chen}},\ }\href@noop {}
  {\bibfield  {journal} {\bibinfo  {journal} {arXiv preprint arXiv:1802.10108}\
  } (\bibinfo {year} {2018})}\BibitemShut {NoStop}%
\bibitem [{\citenamefont {Bulmash}\ and\ \citenamefont
  {Barkeshli}(2018)}]{bulmash2018higgs}%
  \BibitemOpen
  \bibfield  {author} {\bibinfo {author} {\bibfnamefont {D.}~\bibnamefont
  {Bulmash}}\ and\ \bibinfo {author} {\bibfnamefont {M.}~\bibnamefont
  {Barkeshli}},\ }\href@noop {} {\bibfield  {journal} {\bibinfo  {journal}
  {arXiv preprint arXiv:1802.10099}\ } (\bibinfo {year} {2018})}\BibitemShut
  {NoStop}%
\bibitem [{\citenamefont {Slagle}\ and\ \citenamefont
  {Kim}(2017{\natexlab{c}})}]{Slagle2017-la}%
  \BibitemOpen
  \bibfield  {author} {\bibinfo {author} {\bibfnamefont {K.}~\bibnamefont
  {Slagle}}\ and\ \bibinfo {author} {\bibfnamefont {Y.~B.}\ \bibnamefont
  {Kim}},\ }\href@noop {} {\  (\bibinfo {year} {2017}{\natexlab{c}})},\ \Eprint
  {http://arxiv.org/abs/1712.04511} {arXiv:1712.04511 [cond-mat.str-el]}
  \BibitemShut {NoStop}%
\bibitem [{\citenamefont {Shirley}\ \emph {et~al.}(2017)\citenamefont
  {Shirley}, \citenamefont {Slagle}, \citenamefont {Wang},\ and\ \citenamefont
  {Chen}}]{shirley2017fracton}%
  \BibitemOpen
  \bibfield  {author} {\bibinfo {author} {\bibfnamefont {W.}~\bibnamefont
  {Shirley}}, \bibinfo {author} {\bibfnamefont {K.}~\bibnamefont {Slagle}},
  \bibinfo {author} {\bibfnamefont {Z.}~\bibnamefont {Wang}}, \ and\ \bibinfo
  {author} {\bibfnamefont {X.}~\bibnamefont {Chen}},\ }\href@noop {} {\bibfield
   {journal} {\bibinfo  {journal} {arXiv preprint arXiv:1712.05892}\ }
  (\bibinfo {year} {2017})}\BibitemShut {NoStop}%
\bibitem [{\citenamefont {Prem}\ \emph {et~al.}(2018)\citenamefont {Prem},
  \citenamefont {Vijay}, \citenamefont {Chou}, \citenamefont {Pretko},\ and\
  \citenamefont {Nandkishore}}]{prem2018pinch}%
  \BibitemOpen
  \bibfield  {author} {\bibinfo {author} {\bibfnamefont {A.}~\bibnamefont
  {Prem}}, \bibinfo {author} {\bibfnamefont {S.}~\bibnamefont {Vijay}},
  \bibinfo {author} {\bibfnamefont {Y.-Z.}\ \bibnamefont {Chou}}, \bibinfo
  {author} {\bibfnamefont {M.}~\bibnamefont {Pretko}}, \ and\ \bibinfo {author}
  {\bibfnamefont {R.~M.}\ \bibnamefont {Nandkishore}},\ }\href@noop {}
  {\bibfield  {journal} {\bibinfo  {journal} {arXiv preprint arXiv:1806.04148}\
  } (\bibinfo {year} {2018})}\BibitemShut {NoStop}%
\bibitem [{\citenamefont {Slagle}\ \emph {et~al.}(2018)\citenamefont {Slagle},
  \citenamefont {Prem},\ and\ \citenamefont {Pretko}}]{slagle2018symmetric}%
  \BibitemOpen
  \bibfield  {author} {\bibinfo {author} {\bibfnamefont {K.}~\bibnamefont
  {Slagle}}, \bibinfo {author} {\bibfnamefont {A.}~\bibnamefont {Prem}}, \ and\
  \bibinfo {author} {\bibfnamefont {M.}~\bibnamefont {Pretko}},\ }\href@noop {}
  {\bibfield  {journal} {\bibinfo  {journal} {arXiv preprint arXiv:1807.00827}\
  } (\bibinfo {year} {2018})}\BibitemShut {NoStop}%
\bibitem [{\citenamefont {Gromov}(2017)}]{gromov2017fractional}%
  \BibitemOpen
  \bibfield  {author} {\bibinfo {author} {\bibfnamefont {A.}~\bibnamefont
  {Gromov}},\ }\href@noop {} {\bibfield  {journal} {\bibinfo  {journal} {arXiv
  preprint arXiv:1712.06600}\ } (\bibinfo {year} {2017})}\BibitemShut {NoStop}%
\bibitem [{\citenamefont {Pai}\ and\ \citenamefont
  {Pretko}(2018)}]{pai2018fractonic}%
  \BibitemOpen
  \bibfield  {author} {\bibinfo {author} {\bibfnamefont {S.}~\bibnamefont
  {Pai}}\ and\ \bibinfo {author} {\bibfnamefont {M.}~\bibnamefont {Pretko}},\
  }\href@noop {} {\bibfield  {journal} {\bibinfo  {journal} {arXiv preprint
  arXiv:1804.01536}\ } (\bibinfo {year} {2018})}\BibitemShut {NoStop}%
\bibitem [{\citenamefont
  {Pretko}(2017{\natexlab{a}})}]{pretko2017subdimensional}%
  \BibitemOpen
  \bibfield  {author} {\bibinfo {author} {\bibfnamefont {M.}~\bibnamefont
  {Pretko}},\ }\href@noop {} {\bibfield  {journal} {\bibinfo  {journal}
  {Physical Review B}\ }\textbf {\bibinfo {volume} {95}},\ \bibinfo {pages}
  {115139} (\bibinfo {year} {2017}{\natexlab{a}})}\BibitemShut {NoStop}%
\bibitem [{\citenamefont {Ma}\ and\ \citenamefont
  {Pretko}(2018)}]{ma2018higher}%
  \BibitemOpen
  \bibfield  {author} {\bibinfo {author} {\bibfnamefont {H.}~\bibnamefont
  {Ma}}\ and\ \bibinfo {author} {\bibfnamefont {M.}~\bibnamefont {Pretko}},\
  }\href@noop {} {\bibfield  {journal} {\bibinfo  {journal} {arXiv preprint
  arXiv:1803.04980}\ } (\bibinfo {year} {2018})}\BibitemShut {NoStop}%
\bibitem [{\citenamefont {Pretko}(2017{\natexlab{b}})}]{pretko2017generalized}%
  \BibitemOpen
  \bibfield  {author} {\bibinfo {author} {\bibfnamefont {M.}~\bibnamefont
  {Pretko}},\ }\href@noop {} {\bibfield  {journal} {\bibinfo  {journal}
  {Physical Review B}\ }\textbf {\bibinfo {volume} {96}},\ \bibinfo {pages}
  {035119} (\bibinfo {year} {2017}{\natexlab{b}})}\BibitemShut {NoStop}%
\bibitem [{\citenamefont {Pretko}(2017{\natexlab{c}})}]{pretko2017finite}%
  \BibitemOpen
  \bibfield  {author} {\bibinfo {author} {\bibfnamefont {M.}~\bibnamefont
  {Pretko}},\ }\href@noop {} {\bibfield  {journal} {\bibinfo  {journal}
  {Physical Review B}\ }\textbf {\bibinfo {volume} {96}},\ \bibinfo {pages}
  {115102} (\bibinfo {year} {2017}{\natexlab{c}})}\BibitemShut {NoStop}%
\bibitem [{\citenamefont {You}\ \emph {et~al.}(2018{\natexlab{b}})\citenamefont
  {You}, \citenamefont {Devakul}, \citenamefont {Burnell},\ and\ \citenamefont
  {Sondhi}}]{you2018symmetric}%
  \BibitemOpen
  \bibfield  {author} {\bibinfo {author} {\bibfnamefont {Y.}~\bibnamefont
  {You}}, \bibinfo {author} {\bibfnamefont {T.}~\bibnamefont {Devakul}},
  \bibinfo {author} {\bibfnamefont {F.}~\bibnamefont {Burnell}}, \ and\
  \bibinfo {author} {\bibfnamefont {S.}~\bibnamefont {Sondhi}},\ }\href@noop {}
  {\bibfield  {journal} {\bibinfo  {journal} {arXiv preprint arXiv:1805.09800}\
  } (\bibinfo {year} {2018}{\natexlab{b}})}\BibitemShut {NoStop}%
\bibitem [{\citenamefont {Prem}\ \emph {et~al.}(2017)\citenamefont {Prem},
  \citenamefont {Haah},\ and\ \citenamefont {Nandkishore}}]{prem2017-ql}%
  \BibitemOpen
  \bibfield  {author} {\bibinfo {author} {\bibfnamefont {A.}~\bibnamefont
  {Prem}}, \bibinfo {author} {\bibfnamefont {J.}~\bibnamefont {Haah}}, \ and\
  \bibinfo {author} {\bibfnamefont {R.}~\bibnamefont {Nandkishore}},\
  }\href@noop {} {\bibfield  {journal} {\bibinfo  {journal} {Phys. Rev. B
  Condens. Matter}\ }\textbf {\bibinfo {volume} {95}},\ \bibinfo {pages}
  {155133} (\bibinfo {year} {2017})}\BibitemShut {NoStop}%
\bibitem [{\citenamefont {You}\ \emph {et~al.}(2018{\natexlab{c}})\citenamefont
  {You}, \citenamefont {Devakul}, \citenamefont {Burnell},\ and\ \citenamefont
  {Sondhi}}]{you2018subsystem}%
  \BibitemOpen
  \bibfield  {author} {\bibinfo {author} {\bibfnamefont {Y.}~\bibnamefont
  {You}}, \bibinfo {author} {\bibfnamefont {T.}~\bibnamefont {Devakul}},
  \bibinfo {author} {\bibfnamefont {F.}~\bibnamefont {Burnell}}, \ and\
  \bibinfo {author} {\bibfnamefont {S.}~\bibnamefont {Sondhi}},\ }\href@noop {}
  {\bibfield  {journal} {\bibinfo  {journal} {Physical Review B}\ }\textbf
  {\bibinfo {volume} {98}},\ \bibinfo {pages} {035112} (\bibinfo {year}
  {2018}{\natexlab{c}})}\BibitemShut {NoStop}%
\bibitem [{\citenamefont {Devakul}\ \emph
  {et~al.}(2018{\natexlab{a}})\citenamefont {Devakul}, \citenamefont
  {Williamson},\ and\ \citenamefont {You}}]{devakul2018strong}%
  \BibitemOpen
  \bibfield  {author} {\bibinfo {author} {\bibfnamefont {T.}~\bibnamefont
  {Devakul}}, \bibinfo {author} {\bibfnamefont {D.~J.}\ \bibnamefont
  {Williamson}}, \ and\ \bibinfo {author} {\bibfnamefont {Y.}~\bibnamefont
  {You}},\ }\href@noop {} {\bibfield  {journal} {\bibinfo  {journal} {arXiv
  preprint arXiv:1808.05300}\ } (\bibinfo {year}
  {2018}{\natexlab{a}})}\BibitemShut {NoStop}%
\bibitem [{\citenamefont {Devakul}\ \emph
  {et~al.}(2018{\natexlab{b}})\citenamefont {Devakul}, \citenamefont {You},
  \citenamefont {Burnell},\ and\ \citenamefont {Sondhi}}]{devakul2018fractal}%
  \BibitemOpen
  \bibfield  {author} {\bibinfo {author} {\bibfnamefont {T.}~\bibnamefont
  {Devakul}}, \bibinfo {author} {\bibfnamefont {Y.}~\bibnamefont {You}},
  \bibinfo {author} {\bibfnamefont {F.}~\bibnamefont {Burnell}}, \ and\
  \bibinfo {author} {\bibfnamefont {S.}~\bibnamefont {Sondhi}},\ }\href@noop {}
  {\bibfield  {journal} {\bibinfo  {journal} {arXiv preprint arXiv:1805.04097}\
  } (\bibinfo {year} {2018}{\natexlab{b}})}\BibitemShut {NoStop}%
\bibitem [{\citenamefont {{Dubinkin}}\ and\ \citenamefont
  {{Hughes}}(2018)}]{2018arXiv180709781D}%
  \BibitemOpen
  \bibfield  {author} {\bibinfo {author} {\bibfnamefont {O.}~\bibnamefont
  {{Dubinkin}}}\ and\ \bibinfo {author} {\bibfnamefont {T.~L.}\ \bibnamefont
  {{Hughes}}},\ }\href@noop {} {\bibfield  {journal} {\bibinfo  {journal}
  {ArXiv e-prints}\ } (\bibinfo {year} {2018})},\ \Eprint
  {http://arxiv.org/abs/1807.09781} {arXiv:1807.09781 [cond-mat.str-el]}
  \BibitemShut {NoStop}%
\bibitem [{\citenamefont {Xu}\ and\ \citenamefont {Moore}(2004)}]{Xu2004-oj}%
  \BibitemOpen
  \bibfield  {author} {\bibinfo {author} {\bibfnamefont {C.}~\bibnamefont
  {Xu}}\ and\ \bibinfo {author} {\bibfnamefont {J.~E.}\ \bibnamefont {Moore}},\
  }\href@noop {} {\bibfield  {journal} {\bibinfo  {journal} {Phys. Rev. Lett.}\
  }\textbf {\bibinfo {volume} {93}},\ \bibinfo {pages} {047003} (\bibinfo
  {year} {2004})}\BibitemShut {NoStop}%
\bibitem [{\citenamefont {Shirley}\ \emph
  {et~al.}(2018{\natexlab{b}})\citenamefont {Shirley}, \citenamefont {Slagle},\
  and\ \citenamefont {Chen}}]{shirley2018foliated}%
  \BibitemOpen
  \bibfield  {author} {\bibinfo {author} {\bibfnamefont {W.}~\bibnamefont
  {Shirley}}, \bibinfo {author} {\bibfnamefont {K.}~\bibnamefont {Slagle}}, \
  and\ \bibinfo {author} {\bibfnamefont {X.}~\bibnamefont {Chen}},\ }\href@noop
  {} {\bibfield  {journal} {\bibinfo  {journal} {arXiv preprint
  arXiv:1806.08679}\ } (\bibinfo {year} {2018}{\natexlab{b}})}\BibitemShut
  {NoStop}%
\bibitem [{\citenamefont {You}(2019)}]{you2019higher}%
  \BibitemOpen
  \bibfield  {author} {\bibinfo {author} {\bibfnamefont {Y.}~\bibnamefont
  {You}},\ }\href@noop {} {\bibfield  {journal} {\bibinfo  {journal} {arXiv
  preprint arXiv:1908.04299}\ } (\bibinfo {year} {2019})}\BibitemShut {NoStop}%
\bibitem [{\citenamefont {Affleck}\ \emph {et~al.}(1988)\citenamefont
  {Affleck}, \citenamefont {Kennedy}, \citenamefont {Lieb},\ and\ \citenamefont
  {Tasaki}}]{affleck1988valence}%
  \BibitemOpen
  \bibfield  {author} {\bibinfo {author} {\bibfnamefont {I.}~\bibnamefont
  {Affleck}}, \bibinfo {author} {\bibfnamefont {T.}~\bibnamefont {Kennedy}},
  \bibinfo {author} {\bibfnamefont {E.~H.}\ \bibnamefont {Lieb}}, \ and\
  \bibinfo {author} {\bibfnamefont {H.}~\bibnamefont {Tasaki}},\ }in\
  \href@noop {} {\emph {\bibinfo {booktitle} {Condensed matter physics and
  exactly soluble models}}}\ (\bibinfo  {publisher} {Springer},\ \bibinfo
  {year} {1988})\ pp.\ \bibinfo {pages} {253--304}\BibitemShut {NoStop}%
\bibitem [{\citenamefont {Dubinkin}\ and\ \citenamefont
  {Hughes}(2018)}]{dubinkin2018higher}%
  \BibitemOpen
  \bibfield  {author} {\bibinfo {author} {\bibfnamefont {O.}~\bibnamefont
  {Dubinkin}}\ and\ \bibinfo {author} {\bibfnamefont {T.~L.}\ \bibnamefont
  {Hughes}},\ }\href@noop {} {\bibfield  {journal} {\bibinfo  {journal} {arXiv
  preprint arXiv:1807.09781}\ } (\bibinfo {year} {2018})}\BibitemShut {NoStop}%
\bibitem [{\citenamefont {Goldstone}\ and\ \citenamefont
  {Wilczek}(1981)}]{goldstone1981fractional}%
  \BibitemOpen
  \bibfield  {author} {\bibinfo {author} {\bibfnamefont {J.}~\bibnamefont
  {Goldstone}}\ and\ \bibinfo {author} {\bibfnamefont {F.}~\bibnamefont
  {Wilczek}},\ }\href@noop {} {\bibfield  {journal} {\bibinfo  {journal}
  {Physical Review Letters}\ }\textbf {\bibinfo {volume} {47}},\ \bibinfo
  {pages} {986} (\bibinfo {year} {1981})}\BibitemShut {NoStop}%
\bibitem [{\citenamefont {Qi}\ and\ \citenamefont
  {Zhang}(2011)}]{qi2011topological}%
  \BibitemOpen
  \bibfield  {author} {\bibinfo {author} {\bibfnamefont {X.-L.}\ \bibnamefont
  {Qi}}\ and\ \bibinfo {author} {\bibfnamefont {S.-C.}\ \bibnamefont {Zhang}},\
  }\href@noop {} {\bibfield  {journal} {\bibinfo  {journal} {Reviews of Modern
  Physics}\ }\textbf {\bibinfo {volume} {83}},\ \bibinfo {pages} {1057}
  (\bibinfo {year} {2011})}\BibitemShut {NoStop}%
\bibitem [{\citenamefont {Dubinkin}\ and\ \citenamefont
  {Hughes}(2019)}]{dubinkin2019}%
  \BibitemOpen
  \bibfield  {author} {\bibinfo {author} {\bibfnamefont {O.}~\bibnamefont
  {Dubinkin}}\ and\ \bibinfo {author} {\bibfnamefont {T.~L.}\ \bibnamefont
  {Hughes}},\ }\href@noop {} {\bibfield  {journal} {\bibinfo  {journal} {(in
  preparation)}\ } (\bibinfo {year} {2019})}\BibitemShut {NoStop}%
\bibitem [{\citenamefont {Streda}(1982)}]{streda1982}%
  \BibitemOpen
  \bibfield  {author} {\bibinfo {author} {\bibfnamefont {P.}~\bibnamefont
  {Streda}},\ }\href@noop {} {\bibfield  {journal} {\bibinfo  {journal} {J Phys
  C: Solid State Physics}\ }\textbf {\bibinfo {volume} {15}},\ \bibinfo {pages}
  {L717} (\bibinfo {year} {1982})}\BibitemShut {NoStop}%
\bibitem [{\citenamefont
  {Thouless}(1983{\natexlab{b}})}]{thouless1983quantization}%
  \BibitemOpen
  \bibfield  {author} {\bibinfo {author} {\bibfnamefont {D.}~\bibnamefont
  {Thouless}},\ }\href@noop {} {\bibfield  {journal} {\bibinfo  {journal}
  {Physical Review B}\ }\textbf {\bibinfo {volume} {27}},\ \bibinfo {pages}
  {6083} (\bibinfo {year} {1983}{\natexlab{b}})}\BibitemShut {NoStop}%
\bibitem [{\citenamefont {You}\ \emph {et~al.}(2019{\natexlab{b}})\citenamefont
  {You}, \citenamefont {Devakul}, \citenamefont {Sondhi},\ and\ \citenamefont
  {Burnell}}]{you2019fractonic}%
  \BibitemOpen
  \bibfield  {author} {\bibinfo {author} {\bibfnamefont {Y.}~\bibnamefont
  {You}}, \bibinfo {author} {\bibfnamefont {T.}~\bibnamefont {Devakul}},
  \bibinfo {author} {\bibfnamefont {S.}~\bibnamefont {Sondhi}}, \ and\ \bibinfo
  {author} {\bibfnamefont {F.}~\bibnamefont {Burnell}},\ }\href@noop {}
  {\bibfield  {journal} {\bibinfo  {journal} {arXiv preprint arXiv:1904.11530}\
  } (\bibinfo {year} {2019}{\natexlab{b}})}\BibitemShut {NoStop}%
\bibitem [{\citenamefont {Callan~Jr}\ and\ \citenamefont
  {Harvey}(1985)}]{callan1985anomalies}%
  \BibitemOpen
  \bibfield  {author} {\bibinfo {author} {\bibfnamefont {C.~G.}\ \bibnamefont
  {Callan~Jr}}\ and\ \bibinfo {author} {\bibfnamefont {J.~A.}\ \bibnamefont
  {Harvey}},\ }\href@noop {} {\bibfield  {journal} {\bibinfo  {journal}
  {Nuclear Physics B}\ }\textbf {\bibinfo {volume} {250}},\ \bibinfo {pages}
  {427} (\bibinfo {year} {1985})}\BibitemShut {NoStop}%
\bibitem [{\citenamefont {Chu}\ \emph {et~al.}(2011)\citenamefont {Chu},
  \citenamefont {Shi},\ and\ \citenamefont {Shen}}]{chu2011surface}%
  \BibitemOpen
  \bibfield  {author} {\bibinfo {author} {\bibfnamefont {R.-L.}\ \bibnamefont
  {Chu}}, \bibinfo {author} {\bibfnamefont {J.}~\bibnamefont {Shi}}, \ and\
  \bibinfo {author} {\bibfnamefont {S.-Q.}\ \bibnamefont {Shen}},\ }\href@noop
  {} {\bibfield  {journal} {\bibinfo  {journal} {Physical Review B}\ }\textbf
  {\bibinfo {volume} {84}},\ \bibinfo {pages} {085312} (\bibinfo {year}
  {2011})}\BibitemShut {NoStop}%
\bibitem [{\citenamefont {Naculich}(1988)}]{naculich1988}%
  \BibitemOpen
  \bibfield  {author} {\bibinfo {author} {\bibfnamefont {S.~G.}\ \bibnamefont
  {Naculich}},\ }\href@noop {} {\bibfield  {journal} {\bibinfo  {journal}
  {Nucl. Phys. B}\ }\textbf {\bibinfo {volume} {296}},\ \bibinfo {pages} {837}
  (\bibinfo {year} {1988})}\BibitemShut {NoStop}%
\bibitem [{\citenamefont {Stone}(2012)}]{stone2012}%
  \BibitemOpen
  \bibfield  {author} {\bibinfo {author} {\bibfnamefont {M.}~\bibnamefont
  {Stone}},\ }\href@noop {} {\bibfield  {journal} {\bibinfo  {journal} {Phys.
  Rev. B}\ }\textbf {\bibinfo {volume} {85}},\ \bibinfo {pages} {184503}
  (\bibinfo {year} {2012})}\BibitemShut {NoStop}%
\bibitem [{\citenamefont {Fradkin}(2013)}]{fradkin2013field}%
  \BibitemOpen
  \bibfield  {author} {\bibinfo {author} {\bibfnamefont {E.}~\bibnamefont
  {Fradkin}},\ }\href@noop {} {\emph {\bibinfo {title} {Field theories of
  condensed matter physics}}}\ (\bibinfo  {publisher} {Cambridge University
  Press},\ \bibinfo {year} {2013})\BibitemShut {NoStop}%
\end{thebibliography}
\end{document}